\documentclass[twocolumn,times]{aastex61}

\hypersetup{linkcolor=red,citecolor=blue,filecolor=cyan,urlcolor=magenta}

\usepackage{CJKutf8}

\usepackage{amsmath}
\usepackage{amssymb}
\usepackage{amsfonts}
\usepackage{graphicx}
\usepackage{wasysym}
\usepackage{multirow}
\usepackage{mdframed}
\usepackage[T1]{fontenc}

\newcommand{\methanol}{\mbox{CH$_3$OH}}
\newcommand{\twelveco}{\mbox{$^{12}$CO}}

\newcommand{\thirteenco}{\mbox{$^{13}$CO}}
\newcommand{\ceighteeno}{\mbox{C$^{18}$O}}
\newcommand{\fmh}{\mbox{H$_2$CO}}
\newcommand{\mthc}{\mbox{CH$_3$CN}}
\newcommand{\cyacet}{\mbox{HC$_3$N}}
\newcommand{\water}{\mbox{H$_2$O}}
\newcommand{\amm}{\mbox{NH$_3$}}

\newcommand{\kms}{\mbox{km\,s$^{-1}$}}
\newcommand{\kmspc}{\mbox{km\,s$^{-1}$\,pc$^{-1}$}}
\newcommand{\sqc}{\mbox{cm$^{-2}$}}
\newcommand{\cc}{\mbox{cm$^{-3}$}}
\newcommand{\lsol}{\mbox{$L_\odot$}}
\newcommand{\msol}{\mbox{$M_\odot$}}
\newcommand{\msolpyr}{\mbox{$M_\odot$\,yr$^{-1}$}}
\newcommand{\mjypbm}{\mbox{mJy\,beam$^{-1}$}}
\newcommand{\jypbm}{\mbox{Jy\,beam$^{-1}$}}
\newcommand{\hii}{\mbox{H\,{\sc ii}}}
\newcommand{\vlsr}{\mbox{$V_\text{lsr}$}}

\received{- -, --}
\revised{- -, --}
\accepted{- -, --}
\submitjournal{the Astrophysical Journal}

\shorttitle{Filamentary Fragmentation and Accretion}
\shortauthors{Lu et al.}

\begin{document}
\begin{CJK}{UTF8}{gbsn}

\title{Filamentary Fragmentation and Accretion in High-Mass Star-Forming Molecular Clouds}

\correspondingauthor{Xing Lu}
\email{xinglv.nju@gmail.com, xing.lu@nao.ac.jp}

\author{Xing Lu (吕行)}
\affil{National Astronomical Observatory of Japan, 2-21-1 Osawa, Mitaka,Tokyo, 181-8588, Japan}

\author{Qizhou Zhang}
\affiliation{Harvard-Smithsonian Center for Astrophysics, 60 Garden Street, Cambridge, MA 02138, USA}

\author{Hauyu Baobab Liu}
\affiliation{European Southern Observatory, Karl-Schwarzschild-Stra\ss{}e 2, D-85748 Garching, Germany}

\author{Patricio Sanhueza}
\affiliation{National Astronomical Observatory of Japan, 2-21-1 Osawa, Mitaka,Tokyo, 181-8588, Japan}

\author{Ken'ichi Tatematsu}
\affiliation{National Astronomical Observatory of Japan, 2-21-1 Osawa, Mitaka,Tokyo, 181-8588, Japan}

\author{Siyi Feng}
\affiliation{Max-Planck-Institut f\"ur Extraterrestrische Physik, Gie\ss{}enbachstra\ss{}e 1, 85748 Garching bei M\"unchen, Germany}

\author{Howard A. Smith}
\affiliation{Harvard-Smithsonian Center for Astrophysics, 60 Garden Street, Cambridge, MA 02138, USA}

\author{Philip C. Myers}
\affiliation{Harvard-Smithsonian Center for Astrophysics, 60 Garden Street, Cambridge, MA 02138, USA}

\author{T.~K.~Sridharan}
\affiliation{Harvard-Smithsonian Center for Astrophysics, 60 Garden Street, Cambridge, MA 02138, USA}

\author{Qiusheng Gu}
\affiliation{School of Astronomy and Space Science, Nanjing University, Nanjing, Jiangsu 210093, China}

\begin{abstract}
Filamentary structures are ubiquitous in high-mass star-forming molecular clouds. Their relation with high-mass star formation is still to be understood. Here we report interferometric observations toward 8 filamentary high-mass star-forming clouds. A total of 50 dense cores are identified in these clouds, most of which present signatures of high-mass star formation. Five of them are not associated with any star formation indicators, hence are prestellar core candidates. Evolutionary phases of these cores and their linewidths, temperatures, \amm{} abundances, and virial parameters are found to be correlated. In a sub-sample of 4 morphologically well-defined filaments, we find that their fragmentation can not be solely explained by thermal or turbulence pressure support. We also investigate distributions of gas temperatures and non-thermal motions along the filaments, and find a spatial correlation between non-thermal linewidths and star formation activities. We find evidence of gas flows along these filaments, and derive an accretion rate along filaments of $\sim$10$^{-4}$~\msolpyr{}. These results suggest a strong relationship between massive filaments and high-mass star formation, through i) filamentary fragmentation in very early evolutionary phases to form dense cores, ii) accretion flows along filaments that are important for the growth of dense cores and protostars, and iii) enhancement of non-thermal motion in the filaments by the feedback or accretion during star formation.
\end{abstract}


\section{INTRODUCTION}\label{sec:intro}
Galactic wide surveys from \textit{MSX}, \textit{Spitzer} and \textit{Herschel} space telescopes reveal filamentary structures in massive molecular clouds \citep{carey1998,carey2000,churchwell2009,molinari2010higal,andre2010}. These filaments usually extend $\gtrsim$1~pc, contain $\gtrsim$10$^3$~\msol{} of dense gas, and break into massive fragments along their major axes \citep{jackson2010,contreras2013,li2016} where signatures of embedded high-mass ($>$8~\msol) protostars have been found \citep{wang2006,schneider2010,nguyenluong2011,busquet2013}. The ubiquity of such structures has stimulated discussion on how massive filaments are related to high-mass star formation \citep{myers2009,myers2013,myers2017,andre2014,smith2014,smith2016}.

To efficiently investigate high-mass star formation in filaments, it is necessary to resolve filaments to $\sim$0.1~pc, the spatial scale of dense cores, and combine kinematics information from spectral lines with reliable dense gas column densities from dust emission. This will allow us to study star formation activities embedded in dense cores and their relation to the environment in filaments. Such high spatial resolution observations toward massive filaments are still rare and mostly focused on case studies toward infrared dark clouds in very early evolutionary phases \citep[e.g.,][]{wang2011,wang2014,peretto2013,henshaw2014,beuther2015,zhang2015,ohashi2016,busquet2016}.

These observations have found several key features of massive filaments that distinguish them from low-mass ($\lesssim$10$^2$~\msol{}) filaments in nearby molecular clouds. Gas motions in low-mass filaments are in general subsonic \citep{pineda2011,hacar2011}, hence these filaments are velocity coherent and turbulence has been dissipated. Massive filaments, on the other hand, usually present strong turbulence and powerful outflows, suggesting ongoing dynamical processes \citep{wang2011,wang2014,zhang2015}. Low-mass filaments have been well characterized by an isothermal cylinder supported by thermal pressure \citep{pineda2011,hacar2011,hacar2016}. Massive filaments are unlikely to be solely supported by thermal pressure, and additional support by turbulent pressure or magnetic field may be required \citep{wang2011,wang2014,beuther2015,contreras2016}. Signatures of gas flows along filaments have been found, indicating accretion rates orders of magnitude higher than those in low-mass filaments \citep{peretto2013,henshaw2014,zhang2015}. These results suggest that massive filamentary infrared dark clouds, dominated by turbulent motions, are the cradle of high-mass stars and clusters.

Despite the progress on observations of filaments in very early (prestellar) evolutionary phases in the infrared dark clouds, how massive filaments evolve through the protostellar phase is less well understood. At early evolutionary phases, turbulence in the environment may not be dissipated yet and the role of protostellar feedback is not well characterized. Whether protostars would keep accreting from filaments as they evolve is also to be explored. So far, case studies of star-forming massive filaments with high spatial resolution observations have only been done toward several nearby prototypical clouds \citep[e.g., Orion A:][]{takahashi2013,kainulainen2017} and further away massive clouds at late evolutionary phases such as \hii{} regions \citep[e.g.,][]{galvan2010,galvan2013,baobab2012b}. A systematic study of a sample across protostellar phases is still missing.

To investigate the relation between filaments and high-mass star formation across a variety of evolutionary phases, we set out to study a sample of massive filaments mostly in protostellar phases, i.e., when high-mass protostars are already present. This will enable us to examine turbulence, protostellar feedback, and accretion in filaments across the star formation history.

\begin{figure*}[!t]
\centering
\includegraphics[width=0.33\textwidth]{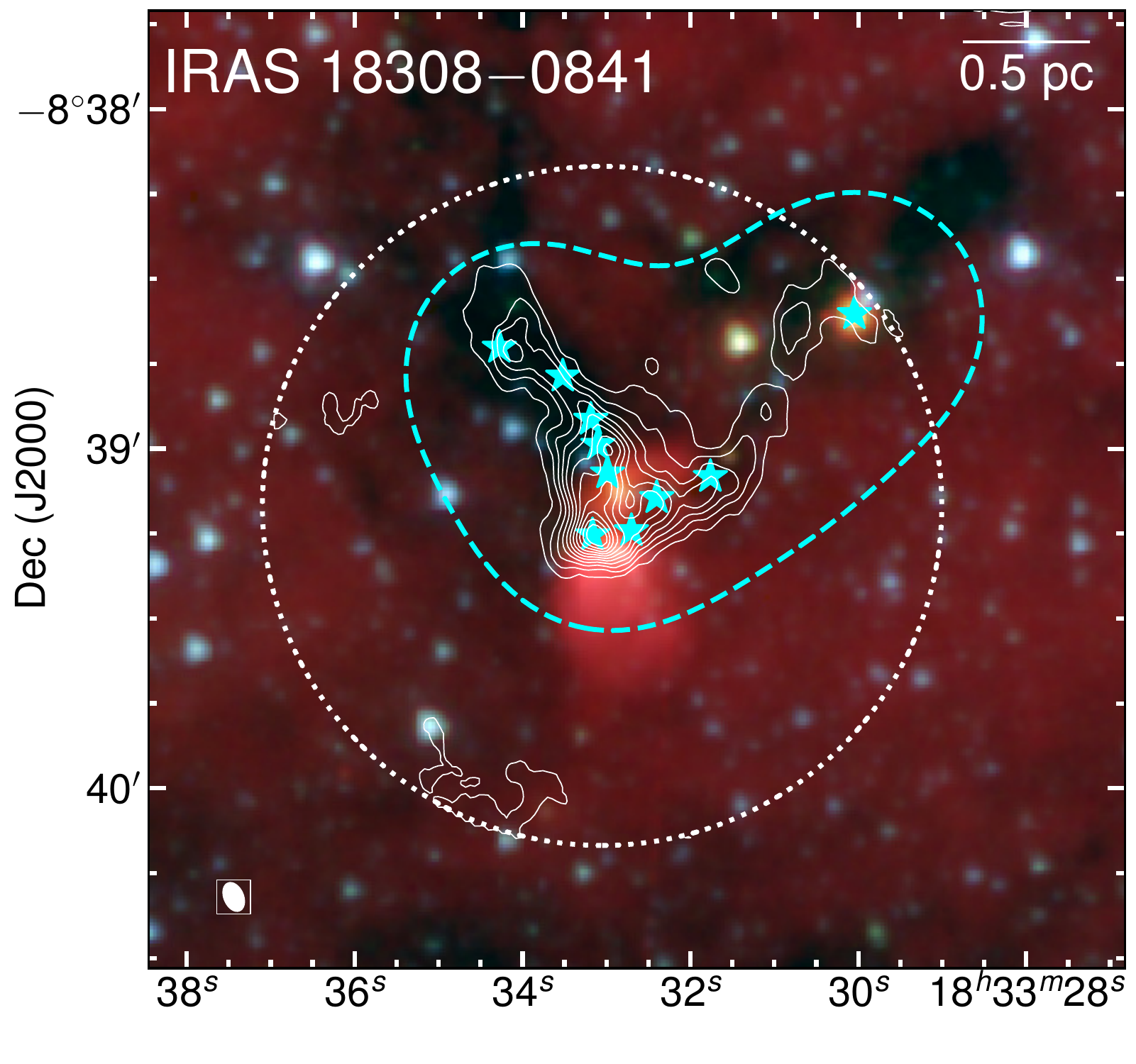}
\includegraphics[width=0.33\textwidth]{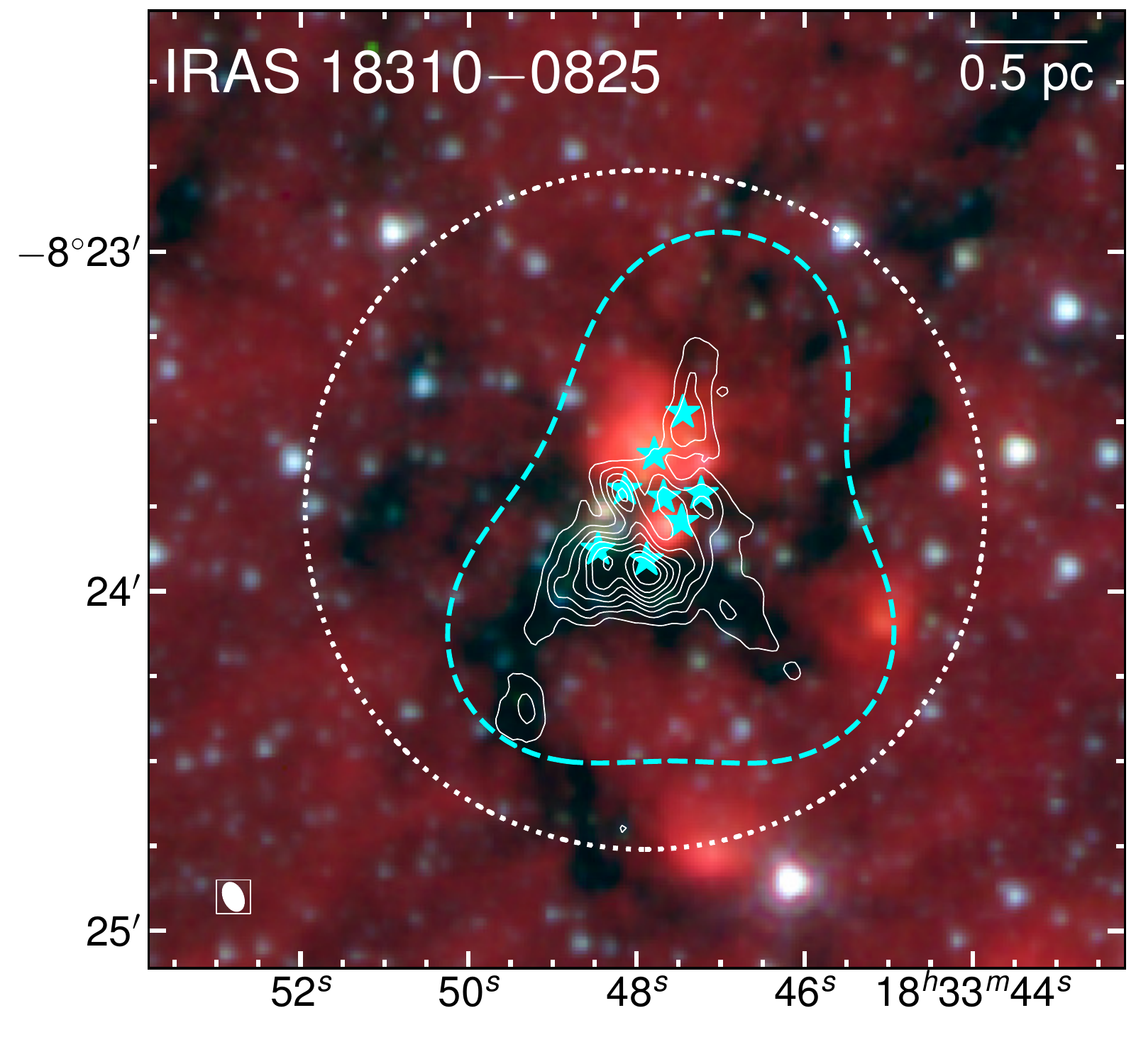}
\includegraphics[width=0.33\textwidth]{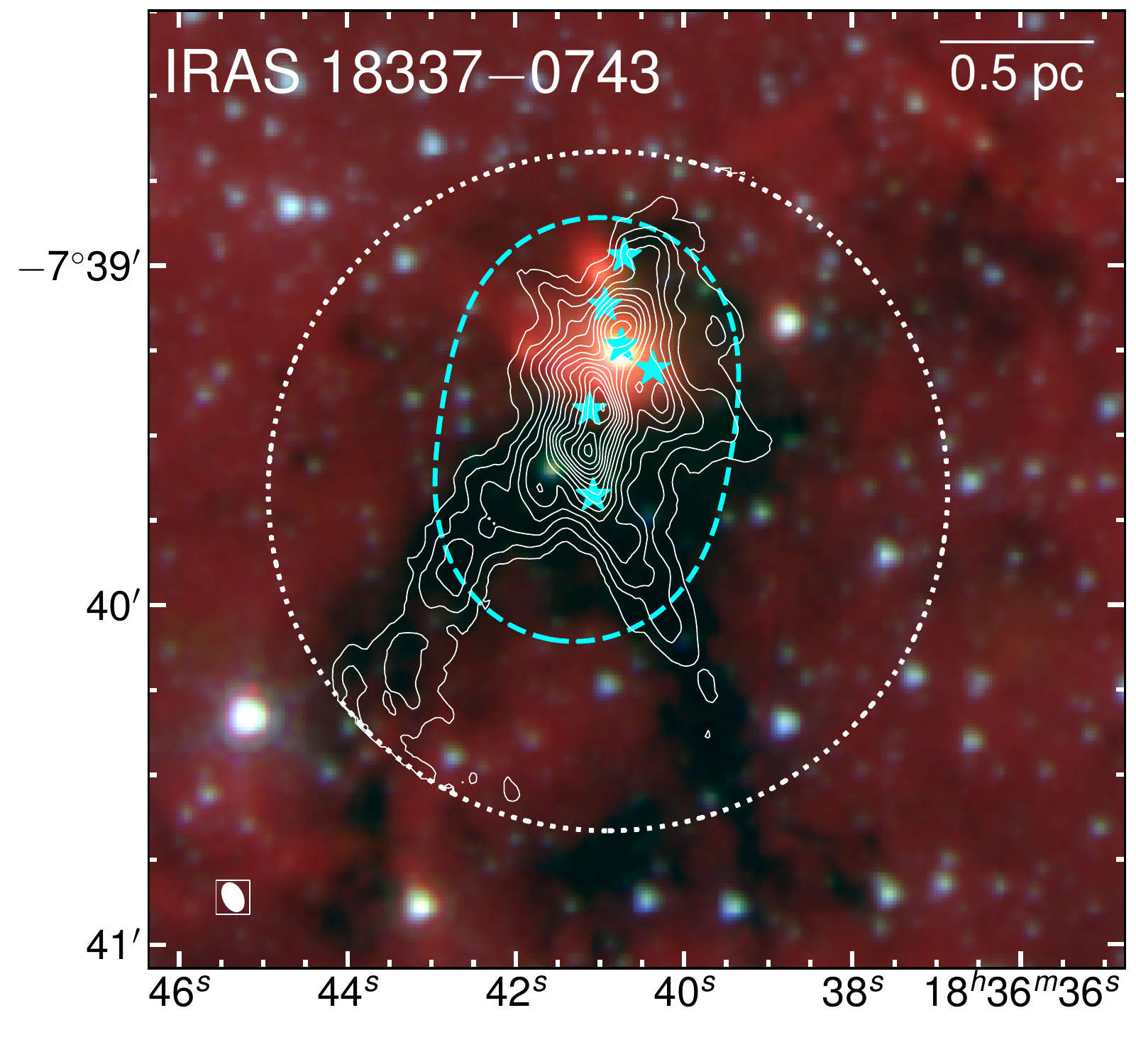} \\
\includegraphics[width=0.33\textwidth]{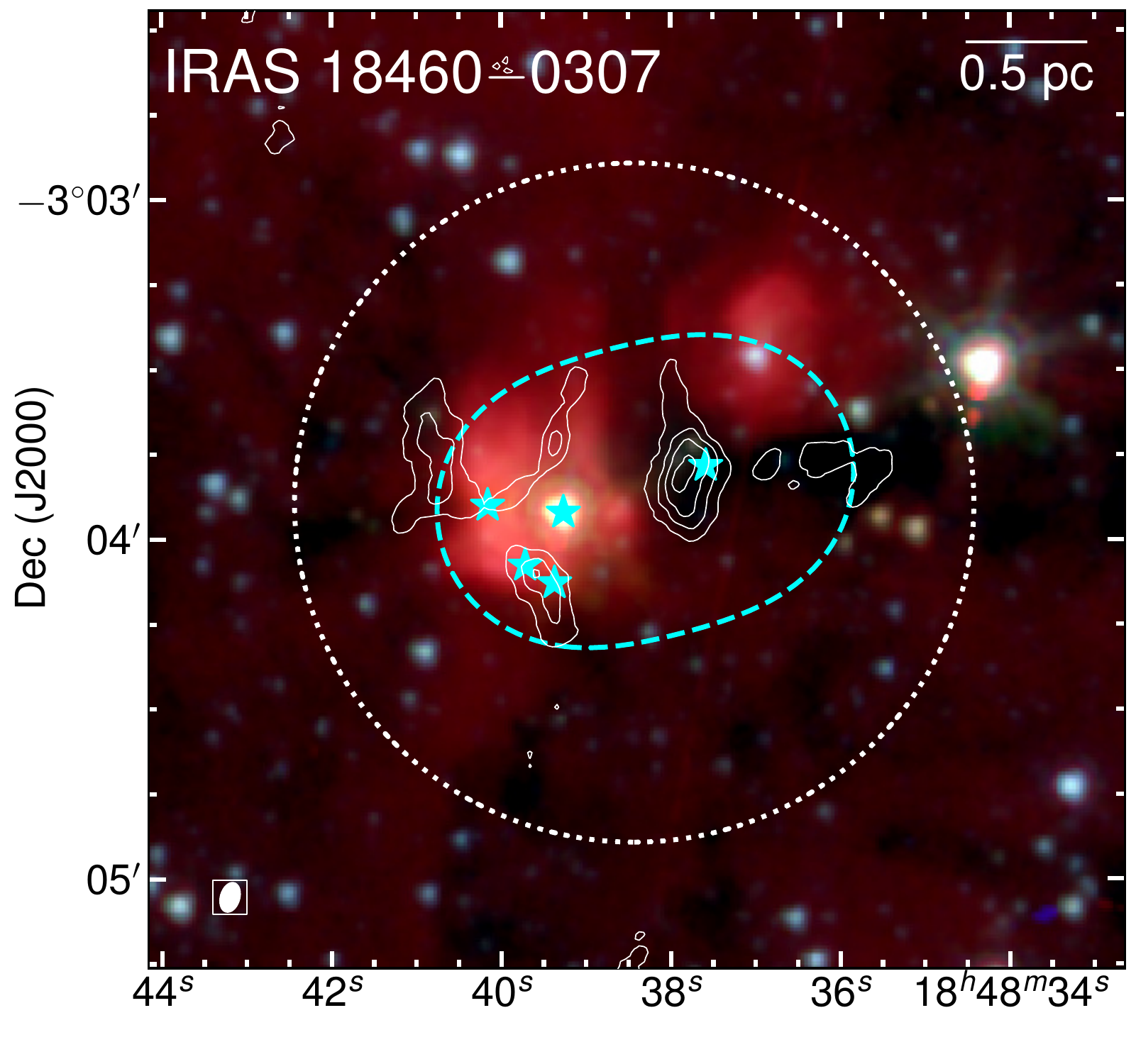}
\includegraphics[width=0.33\textwidth]{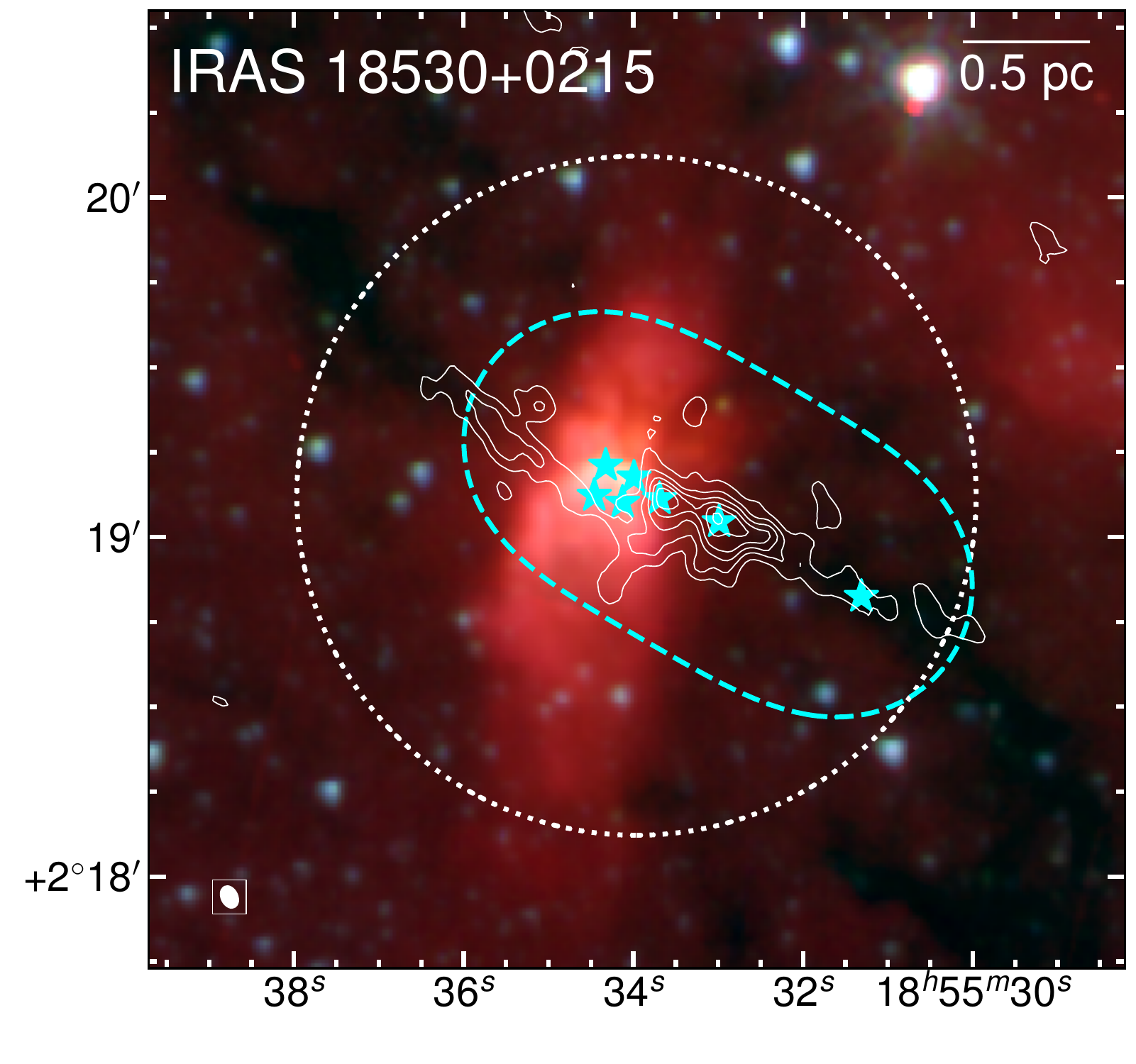}
\includegraphics[width=0.33\textwidth]{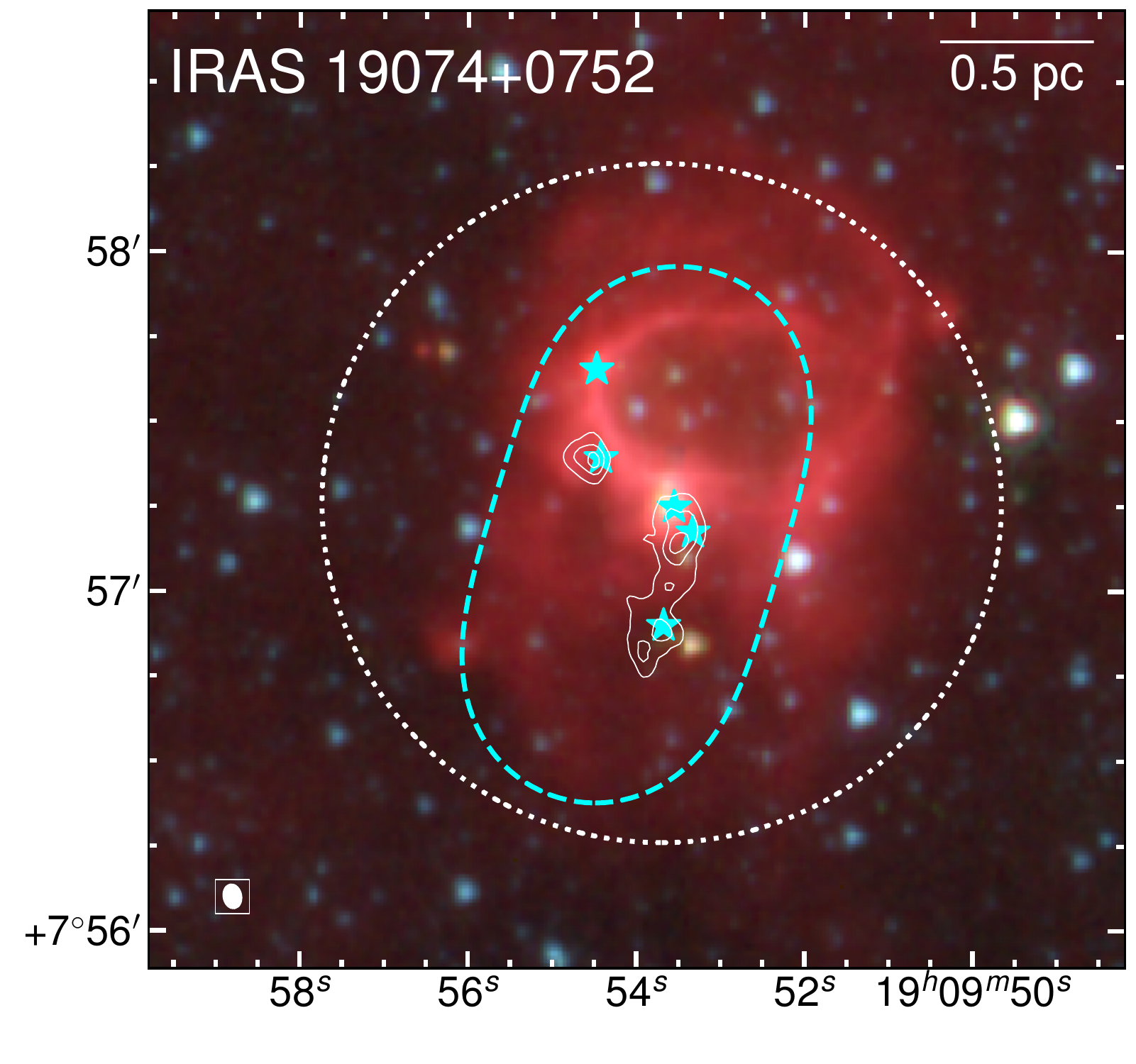} \\
\includegraphics[width=0.33\textwidth]{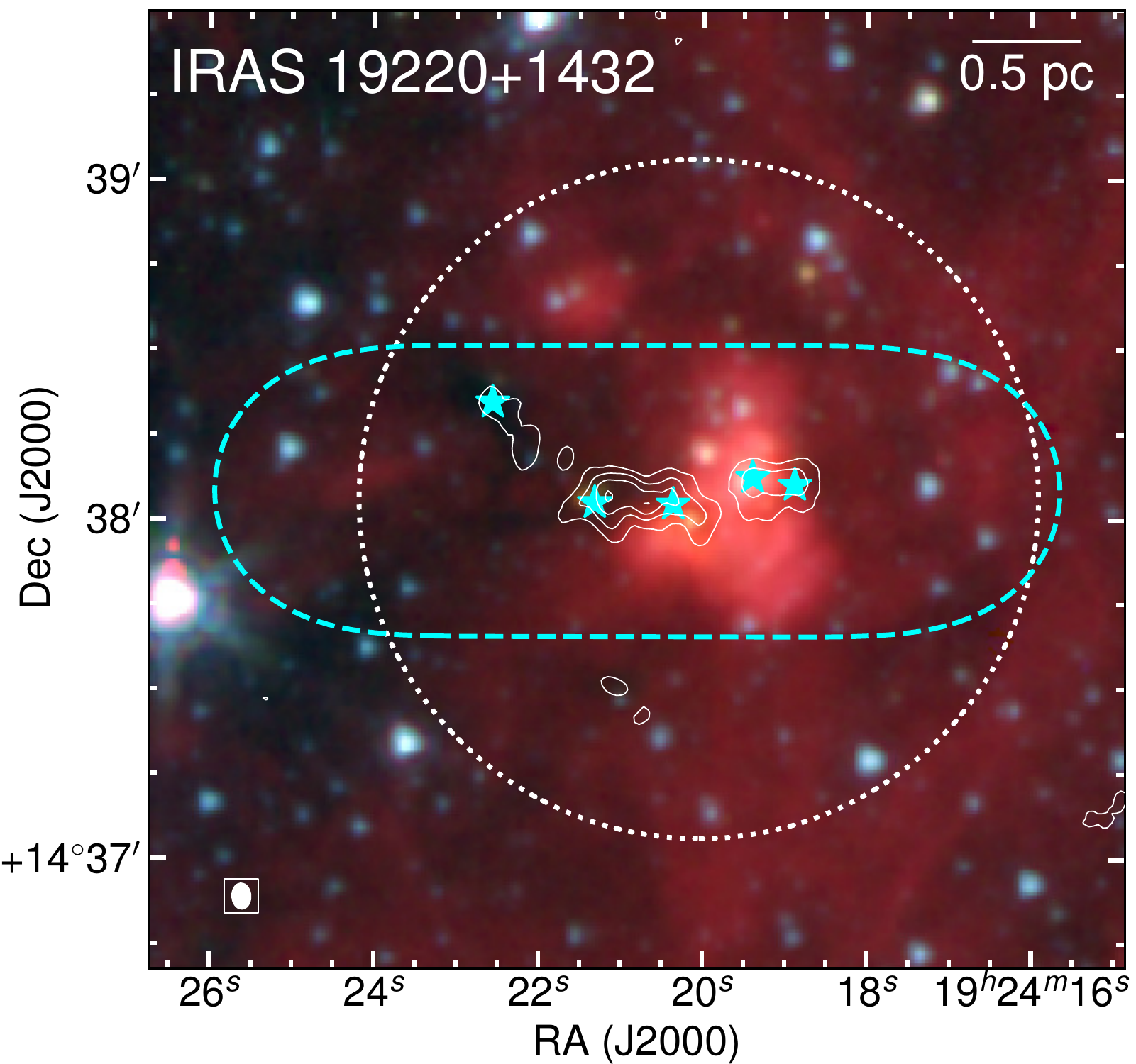}
\includegraphics[width=0.33\textwidth]{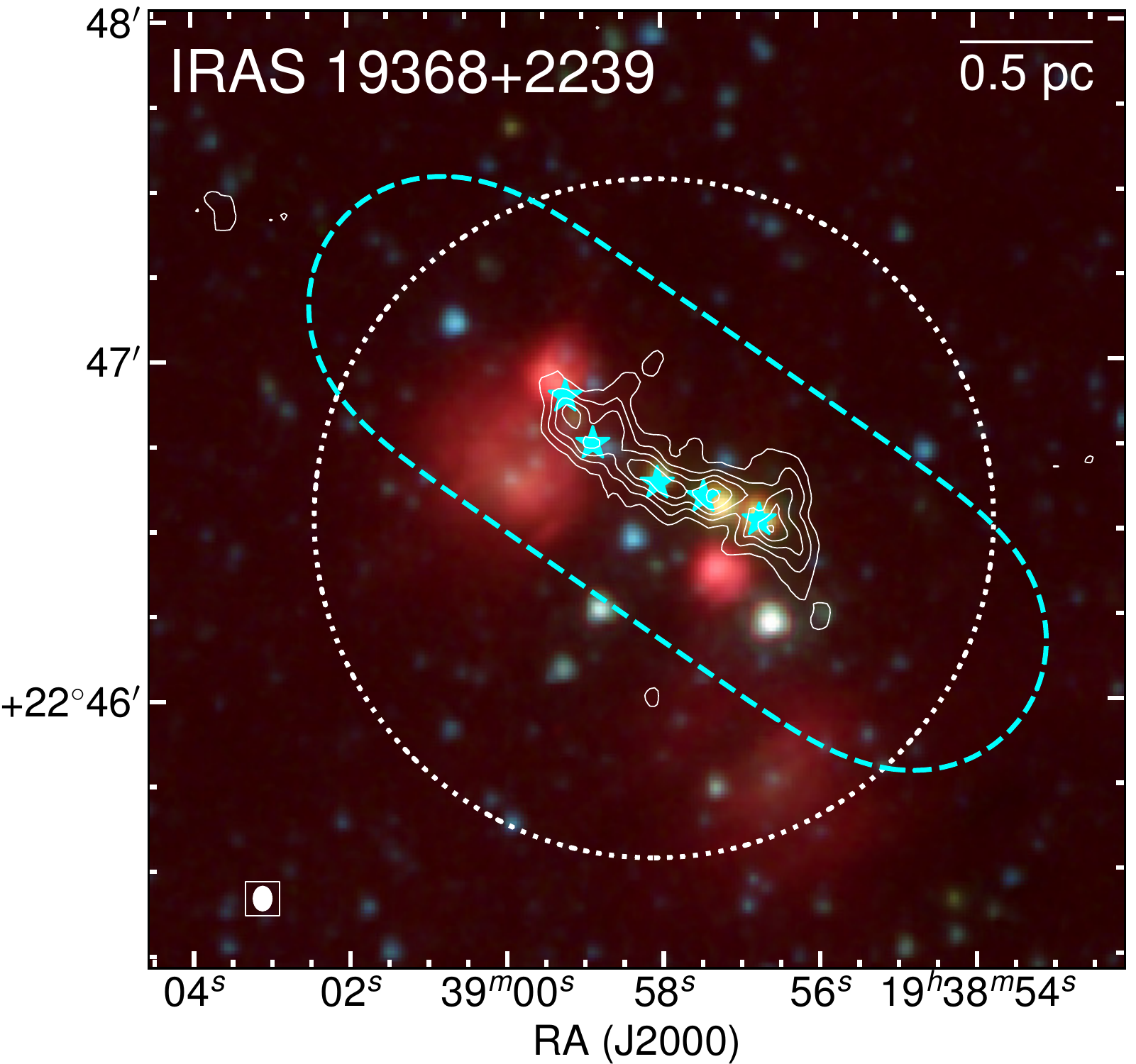} \\
\caption{VLA \amm{} maps of the 8 clouds. Background three-color images show \textit{Spitzer} 3.6~\micron{} (blue), 4.5~\micron{} (green), and 8.0~\micron{} (red) emission. Contours show integrated intensities of the VLA \amm{} (1,~1) main hyperfine component, with the white dotted loop in each panel showing the VLA primary beam size \citep{lu2014}. Contour levels are in steps of 20~\mjypbm\,\kms{}. Contours outside of the primary beam are suppressed. The cyan stars mark locations of dense cores detected with the SMA in this paper, with the cyan dashed loop showing FWHM of the SMA mosaic field. Synthesized beams of the VLA observations are shown in the lower-left corner of each panel.}
\label{fig:sample}
\end{figure*}

\subsection{Target Selection}
Our recent Very Large Array (VLA) \amm{} survey towards 62 high-mass star-forming regions resolved ubiquitous filamentary structures \citep{lu2014}. The filaments have typically lengths of $\gtrsim$1~pc, widths of $\sim$0.2~pc, and show coherent velocities within ranges of $\pm$2~\kms{}. They often break into a chain of \amm{} emission peaks along their major axes, some of which are associated with massive protostellar objects revealed by infrared point sources or masers. 
We compiled a sample of 8 molecular clouds from this survey, as listed in \autoref{tab:sample} and shown in \autoref{fig:sample}. These 8 clouds are optimized for our study because:

1) The VLA \amm{} emission exhibits filamentary structures with aspect ratios of $>$5, hence they are all typical filaments.

2) They are all luminous ($\sim$10$^4$~\lsol{}) and massive ($\gtrsim$10$^3$~\msol{}), with sufficient mass reservoir for high-mass star formation. Given the large luminosities, they should already harbor protostellar objects, although prestellar cores may still exist inside the clouds.

3) Their distances are $\sim$4--5~kpc, allowing for uniform spatial resolutions in a single observational setup.

In the rest of the paper, we use the abbreviated names of clouds, e.g., I18308 for IRAS 18308$-$0841.

\subsection{Structure of the Paper}
We introduce our interferometer observations as well as archived single-dish dust continuum observations in \autoref{sec:obs}. Then in \autoref{sec:results}, we outline the results of the observations, including the SMA dust emission, the VLA centimeter continuum emission, and molecular lines obtained by the SMA and VLA. In \autoref{sec:cores}, we identify dense cores based on the SMA dust emission and derive their physical properties, and discuss their high-mass star formation activities in \autoref{sec:cores_sf}. In \autoref{sec:filaments}, we select 4 morphologically well-defined filaments and discuss their fragmentation, longitudinal gas properties, and accretion. We summarize the paper and reach our conclusions in \autoref{sec:conclusions}.

\section{OBSERVATIONS AND DATA REDUCTION}\label{sec:obs}

\subsection{Submillimeter Array (SMA) Observations}\label{subsec:obs_sma}
The Submillimeter Array \citep[SMA;][]{ho2004}\footnote{The SMA is a joint project between the Smithsonian Astrophysical Observatory and the Academia Sinica Institute of Astronomy and Astrophysics, and is funded by the Smithsonian Institution and the Academia Sinica.} was used to observe the 8 clouds at 1.3~mm in the compact and subcompact configurations. The baseline lengths range from 9.5~m to 87~m. Each cloud was mosaiced with 2 to 5 pointings to cover the filamentary structures seen in the VLA \amm{} maps (see dashed loops in \autoref{fig:sample}).

As shown in \autoref{tab:smaobs}, the observation campaign was carried out in four years, during which the correlators of the SMA kept evolving. We used a fixed local oscillator (LO) frequency of 224.9~GHz for all observations. The intermediate frequencies (IFs) of the observations are listed in the note under \autoref{tab:smaobs}. In the observations up to 2015 Apr 10, the Application Specific Integrated Circuit (ASIC) correlator covers 216.9--220.9~GHz and 228.9--232.9 GHz in the lower and upper sidebands, respectively, leading to a total bandwidth of 8~GHz. In the observations from 2015 Apr 13 onward, the SMA Wideband Astronomical ROACH2 Machine (SWARM) correlator operates in parallel with the ASIC correlator. In the last observation taken on 2016 May 19 with the largest bandwidth (ASIC plus SWARM, but half of SWARM bandwidth cover the same IF range as ASIC), the correlators cover 208.9--220.9~GHz and 228.9--240.9~GHz, leading to a total bandwidth of 16~GHz. Note that by the time of this submission, the ASIC correlator on the SMA has been decommissioned and the SWARM correlator is in full operation to provide a maximum bandwidth of 32~GHz in the dual-receiver mode.

The SMA data were calibrated using the IDL superset MIR\footnote{\href{https://www.cfa.harvard.edu/~cqi/mircook.html}{https://www.cfa.harvard.edu/~cqi/mircook.html}}. The calibrated data were inspected with MIRIAD \citep{sault1995}, and imaged with CASA~4.7.2 \citep{mcmullin2007}. The ASIC data have a uniform channel width of 0.812~MHz (1.1~\kms{} at 1.3~mm). The SWARM data, with a uniform channel width of 76.1~KHz (on 2015 Apr 13), or 101.5~kHz (between 2015 Jun and 2016 Feb), or 140~kHz (in 2016 May), were smoothed before calibration by a factor of 8 to a final channel width of 0.609~MHz, 0.812~MHz, or 1.120~MHz to roughly match with that of the ASIC data.

We used multi-frequency synthesis to select all line-free channels, including the SWARM data when present, to image the 1.3~mm continuum. Spectral lines were split from the continuum-subtracted visibility data and were imaged separately with a uniform channel width of 1.1~\kms{}. To better recover the diffuse emission, we applied the multiscale CLEAN algorithm in CASA to both continuum and spectral lines. After trying several combinations, we selected the multiscale parameters of pixel numbers [0, 4, 12, 40], corresponding to angular scales of [0\arcsec, 3\farcs{2}, 9\farcs{6}, 32\arcsec]. For both continuum and spectral lines, in particular the \twelveco{} and \thirteenco{} lines, we noted that significant negative clean components could be derived in the largest angular scale (32\arcsec), so we set the parameter \textit{negcomponent} in the CLEAN task in CASA to 1, to stop the iteration when any negative clean component was detected in the largest angular scale. We used the Briggs weighting with a robustness of 0.5 for CLEAN.

The image root-mean-square (RMS) and beams are listed in \autoref{tab:images}. The on-source time for the first 4 clouds is longer than that for the next 4 clouds, leading to better spectral line sensitivity at the same channel width. Meanwhile, the continuum bandwidth for the next 4 clouds is in general larger, thanks to the addition of the SWARM correlator, leading to similar continuum sensitivities for all the 8 clouds. Typical beam size of continuum images is 3\farcs{5}$\times$3\farcs{0} (equivalent to 0.07~pc$\times$0.06~pc at a distance of 4~kpc) and typical RMS is 1.0~\mjypbm{} measured toward regions with no continuum emission. Typical RMS of spectral line images is 40 (for the first 4 clouds) or 80 (for the next 4 clouds)~\mjypbm{} in 1.1~\kms{} measured in line-free channels. For both the continuum and spectral line images, the RMS becomes larger in the vicinity of the bright sources due to limited dynamic range.

All the images presented in figures below are without primary beam corrections to have uniform RMS levels across maps. When calculating column densities and masses (e.g., in \autoref{subsec:cores_prop}), however, we use images after correcting for primary beam responses in order to obtain the correct fluxes.

\begin{figure*}[!t]
\centering
\includegraphics[width=1.0\textwidth]{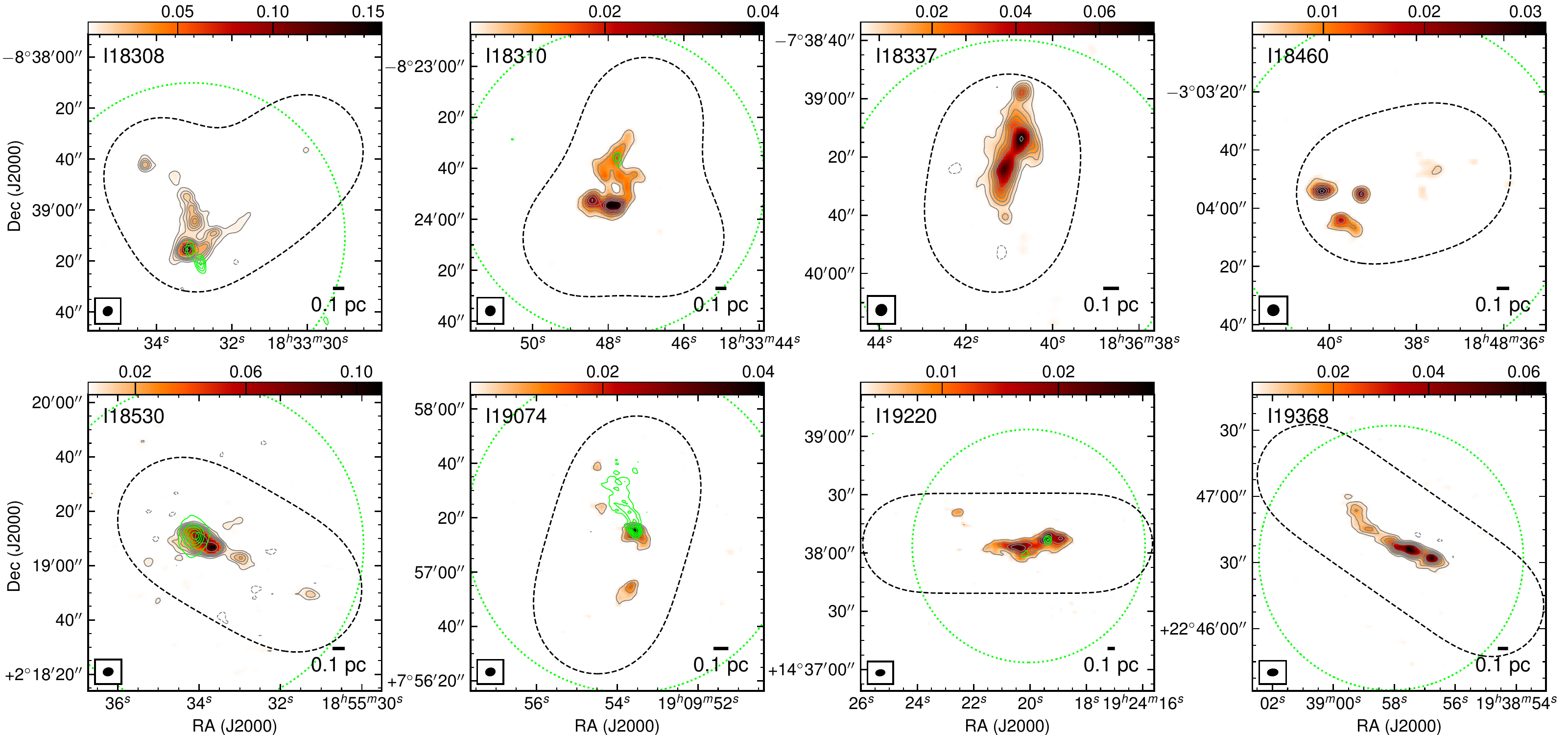}
\caption{Images and black contours show the SMA compact+subcompact array 1.3~mm continuum emission. Contour levels are between 5$\sigma$ and 30$\sigma$ in steps of 5$\sigma$ (1$\sigma$=1~\mjypbm{}), and above 30$\sigma$ in steps of 40$\sigma$. A dashed contour is also displayed to show the $-$5$\sigma$ level. Green contours show the VLA 1.3~cm continuum emission, starting at 5$\sigma$, in steps of 2$\sigma$ (for all clouds except I18530; 1$\sigma$=0.3--0.5~\mjypbm{}) or 20$\sigma$ (for I18530; 1$\sigma$=0.9~\mjypbm{}). Units of color scales are \jypbm{}. The dashed loop in each panel shows the SMA mosaic field, and the dotted loop shows the VLA primary beam.}
\label{fig:smacont}
\end{figure*}

\subsection{Very Large Array (VLA) \amm{} Observations}\label{subsec:obs_vla}
The NRAO\footnote{The National Radio Astronomy Observatory is a facility of the National Science Foundation operated under cooperative agreement by Associated Universities, Inc.} Very Large Array (VLA) \amm{} (1,~1) and (2,~2) observations have been used to derive gas temperatures in \cite{lu2014}. Details of the observations, including observation dates and calibrators can be found therein. Typical achieved RMS is 2--3~\mjypbm{} in 0.62~\kms{}, with a beam size of 5\arcsec$\times$3\arcsec{}.

Here we took the \amm{} (1,~1) and (2,~2) data of the 8 clouds, to study kinematics in filaments and derive temperatures of dense cores. In addition, we fit zeroth-order continuum from the \amm{} (2,~2) images, which have more line-free channels than \amm{} (1,~1), using the \textit{imcontsub} task in CASA, to obtain centimeter continuum images of the 8 clouds. Typical bandwidth included in the fitting is $\sim$2~MHz. The resulting RMS is 0.3--0.9~\mjypbm{}, depending on how large the included bandwidth is and whether the image is dynamic-range limited.

Properties of these images are listed in \autoref{tab:images}. The angular resolutions match well with those of the SMA images.

\subsection{Archival Millimeter Continuum Data}\label{subsec:obs_archival}
In order to obtain total masses of filaments later in \autoref{subsec:filaments_fragmentation}, we need dust emission observations from single-dish telescopes that do not suffer from missing the extended emission that is missed by interferometers. Here we used the 1.2~mm dust continuum data from the MAMBO bolometer array mounted on the IRAM 30~m telescope, which are published in \citet{beuther2002mambo}. All clouds except I19368 are included in this survey. The beam size is 11\arcsec{}, which resolves filamentary structures in the clouds. For I19368, we chose the 850~\micron{} dust continuum data from the SCUBA2 bolometer array on the JCMT, published in \citet{eden2017}, which have a slightly larger beam size of 14\farcs{4}. Note that bolometer arrays filter out extended emission when spatial scales of structures are comparable to the array field of view \citep{kauffmann2008}, but this is not a significant issue for structures in our sample that are in general of <1\arcmin{}.

\section{RESULTS}\label{sec:results}

\subsection{SMA Dust Emission}
The SMA 1.3~mm continuum emission maps of the 8 clouds are shown in \autoref{fig:smacont}. Compact emission at $\sim$3\arcsec{} angular scales are detected in all clouds. Before using the 1.3~mm continuum emission to trace dense gas, we first discuss its nature, i.e., whether it is from thermal dust emission or other sources.

The 1.3~mm continuum emission could have contribution from free-free emission of \hii{} regions. As shown in \autoref{fig:smacont}, in 5 out of the 8 clouds, centimeter continuum emission is seen in the VLA observations above the 5$\sigma$ level, suggesting the presence of free-free emission. For 1.3~mm emission peaks with likely contribution from free-free emission, we assumed a flat spectral index from centimeter to 1.3~mm, and subtracted the flux from their 1.3~mm continuum emission. The free-free contribution subtracted fluxes are listed in parentheses in \autoref{tab:cores}, which are used to derive dense core masses in \autoref{subsec:cores_prop}. Weak free-free emission below the sensitivity of the VLA observations (0.3--0.9~mJy) could exist, but it will not significantly affect the estimate of dense core masses given that their 1.3~mm continuum fluxes are all $\geq$15~mJy.

One of the 1.3~mm emission peaks in I18308 (marked as I18308-c10 in \autoref{fig:cores}) is associated with a bright infrared point source which is identified as an Asymptotic Giant Branch (AGB) star in \citet{robitaille2008} (named as G023.2022+00.0151). We listed its parameters in \autoref{tab:cores}, but excluded it in following discussion.

None of the other 1.3~mm emission peaks have detectable free-free emission contamination or AGB star counterparts, therefore they are likely dominated by thermal dust emission.

\subsection{VLA Centimeter Continuum Emission}

Centimeter continuum emission derived from spectral baseline fitting of the VLA \amm{} data is displayed as contours in \autoref{fig:smacont}. In 5 clouds, centimeter continuum emission is detected above 5$\sigma$ levels. Strong emission is found in two clouds, I18530 and I19074, both of which are known to harbor \hii{} regions \citep{white2005}. The other three clouds, I18308, I18310, and I19220, show compact emission associated with 1.3~mm emission peaks. Later in \autoref{subsec:cores_sf_hii} we will use the centimeter continuum emission to trace \hii{} regions embedded in dense cores.

\begin{figure*}[!t]
\centering
\includegraphics[width=0.85\textwidth]{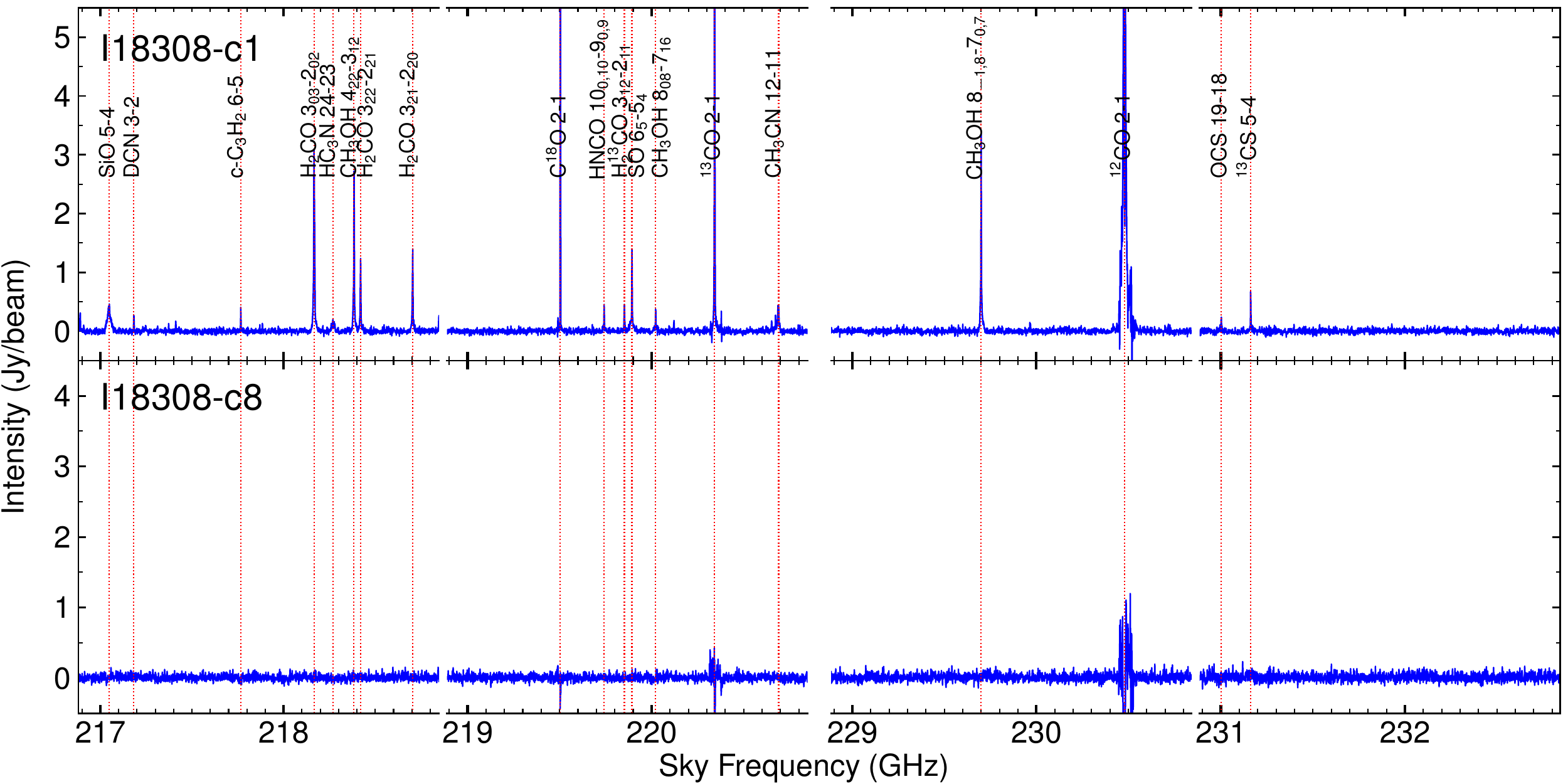}
\caption{Two examples of the SMA 1.3~mm spectra. The upper panel shows a chemical-rich region (I18308-c1, see \autoref{subsec:cores_id}), in which sulfur-bearing molecules (OCS, $^{13}$CS, SO), deuterated species (DCN), and complex organic molecules (\methanol{}, \mthc{}, \cyacet{}, c-C$_3$H$_2$) are detected. The vertical scale is truncated at 5.5~\mjypbm{}, while the three CO isotopologues (\twelveco{}, \thirteenco{}, \ceighteeno{}) are brighter. The lower panel shows a chemical-quiescent region in the same cloud (I18308-c8, see \autoref{subsec:cores_id}), in which only the three CO isotopologues are detected. Vertical dotted lines mark sky frequencies of the line species assuming \vlsr{}=77~\kms{}, with transition names noted in the upper panel. In \twelveco{}, \thirteenco{}, and \ceighteeno{} lines in the lower panel, clear absorption features are seen due to strong side lobes of the brighter source I18308-c1 nearby.}
\label{fig:smaspec}
\end{figure*}

\subsection{SMA 1.3~mm Molecular Lines}
A number of molecular lines were detected by the SMA towards the 8 clouds. Typical lines detected with the ASIC correlator toward a chemical-rich region and a chemical-quiescent region (I18308-c1 and I18308-c8, respectively, see \autoref{subsec:cores_id}) are displayed in \autoref{fig:smaspec}. A few spectral lines were also detected in the extra bandwidths covered by the SWARM correlator, e.g., two \methanol{} lines in the lower sideband and an H$_2$S line in the upper sideband toward a dense core in I18530. However, as shown in \autoref{tab:smaobs}, we did not achieve a uniform coverage of the 8 clouds with the SWARM correlator. Therefore, to obtain a uniform spectral line dataset across the 8 clouds, we only considered the frequencies covered by the ASIC correlator.

Among the detected lines, we focus on two groups in this paper: \twelveco{}, SiO, SO, \methanol{}, and \fmh{} are used to trace potential protostellar outflows (\autoref{subsec:cores_sf_outflows}), and \fmh{} and \methanol{} are used to define velocities of dense cores given their large critical densities (\autoref{subsec:filaments_filaccretion}). The other lines are also useful for chemistry and kinematic analysis, but are left for future studies.

\begin{figure*}[!t].pdf
\centering
\includegraphics[width=1.0\textwidth]{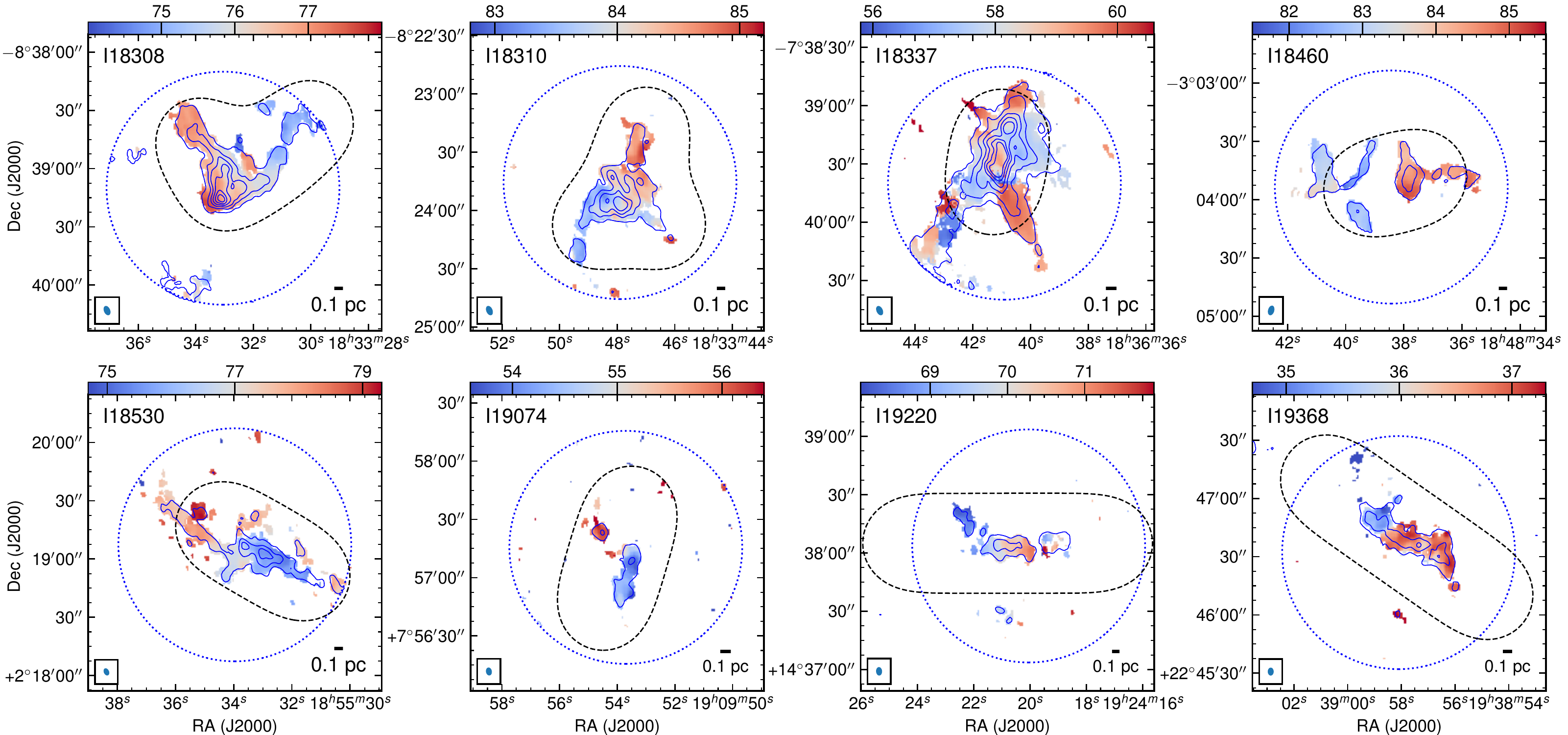} 
\caption{Color scales show centroid LSR velocities derived from simultaneous fitting of VLA \amm{} (1,~1) and (2,~2) lines. The unit of color scales is \kms{}. Contours show the \amm{} (1,1) main hyperfine component integrated intensity, in steps of 15~\mjypbm{}. Dotted and dashed loops are the same as in \autoref{fig:smacont}.}
\label{fig:nh3}
\end{figure*}

\subsection{VLA \amm{} Lines}
Integrated intensities of the \amm{} (1,~1) main hyperfine line in the 8 clouds are presented as contours in \autoref{fig:sample}. In \autoref{subsec:cores_temp}, we will use the \amm{} (1,~1) and (2,~2) lines to derive gas temperatures in dense cores. We will also use them to resolve kinematics in filaments, as shown in the centroid velocity maps in \autoref{fig:nh3} while detailed analysis can be found in Sections~\ref{subsec:filaments_filprop}~\&~\ref{subsec:filaments_filaccretion}.

\section{PROPERTIES OF DENSE CORES}\label{sec:cores}

\begin{figure*}[!t]
\centering
\includegraphics[width=1.0\textwidth]{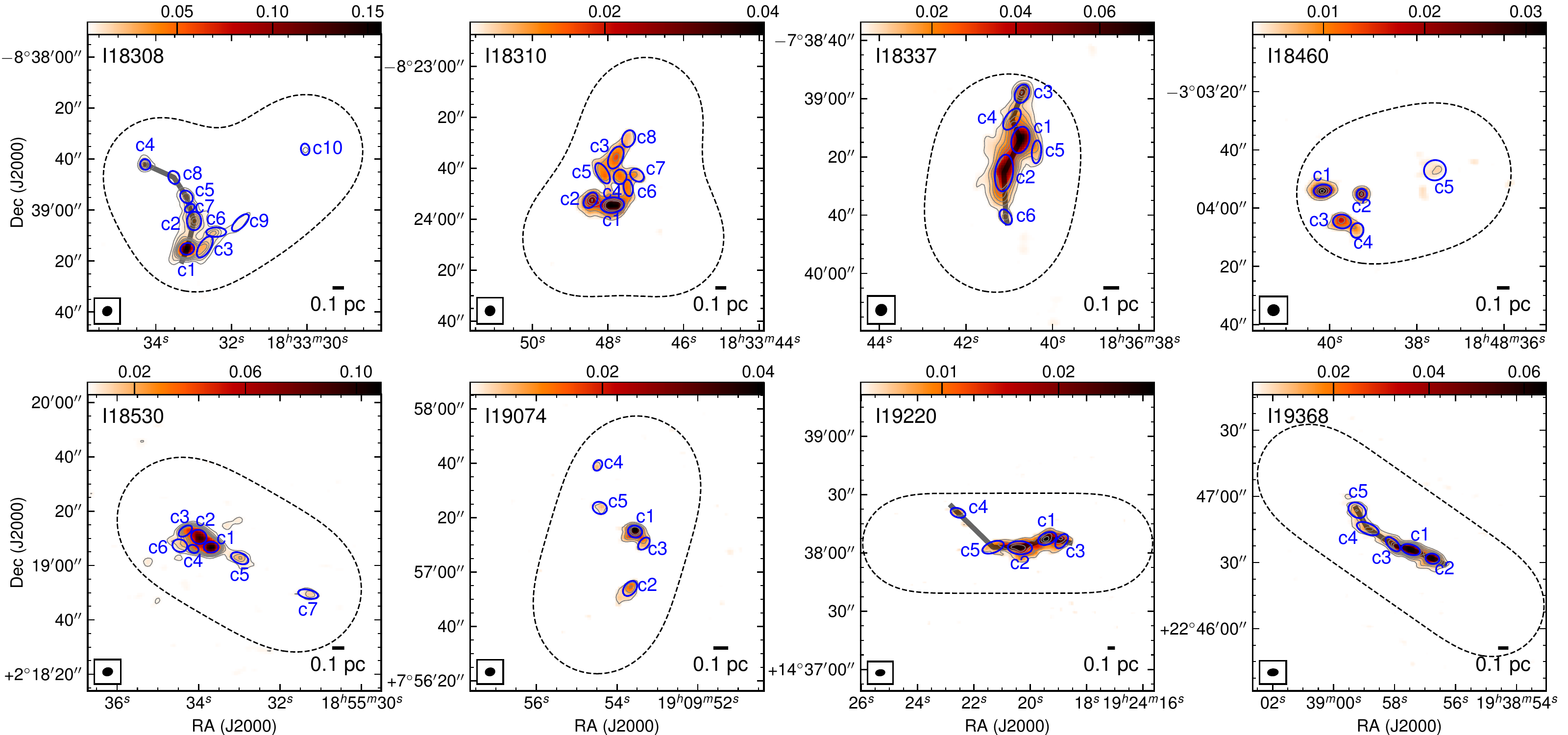}
\caption{Dense cores identified in the 8 clouds. Images and contours show the same 1.3~mm continuum emission as in \autoref{fig:smacont}. Contour levels are identical to those in \autoref{fig:smacont}, but only the first 6 levels are drawn to avoid obscuring other markers. The blue ellipses mark FWHMs of 2-D Gaussians fit to the dense cores, which are convolved with synthesized beams. In I18308, I18337, I19220, and I19368, shadowed bars illustrate the definition of filaments in \autoref{subsec:filaments_fragmentation}.}
\label{fig:cores}
\end{figure*}

\subsection{Identification of Dense Cores}\label{subsec:cores_id}
We identified dense cores based on the dust emission maps. The maps were not corrected for primary beam responses, in order to have uniform RMS levels across maps, shown in \autoref{fig:cores}. Then we obtained fluxes of the identified dense cores from the maps corrected for primary beam responses.

We identified dense cores at 0.1 pc scales by eyes of the dust emission, and fit two-dimensional (2-D) Gaussians to the dense cores using the interactive fitting tool in the CASA viewer. The identification was done in an iterative way: the strongest dust emission peak was fit and subtracted; in the residual image, the remaining strongest dust emission peak was fit and subtracted, until all peaks above 5$\sigma$ level and larger than the synthesized beam were fit. Gaussian fittings to the identified dense cores are marked by circles in \autoref{fig:cores}. The dense cores are named as cloud name plus `c' and a number in the order from high to low peak fluxes, e.g., I18308-c1.

We noted that toward several dust emission peaks (e.g., I18308-c1, I18337-c1), a single 2-D Gaussian is unable to fit the emission well, and residual emission of the $\sim$5$\sigma$ level that is spatially offset from the Gaussian peak is seen. This may suggest that substructures exist within the identified cores, and if they are resolved under higher angular resolutions, multiple Gaussians should be fit to each of them. With the current angular resolution of 3\arcsec, we cannot resolve any substructures, therefore we only fit single Gaussians.

50 dense cores were identified in the 8 clouds, excluding the AGB star counterpart I18308-c10. Their coordinates, deconvolved FWHM sizes, and fluxes after primary-beam correction are listed in \autoref{tab:cores}. As shown later in \autoref{sec:cores_sf}, 45 of the identified dense cores turn out to be bona fide cores which are associated with (UC) \hii{} regions, infrared sources, or molecular outflows. The remaining 5 identified dense cores are not yet detected with any star-forming signature, thus might be candidates of prestellar cores or simply over-densities of molecular gas flows.

\begin{figure}[!t]
\centering
\includegraphics[width=0.48\textwidth]{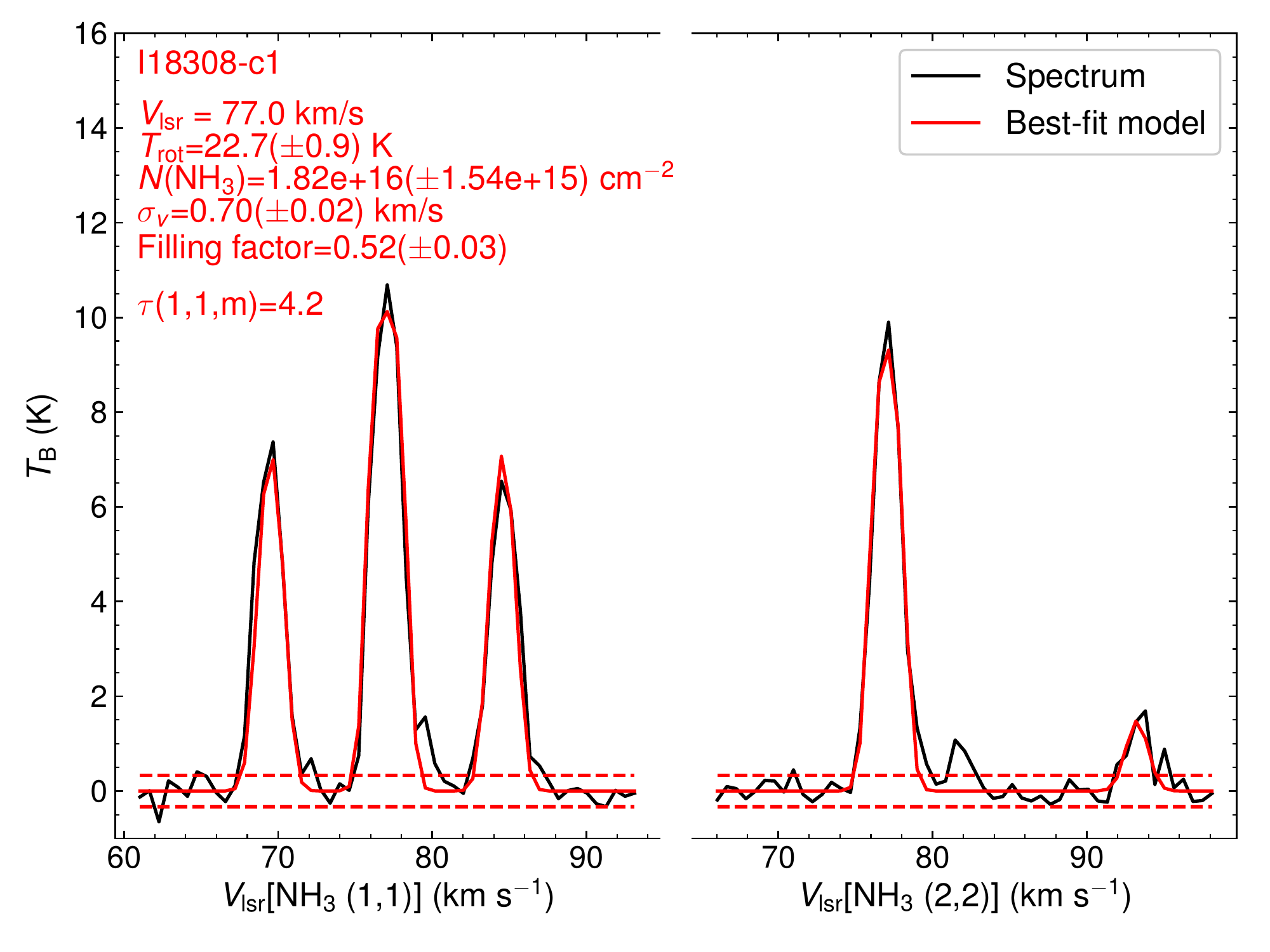}
\caption{Fitting results of gas temperatures in I18308-c1 using the VLA \amm{} (1,~1) and (2,~2) lines. Red curve shows the best-fit model, with the best-fit parameters displayed in the panel.}
\label{fig:trot}
\end{figure}

\subsection{Gas Temperatures in Dense Cores}\label{subsec:cores_temp}
We estimated gas temperatures in dense cores using the VLA \amm{} (1,~1) and (2,~2) lines. The temperatures were then used to calculate core masses below. Linewidths, \amm{} column densities, and \amm{} optical depths were derived at the same time.

We extracted the mean \amm{} (1,~1) and (2,~2) spectra within the FWHM of the 2-D Gaussian fit of each core, and fit them with a customized code \textit{pyAmor}\footnote{\href{https://github.com/xinglunju/pyAmor}{https://github.com/xinglunju/pyAmor}}. We assumed local-thermodynamic-equilibrium (LTE) conditions and the same filling factor for all hyperfine components. The free parameters in the fit are the core velocity \vlsr{}, the rotational temperature $T_\text{rot}$, the \amm{} column density $N$(\amm{}), the velocity dispersion $\sigma_v$, and the filling factor. The core velocity \vlsr{} was first selected by eyes, then was varied within $\pm$0.62~\kms{} (one channel) of the selected value. Model spectra were constructed with these parameters. The non-linear least-square minimization of the difference between the model spectra and the observed spectrum was implemented with the \textit{lmfit} package \citep{newville2014}, then the parameters of the best-fit model spectrum were taken to be the best-fit results. The optical depth at the line center of the main hyperfine component of \amm{} (1,~1), $\tau(1,1,m)$, was derived afterwards using the best-fit $T_\text{rot}$, $N$(\amm{}), and $\sigma_v$. An example of the fitting is shown in \autoref{fig:trot}.

Toward several dense cores, multiple velocity components in \amm{} lines were found. Then we fit two independent models, each of which was constructed with the parameters listed in the last paragraph. Two sets of best-fit parameters were derived, from which we chose the one that has similar \vlsr{} to dense gas tracers in cores (e.g., \fmh{}, \methanol{}) to calculate dense core properties.

\subsection{Masses, Densities, \amm{} Abundances, Virial States}\label{subsec:cores_prop}

The masses of the dense cores were derived using the following equation:
\begin{equation}\label{equ:mass}
M_\text{core}=R_\text{g/d}\frac{S_\nu d^2}{B_\nu(T_\text{dust})\kappa_\nu},
\end{equation}
where $R_\text{g/d}$ is the gas-to-dust mass ratio, $S_\nu$ is the dust emission flux, $d$ is the distance, $B_\nu(T_\text{dust})$ is the Planck function at temperature $T_\text{dust}$, and $\kappa_\nu$ is the dust opacity. We assumed $R_\text{g/d}$=100, and $\kappa_\nu$=0.899~\sqc\,g$^{-1}$ \citep[MRN model with thin ice mantles, after 10$^5$ years of coagulation at 10$^6$~\cc{};][]{ossenkopf1994}. The kinematic distances were derived using the rotation curve model in \citet{reid2009} (see \citealt{lu2014} for details). The near distance was taken in case of kinematic distance ambiguity. Under the assumption of LTE, the dust temperatures $T_\text{dust}$ were taken to be the gas temperatures derived from \amm{}, $T_\text{rot}$, in \autoref{subsec:cores_temp}. We did not use dust temperatures derived from \textit{Herschel} data because their angular resolution is not sufficient to resolve the dense cores, nor did we use gas temperatures based on the SMA \fmh{} or \mthc{} lines because they usually only probe the inner, warmer part of dense cores.

After obtaining core masses, we derived mean H$_2$ number densities in each core, $n$(H$_2$), assuming that they are spheres with a radius $R$ estimated from the deconvolved sizes of 2-D Gaussian fittings using a geometric mean of the two axes. Then
\begin{equation}
n(\text{H}_2)=\frac{3M_\text{core}}{4\pi R^3}\frac{1}{\mu_\text{H$_2$} m_\text{H}},
\end{equation}
where $\mu_\text{H$_2$}$=$2.8$ is the molecular weight per H$_2$.

The mean H$_2$ column densities of the dense cores are:
\begin{equation}
N(\text{H}_2)=R_\text{g/d}\frac{S_\nu}{B_\nu(T_\text{dust})\kappa_\nu}\frac{1}{\Omega}\frac{1}{\mu_\text{H$_2$} m_\text{H}},
\end{equation}
where $\Omega$ is the solid angle of dense cores based on the deconvolved sizes.

The \amm{} abundances, $X$(\amm{}), are simply the \amm{} column densities divided by the H$_2$ column densities. Note that we multiplied the \amm{} column densities shown in \autoref{fig:trot} by the filling factors, and scaled the H$_2$ column densities listed in \autoref{tab:cores} with the dense core sizes convolved with synthesized beams, to obtain the source-averaged column densities.

Then we derived virial parameters of the dense cores \citep{bertoldi1992}, defined as
\begin{equation}
\alpha=\frac{5\sigma_\text{in}^2R}{GM_\text{core}},
\end{equation}
where $\sigma_\text{in}$ is the intrinsic one-dimensional velocity dispersion of the molecule of mean mass, and $R$ is the core radius. The VLA \amm{} lines were used to measure the velocity dispersion $\sigma_v$ (see \autoref{subsec:cores_temp}), then the channel width of 0.62~\kms{} is quadratically subtracted to obtain the deconvolved velocity dispersion of \amm{}:
\begin{equation}
\sigma_\text{deconv,\amm{}}=\sqrt{\sigma_v^2-(0.62~\kms{})^2/(2\sqrt{2\ln2})^2}.
\end{equation}
A $2\sqrt{2\ln2}$ factor is included to convert the FWHM width to the velocity dispersion. Finally $\sigma_\text{in}$ is derived as:
\begin{equation}\label{equ:sigma_in}
\begin{split}
\sigma_\text{in}&=\sqrt{\sigma_\text{nth}^2+c_s^2} \\
&=\sqrt{\sigma_\text{deconv,\amm{}}^2-\frac{k_BT_\text{rot}}{17m_p}+\frac{k_BT_\text{rot}}{\mu_p m_p}},
\end{split}
\end{equation}
where $c_s$=$\sqrt{\frac{k_BT_\text{rot}}{\mu_p m_p}}$ is the isothermal sound speed, $\mu_p$=2.33 is the mean molecular weight assuming 90\% H and 10\% He, and 17 is the molecular weight of \amm{}.
As discussed in \citet{kauffmann2013b}, for a self-gravitating, non-magnetized core, a virial parameter above 2 suggests the core is unbound and may expand, while one below 2 suggests the core is bound and may collapse. Here we have ignored any potential rotation of the cores and assumed a constant radial density profile \citep[see discussion in][]{baobab2012a,baobab2015,feng2016c}.

The derived core masses, densities, column densities, and virial parameters are listed in \autoref{tab:cores}.

\subsection{Uncertainties of Dense Core Properties}\label{subsec:cores_error}
We quantitatively characterized uncertainties in the derived temperatures, masses, densities, linewidths, and virial parameters of the dense cores.

The fitting errors of gas temperatures derived from \amm{} lines in \autoref{subsec:cores_temp} are displayed in \autoref{fig:trot}. These errors take the noise in \amm{} spectra and uncertainties in the fitting processes into account. As discussed in \citet{lu2014}, the flux density scale of the VLA data and the missing fluxes of the VLA as an interferometer do not affect the derived gas temperatures, hence the systematic errors are minor. In general, the temperatures have uncertainties of $<$5\%.

The dense core masses depend on dust opacity, gas-to-dust mass ratio, temperatures, dust emission fluxes, and distances. Discussion on how these factors affect the derived masses can be found in \citet{sanhueza2017}. Here we followed \citet{sanhueza2017} to adopt uncertainties of 28\%, 23\%, and 15\%, respectively, for the dust opacity, gas-to-dust mass ratio, and dust emission fluxes. The uncertainty in temperatures is adopted to be 5\%, as mentioned above, provided no discrepancy between gas and dust temperatures. The uncertainty in kinematic distances is typically 10\% but can be orders of magnitudes larger \citep{reid2009}. At last, we estimated an uncertainty of 44\% for the masses, after propagating all errors, which should be a lower limit given potentially larger systematic errors with the assumed dust opacity, gas-to-dust mass ratio, and distances. 

The H$_2$ number density $n$(H$_2$) has a similar dependence on the factors as the mass, therefore we adopted the same uncertainty of 44\%. The column density $N$(H$_2$) does not depend on distances. Its uncertainty is therefore 40\%. Uncertainties for both number density and column density are lower limits.

The fitting errors of linewidths are displayed in \autoref{fig:trot}. Depending on S/N ratios and numbers of velocity components, the uncertainties in linewidths vary from $<$3\% to more than 10\%.

When considering uncertainties in virial parameters, we adopted a conservative 10\% for all linewidths. Then the virial parameters have an uncertainty of 50\%.

\section{High-mass Star Formation in Dense Cores}\label{sec:cores_sf}

We find signatures of high-mass star formation associated with the dense cores in the 8 clouds. In addition, based on their infrared emission and chemical features, high-mass star formation in the dense cores is at a variety of evolutionary phases, from prestellar to protostellar and even \hii{} regions.

\subsection{\hii{} Regions}\label{subsec:cores_sf_hii}
As shown in \autoref{fig:smacont}, toward several dense cores, compact centimeter continuum emission is detected in the VLA observations, indicating that the protostars have evolved to such a late phase that ambient gas is ionized by young stars, and ultra-compact (UC) \hii{} regions are formed.

Even more evolved phase is found in I19074, where a pc scale \hii{} region traced by an infrared bubble is detected \citep[][see \autoref{fig:sample}]{churchwell2006}. Four dense cores seem to surround the \hii{} region. Their star formation may have been influenced by the \hii{} region.

Here we assume optically thin centimeter continuum emission and an electron temperature of 10$^4$~K, and derive the ionizing photon rate of the (UC) \hii{} regions in these clouds, following \citet{mezger1974}. Fluxes of the compact emission in I18308, I18310, and I19220 are $\sim$3--5~mJy, corresponding to ionizing photon rates of 0.7--1.5$\times$10$^{46}$~s$^{-1}$. Such photon rates are consistent with that of a B0.5 type ZAMS \citep{panagia1973}. The centimeter continuum emission in I18530 and I19074 is stronger, up to 0.25 and 0.10~Jy, respectively, resulting in photon rates of 5.5$\times$10$^{47}$~s$^{-1}$ and 1.5$\times$10$^{47}$~s$^{-1}$. These are consistent with those of a B0 or O9.5 type ZAMS  \citep{panagia1973}. Note that these \hii{} regions do not correspond one-to-one to the dense cores. Some of them have multiple dense core counterparts or are offset from any dense cores.

\begin{figure*}[!t]
\begin{tabular}{p{8.7cm}p{8.7cm}}
\includegraphics[width=0.495\textwidth]{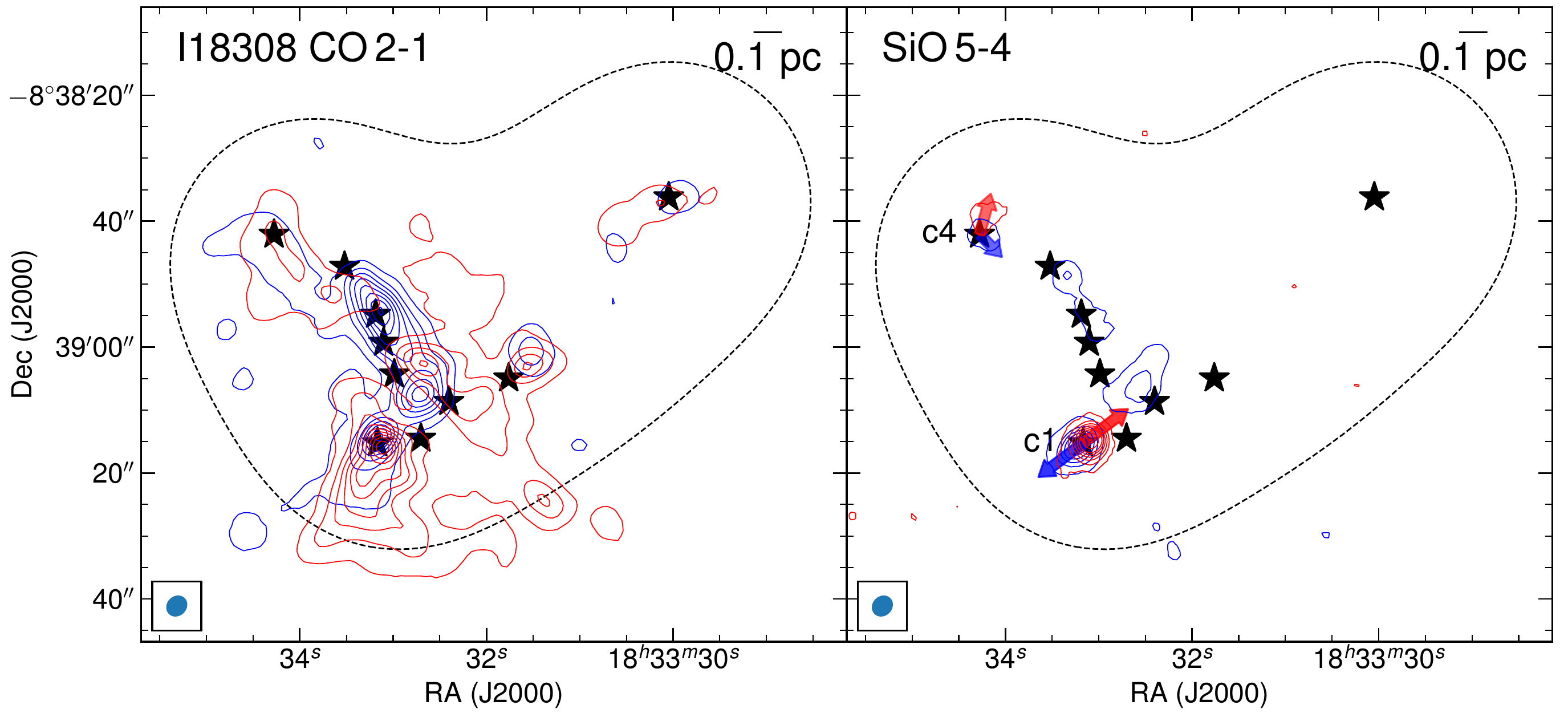} & \includegraphics[width=0.495\textwidth]{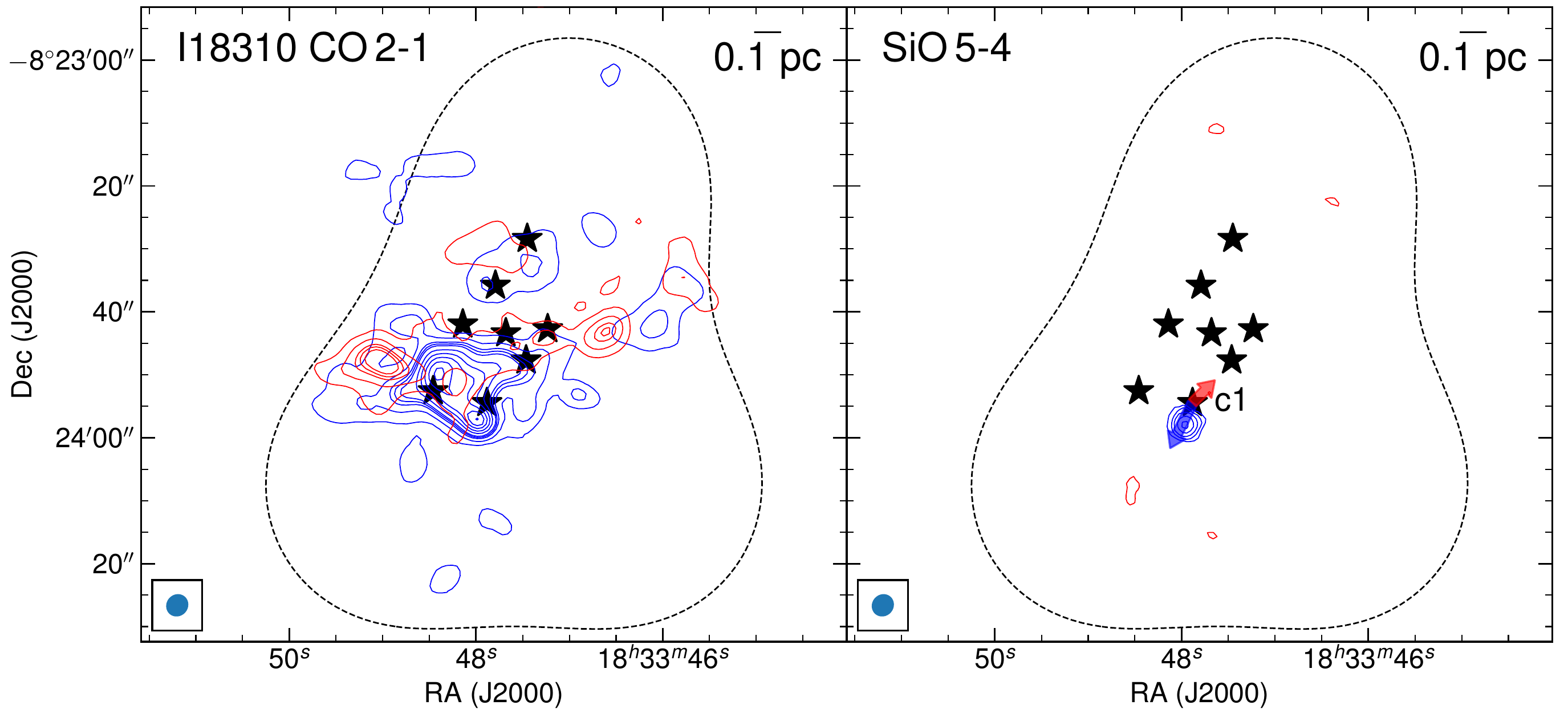} \\
\includegraphics[width=0.495\textwidth]{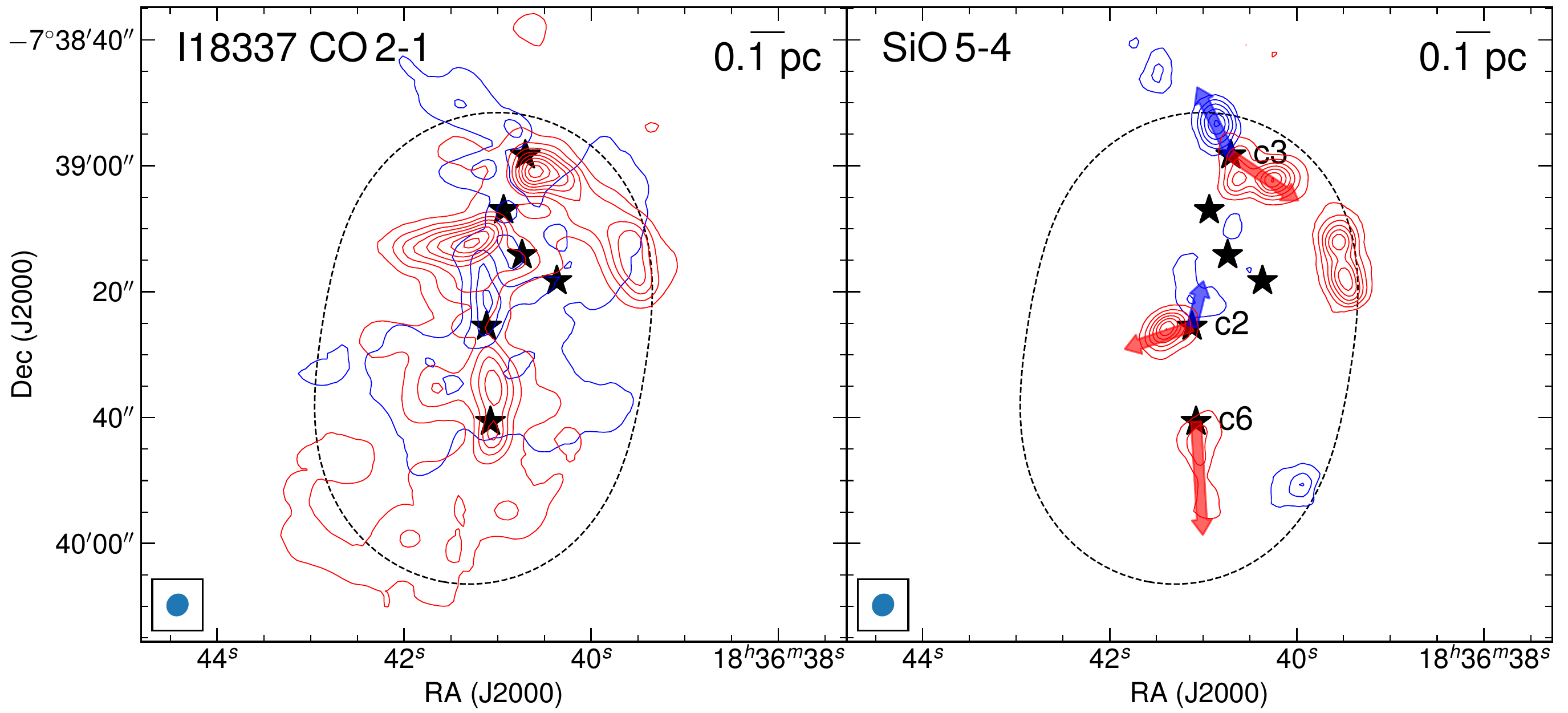} & \includegraphics[width=0.495\textwidth]{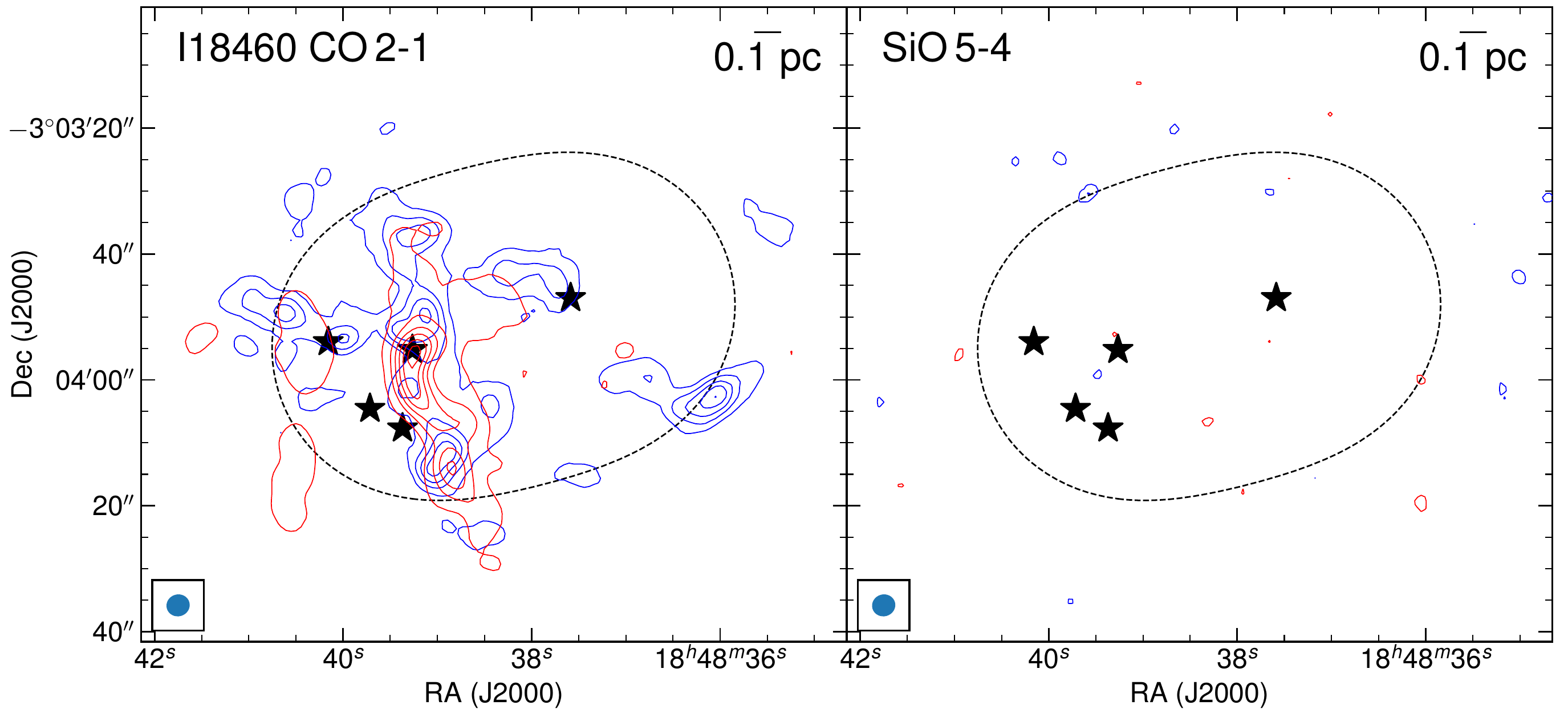} \\
\includegraphics[width=0.495\textwidth]{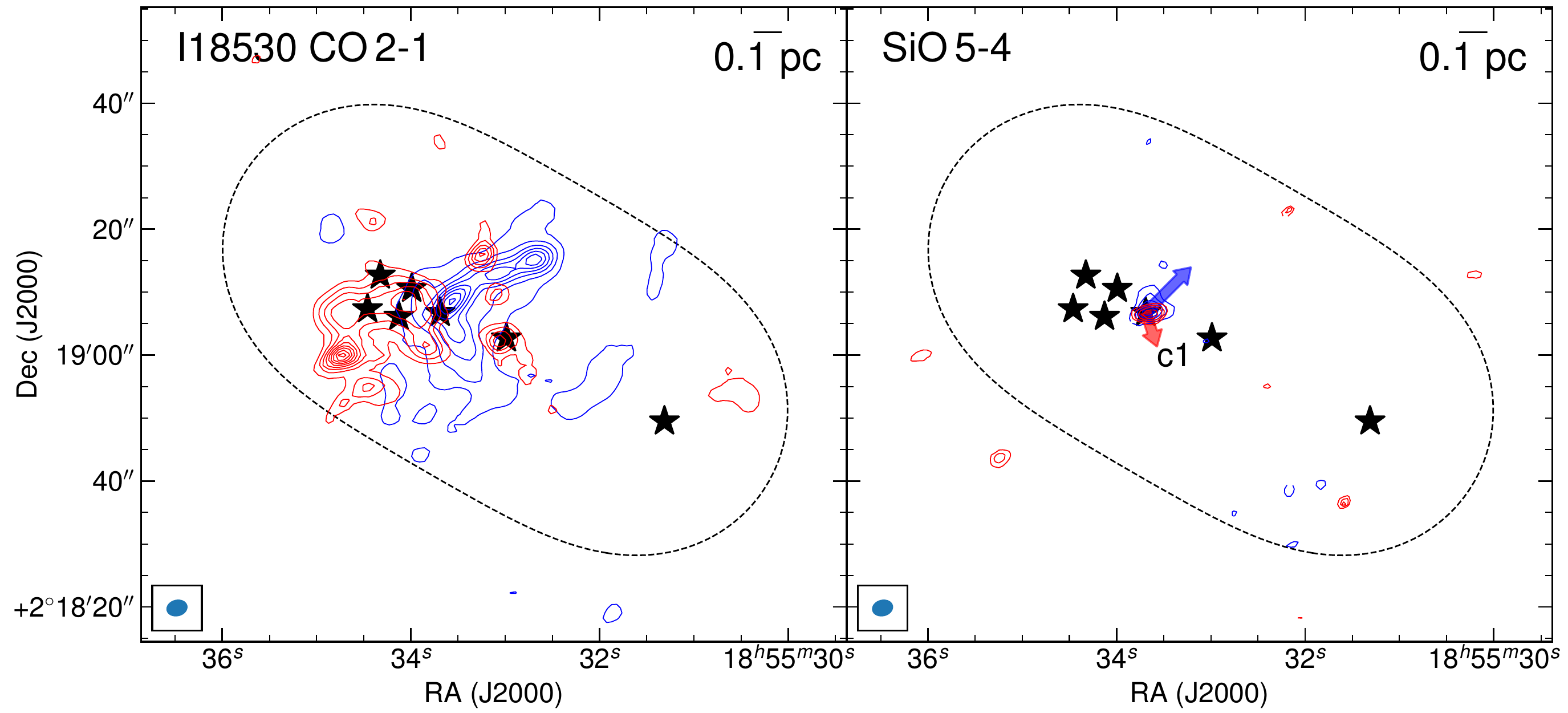} & \includegraphics[width=0.495\textwidth]{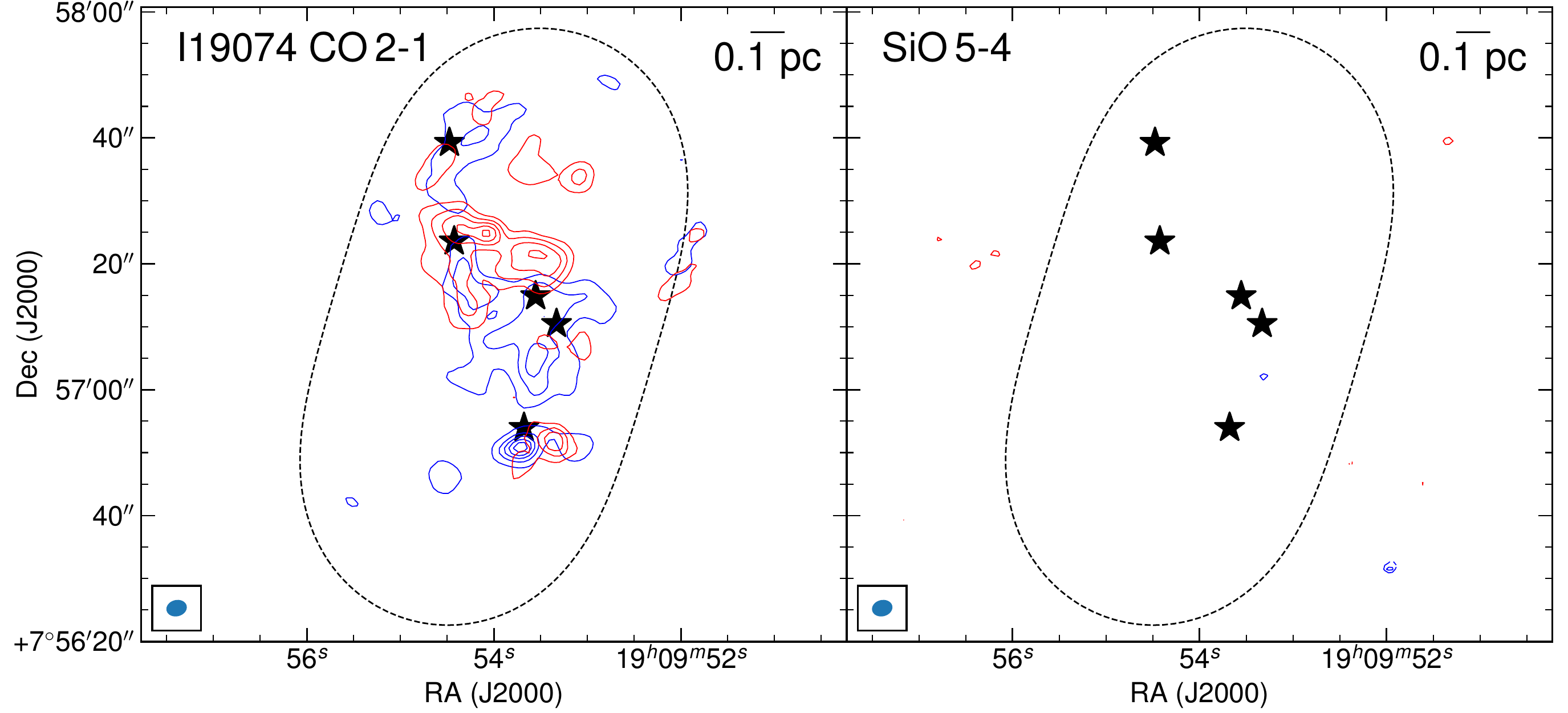} \\
\includegraphics[width=0.495\textwidth]{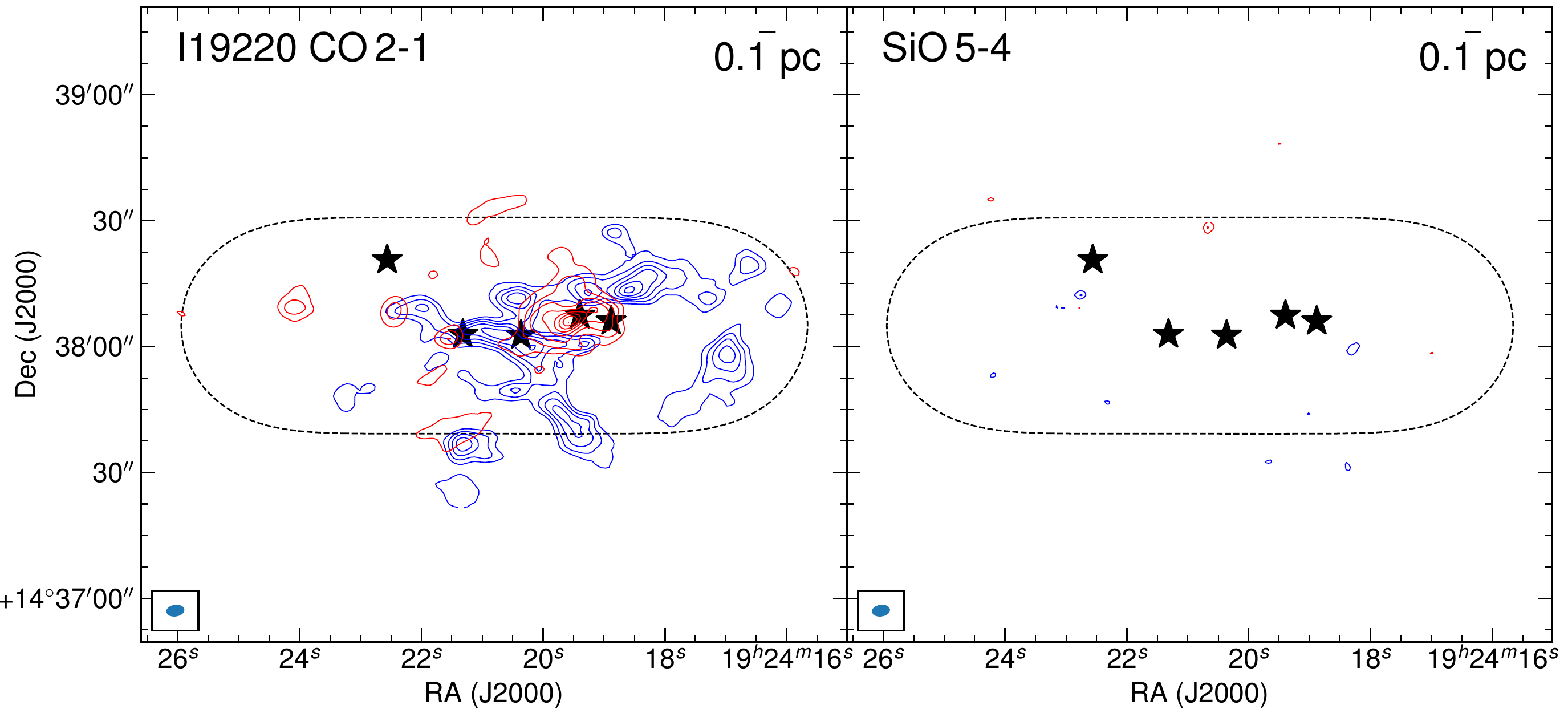} & \includegraphics[width=0.495\textwidth]{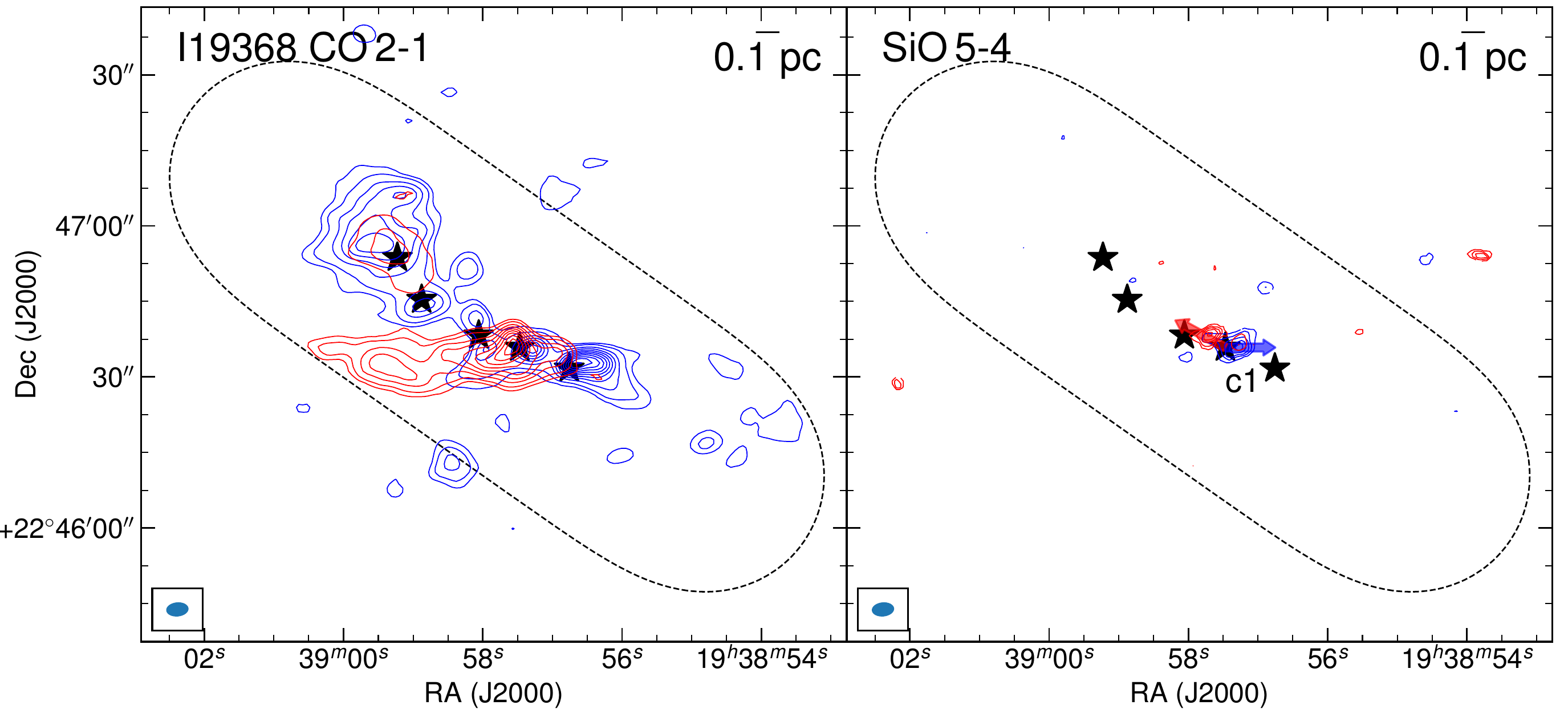} \\
\end{tabular}
\caption{Potential protostellar outflows traced by \twelveco{} and SiO emission. In each panel, the red and blue contours show integrated intensities of the red and blue-shifted gas components within the velocity ranges listed in \autoref{tab:outflow}. For I18310 and I18337 where significant foreground or background \twelveco{} contamination at \vlsr{}$\sim$100~\kms{} is seen, those velocities are excluded when making the \twelveco{} integrated intensity maps. Red and blue arrows in SiO maps of I18308, I18310, I18337, I18530, and I19368 show potential outflows. The black stars mark positions of the dense cores in each cloud. The dashed loops mark the SMA mosaic field. An interactive version of this figure is available in the online journal and at \href{https://xinglunju.github.io/outflows.html}{https://xinglunju.github.io/outflows.html}, where 3-D datacubes of \twelveco{} and SiO are presented. In the interactive modules, the 3-D contours of the \twelveco{} data are plotted at the 20$\sigma$ level, while those of the SiO data are at the 4$\sigma$ level, where 1$\sigma$ is 40~\mjypbm{} for the first 4 sources and 80~\mjypbm{} for the next 4 sources. The 3-D contours are color coded, to show the blue and red-shifted gas components. Contours on the back sides of the cubes show the SMA 1.3~mm continuum emission, with levels at 5, 10, 15, 20, 25, 30, 50, 100, and 150$\sigma$ levels, where 1$\sigma$ is 1~\mjypbm{}. White loops on the back sides mark the SMA mosaic field. Black opaque spheres inside the cubes mark the spatial coordinates and \vlsr{} (defined by a single Gaussian fitting to the mean \fmh{}, \methanol{}, or \ceighteeno{} spectra of the dense cores) of the dense cores. The interactive modules can be mouse controlled with pan, zoom, and rotate.}
\label{fig:outflows}
\end{figure*}

\subsection{Outflows}\label{subsec:cores_sf_outflows}
We use SiO, \twelveco{}, SO, \methanol{}, and \fmh{} lines to trace protostellar outflows around dense cores. \autoref{fig:outflows} shows maps of the \twelveco{} and SiO emission toward the 8 clouds, while an interactive version that demonstrates the 3-D position-position-velocity datacubes is available online to better illustrate potential outflows, which was made with a customized code TDViz\footnote{\href{https://github.com/xinglunju/tdviz}{https://github.com/xinglunju/tdviz}} utilizing Mayavi for 3-D visualization. Although the \twelveco{} line is brighter than SiO, its morphology is more complicated due to contributions from foreground and background gas. For example, as shown in the interactive version of \autoref{fig:outflows} available online, significant foreground or background gas components at \vlsr{}$\sim$100~\kms{} can be seen in the \twelveco{} data toward I18310 and I18337, which is likely attributed to the Scutum or the 4-kpc arm \citep{reid2014}. The SiO emission is weaker but presents a simple morphology, hence revealing potential bipolar outflows more clearly than \twelveco{}. In \autoref{fig:outflows}, outflow candidates traced by SiO are marked toward I18308-c1, I18308-c4, I18310-c1, I18337-c2, I18337-c3, I18337-c6, I18530-c1, and I19368-c1. These outflows are also seen in \twelveco{}. SiO emission seen only in blue-shifted gas toward several dense cores in I18308, or associated with no dust emission in I18337, is uncertain for tracing outflows, although possible counterpart in \twelveco{} emission is observed.

The energetics of these SiO outflows, including their masses, dynamic ages, and outflow rates, are derived assuming LTE and optically thin emission after correcting for primary beam responses of the SMA. The fractional abundance of SiO to H$_2$ is assumed to be 5$\times$10$^{-10}$ \citep{sanhueza2012}. We also assume that all outflows are parallel to the plane of the sky, in which case the derived dynamic ages are lower limits while the derived outflow rates are upper limits. The derived parameters are listed in \autoref{tab:outflow}. The masses of these outflows are a few \msol{} and outflow rates are of order 10$^{-4}$~\msol{}\,yr$^{-1}$, both typical for massive outflows in high-mass star-forming regions \citep{beuther2002outflow,zhang2005}. Their dynamic ages, on the other hand, are of order 10$^3$--10$^4$ years, suggesting that the protostellar objects launching these outflows are young. Note that a variety of SiO abundances from 10$^{-10}$ to 10$^{-7}$ has been observed toward protostellar outflows \citep[see discussion in][]{feng2016a}. If a larger SiO abundance is adopted, the outflow masses will be smaller while the dynamic ages are unaffected. Therefore, the outflow masses as well as the outflow rates we derived are likely upper limits.

We also note that the SiO outflows seem to be preferably detected in the most massive cores in each cloud. This may be explained if the most massive cores evolve faster than the other cores. Then the more evolved cores are able to drive high-velocity outflows that can destroy dust grains and produce SiO molecules.

In several cases, potential bipolar outflows are seen in \twelveco{} but not in SiO, possibly because emission of the latter is dimmer. Examples can be found in I18460, I19074, and I19220 in \autoref{fig:outflows}. However, close to the systemic velocities of the sources (e.g., $\pm$5~\kms{}), images of \twelveco{} are largely subject to missing flux and self-absorption. Therefore, within these velocity ranges, \twelveco{} images cannot achieve a high intensity dynamic range, and do not provide a high intensity contrast in between outflows and the ambient material. Outflows which have lower velocities, or those of which the axes are approximately perpendicular to our line-of-sights, may not be identified from the observations of \twelveco{}. Alternatively, by observing some molecules which are abundant in warm environment (e.g., SO, \methanol{}, \fmh{}), it may be possible to spatially resolve outflows which have lower line-of-sight velocities, or can pick out outflows from the locally enhanced linewidths \citep{wright1996,baobab2010}.

In the following \autoref{subsec:cores_sf_IR}, we look for potential outflows associated with the infrared dark dense cores traced by \twelveco{}, SiO, SO, \methanol{}, or \fmh{} lines, to rule out the possibility of prestellar cores. As shown in \autoref{subsec:cores_sf_IR} two infrared dark dense cores do show signatures of outflows hence are likely protostellar.

In addition to the SMA lines, an \water{} maser at 22.2~GHz, usually excited by outflows in star-forming regions \citep{elitzur1989}, is detected towards I18308-c1 \citep{beuther2002maser}. No other masers have been detected in previous observations toward these 8 clouds \citep[see Table~3 of][]{lu2014}.

\subsection{Hot Molecular Cores}\label{subsec:cores_sf_hmc}
Several dense cores present a variety of 1.3~mm spectral lines in the SMA observations, including several complex organic molecules (\methanol, \mthc, \cyacet). An example has been shown in \autoref{fig:smaspec}. These species are usually seen in hot molecular cores, where feedback from protostars heat ambient gas in a radius of $\sim$0.01~pc to temperatures of $>$100~K and evaporate dust grain surface to release the molecules \citep[e.g.,][]{beuther2006,beuther2007c,silva2017}. The most evident hot molecular cores in our sample include I18308-c1 and I18337-c1. Other dense cores may show a few \methanol{} lines but are not detected in weaker lines such as \mthc{}. However, \methanol{} has also been detected in quiescent regions \citep[e.g.,][]{sanhueza2013}, hence does not uniquely trace hot molecular cores. Therefore, we apply a simplified criteria of the detection of any characteristic hot core lines other than \methanol{} (e.g., \cyacet, \mthc{}), and mark the hot molecular core candidates in \autoref{tab:sf}. Some of the hot molecular cores (e.g., I18308-c1) are already embedded with UC \hii{} regions.

\subsection{Infrared Emission}\label{subsec:cores_sf_IR}
Embedded protostars heat ambient dust, which then emits at infrared wavelengths. Infrared emission hence is often used to characterize star formation activities in molecular clouds. Here we focus on three infrared wavelengths: the 4.5~\micron{} and 24~\micron{} emission from the \textit{Spitzer} space telescope, and the 70~\micron{} emission from the \textit{Herschel} space telescope. The 4.5~\micron{} and 24~\micron{} maps are retrieved from the Spitzer Heritage Archive hosted in the NASA/IPAC Infrared Science Archive. The 70~\micron{} maps are retrieved from the Hi-GAL Catalogs and Image Server hosted in the ASI Science Data Center\footnote{\href{http://tools.asdc.asi.it/HiGAL.jsp}{http://tools.asdc.asi.it/HiGAL.jsp}}.

Extended 4.5~\micron{} emission has been suggested to trace shocked gas in outflows \citep{qiu2008,cyganowski2008,cyganowski2011,takami2010,takami2012}. 24~\micron{} emission usually suggest thermal dust emission, which can be heated by protostars \citep{carey2009,gutermuth2015}. 70~\micron{} data are complementary to 24~\micron{}, since they could reveal deeply embedded protostars that are undetected at 24~\micron{} \citep{guzman2015}.

Dense cores appearing dark at the three wavelengths unlikely harbor any massive protostars, therefore are candidates of prestellar cores before embedded protostars are born. Note that they are still `candidates' of prestellar cores because several recent studies pointed out that there may be embedded low or intermediate-mass protostars or high-mass protostars at very early evolutionary phases in such dense cores \citep{tan2016,feng2016a,traficante2017}. Indeed, we look for potential outflows associated with these infrared dark dense cores as described in \autoref{subsec:cores_sf_outflows} and \autoref{fig:outflows}, and find two dense cores, I18337-c6 and I19074-c2, which appear to be infrared dark but are associated with SiO or \twelveco{} outflows, hence are more likely protostellar instead of prestellar.

The results of the infrared emission are listed in \autoref{tab:sf}. We find 7 dense cores that are dark at all three wavelengths, including the two that are associated with molecular outflows. The other dense cores show bright emission at one or more wavelengths, hence already have embedded protostars.

\begin{figure*}[!t]
\centering
\includegraphics[width=0.98\textwidth]{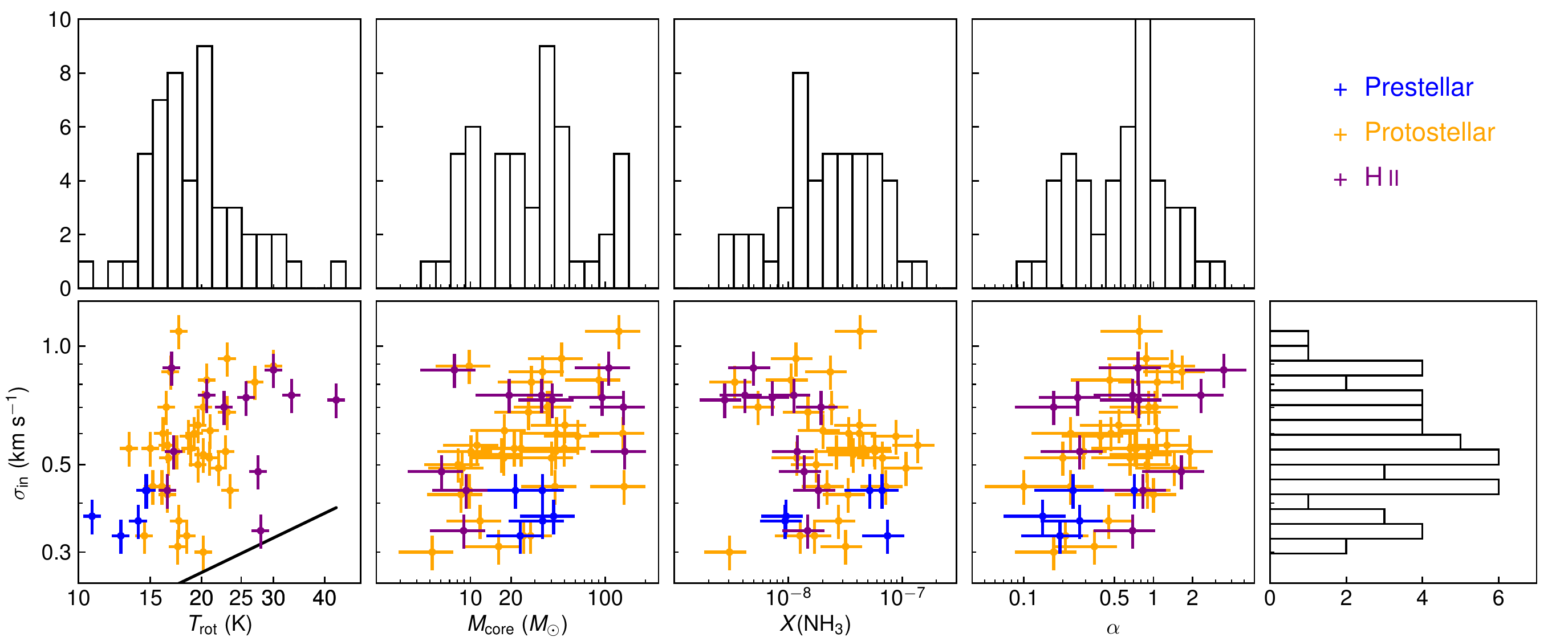}
\caption{Physical properties of dense cores. The $x$-axes of the four scatter plots, from left to right, are \amm{} rotational temperature, core mass, \amm{} fractional abundance, and virial parameter, respectively, while the $y$-axes are all intrinsic linewidth measured with \amm{} lines. The scatter plots are color coded to show the three evolutionary phases. Error bars correspond to the uncertainties discussed in \autoref{subsec:cores_error}. In the left most panel, where linewidths are plotted again temperatures, the solid line shows the expected isothermal sound speed at a given temperature. Histograms attached to the scatter plots show distributions of these properties with all three evolutionary phases together.}
\label{fig:core_scatter}
\end{figure*}

\subsection{Evolutionary Phases of Dense Cores}\label{subsec:cores_sf_evolution}
We tentatively assign evolutionary phases to the dense cores based on their star formation signatures discussed above. We identify three phases: prestellar, in which protostars have not formed hence no star formation activities are detected; protostellar, in which protostars have formed and star formation has been detected; and (UC) \hii{}, in which young stars ionize ambient gas and free-free emission is detected. The phases are marked in \autoref{tab:sf}.

Among the 50 dense cores, 5 cores are identified as candidates of prestellar cores. They show no infrared emission, centimeter continuum emission, or outflows. Other than the three CO isotopologues, they are usually free of 1.3~mm spectral lines (\autoref{fig:smaspec}). A total of 11 dense cores are associated with compact free-free emission above 5$\sigma$ levels, as marked in \autoref{tab:sf}. They likely have embedded (UC) \hii{} regions. The remaining 34 dense cores show various signatures of star formation but no detectable free-free emission above the 3$\sigma$ level, hence are classified as protostellar.

\subsection{Distribution of Dense Core Properties and Correlation with Evolutionary Phases}\label{subsec:cores_stats}
Physical properties of the 50 identified dense cores, including their linewidths, \amm{} rotational temperatures, masses, \amm{} abundances, and virial parameters, are plotted in \autoref{fig:core_scatter}. The markers are color coded to show the three evolutionary phases: prestellar, protostellar, and (UC) \hii{} regions. Mean and median values of the properties are summarized in \autoref{tab:cores_stats}.

Overall distributions of the properties are illuminated by the histograms in \autoref{fig:core_scatter}, while those of each evolutionary phases can be seen in the scatter plots in \autoref{fig:core_scatter}. The intrinsic linewidths lie in the range of 0.3--1.0~\kms{}, while the \amm{} rotational temperatures are 10--40~K, corresponding to isothermal sound speeds of 0.19--0.38~\kms{}. As indicated by the solid line in \autoref{fig:core_scatter}, all the dense cores have linewidths larger than the expected isothermal sound speeds, hence are dominated by non-thermal motions. The mean linewidth and temperature of the dense cores are 0.58~\kms{} and 20.0~K, respectively, in agreement with previous interferometric studies of dense cores \citep{sanchezmonge2013,lu2014}.

The core masses lie between 5 and 140~\msol{}. The smallest core mass, 5~\msol{}, corresponds to 15 times the 1$\sigma$ mass sensitivity assuming a dust continuum RMS of 1~mJy, dust temperature of 20~K, and distance of 4~kpc (see \autoref{equ:mass}). However, as mentioned in \autoref{subsec:obs_sma}, the continuum RMS will be much larger than 1~mJy around bright sources. Given that massive cores of $\gtrsim$100~\msol{} are detected in these clouds, less massive ($<$5~\msol{}) dense cores around the massive ones may be missed in the SMA observations simply due to limited dynamic ranges. The most massive cores in each cloud, which are usually of $\gtrsim$100~\msol{}, lead to a secondary peak in the histogram of core masses. As mentioned in \autoref{subsec:cores_id} some of these massive cores may be resolved into multiple substructures with higher angular resolution. We also note that 3 prestellar core candidates have sufficient mass ($>$30~\msol{}) to form high-mass stars assuming a typical star formation efficiency of 30\% at the dense core level \citep{alves2007,konyves2015}. However, higher angular resolution observations are necessary to discard further fragmentation.

The \amm{} abundances vary between 3$\times$10$^{-9}$ and 2$\times$10$^{-7}$, with a mean of 3$\times$10$^{-8}$, consistent with previous studies toward high-mass star-forming regions \citep[e.g.,][]{harju1993,dunham2011,wienen2012}.

Virial parameters of all the dense cores, except for 2 (UC) \hii{} regions, are $<$2, suggesting that they are gravitationally bound when ignoring rotation and magnetic field and assuming constant radial density profiles. In particular, the 5 prestellar core candidates have virial parameters of $\lesssim$0.5, hence are strongly self-gravitating. Therefore, they are likely to keep collapsing and form protostars. These small virial parameters are consistent with those found in other prestellar core candidates embedded in massive clumps \citep[e.g.,][]{zhang2015,lu2015a,ohashi2016,sanhueza2017}.

We discuss potential correlations between dense core properties and the three evolutionary phases. Based on the scatter plots in \autoref{fig:core_scatter} and the statistics in \autoref{tab:cores_stats}, the linewidths, \amm{} rotational temperatures, \amm{} abundances, and virial parameters show strong correlation with the three evolutionary phases.

The 5 prestellar core candidates show lower temperatures of $\lesssim$15~K and smaller linewidth of $\lesssim$0.4~\kms{}, the (UC) \hii{} regions show higher temperatures of $\gtrsim$25~K and larger linewidth of $\gtrsim$0.6~\kms{}, while the protostellar cores show intermediate properties between the former two phases. The difference in temperatures and linewidths is likely attributed to feedback of embedded protostars, i.e., radiation and outflows. Similar correlations have been reported in e.g., \citet{sanchezmonge2013} and \citet{lu2014}.

The \amm{} abundances in the (UC) \hii{} regions tend to be lower as compared to the pre/protostellar cores. This may be a consequence of the destruction of \amm{} in (UC) \hii{} regions \citep[e.g., photodissociation,][]{lee1984}, or the underestimate of \amm{} column densities in (UC) \hii{} regions: the (1,~1) and (2,~2) lines are no longer a good thermometer when temperatures are $>$30~K, especially when they become optically thick \citep{ho1983}. Being optically thick also brings in additional uncertainty for estimating column densities of each transition. Temperatures of $>$30~K and optically thick \amm{} lines are found in the (UC) \hii{} regions, in which case column densities could be underestimated.

At last, the mean values of the virial parameter for prestellar core candidates, protostellar cores, and (UC) \hii{} regions, increase from 0.31, 0.73, to 1.08. As shown in \autoref{fig:core_scatter} and \autoref{tab:cores_stats}, the core masses of different evolutionary phases do not show significant discrepancy, most of which are $\sim$20--50~\msol{}. The difference in virial parameters across the three evolutionary phases is mainly due to the variation of linewidths. Taking no account of the effect of rotation, magnetic field, or inhomogeneous density profile in cores, this suggests that the cores evolve from a strongly self-gravitating state in the prestellar phase to virial equilibrium in the (UC) \hii{} phase, and this process is regulated by the increase of linewidth which is dominated by turbulence.

\section{Fragmentation and Accretion in Filaments}\label{sec:filaments}

\subsection{Gravitational Fragmentation of Filaments}\label{subsec:filaments_fragmentation}
It has been suggested that in high-mass star-forming filaments thermal pressure is far from enough to support the fragmentation. Additional support by turbulent pressure may be necessary to bring the filaments to equilibrium in the radial directions, so that instability along the filaments and subsequent fragmentation can grow \citep{jackson2010,lu2014,wang2014,beuther2015,feng2016c,contreras2016}. To investigate the fragmentation of filaments hence the formation of dense cores, we compare our observations to the model of a collapsing isothermal cylinder, supported by either thermal or non-thermal motions. We select four clouds in the sample that show morphologically well-defined filaments and fragmentation in the SMA dust emission: I18308, I18337, I19220, and I19368. As shown in the maps in \autoref{fig:cores}, the aspect ratios of the filaments revealed by the SMA dust emission are $\gtrsim$10. Dense cores embedded in these filaments tend to be aligned along their major axes and regularly spaced \citep[also see example in IRDC G28.34+0.06,][]{zhang2009}. The other 4 clouds do not present a clear filamentary morphology in the SMA dust emission, therefore are not included in the analysis.

\begin{figure}[!t]
\centering
\includegraphics[width=0.475\textwidth]{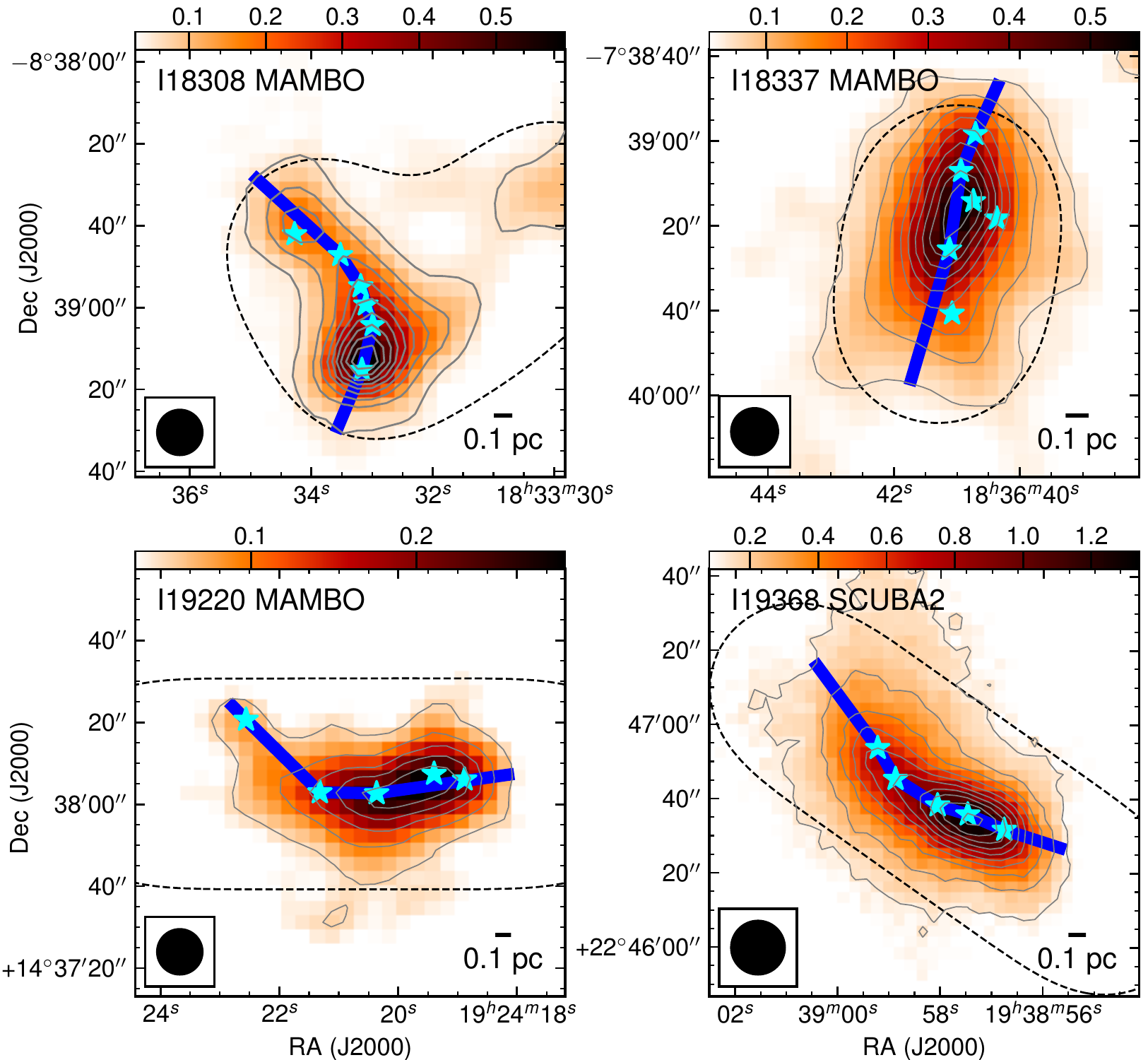} 
\caption{Single-dish dust emission maps of the 4 filaments. Both background images and contours show dust emission mapped by bolometer arrays. Contour levels start at 5$\sigma$ in steps of 5$\sigma$, where 1$\sigma$ is $\sim$11~\mjypbm{} for MAMBO \citep{beuther2002mambo} and 26.4~\mjypbm{} for SCUBA2 \citep{eden2017}. The unit of color scales is \jypbm{}. The definition of `filaments' in \autoref{subsec:filaments_fragmentation} is illustrated with blue bars in each panel. Dense cores detected by the SMA are marked by cyan stars, with dashed loops showing the SMA mosaic fields. Note that dense cores located outside the filament in I18308 are not marked.}
\label{fig:bolometer}
\end{figure}

By applying the cylindric collapse models \citep[e.g.,][]{chandra1953,ostriker1964,nagasawa1987,fiege2000,kirk2015}, we compare observations with models that take thermal or turbulent (or magnetic field) support into account, and determine which mechanisms are responsible for the fragmentation of filaments.  Two characteristic parameters can be easily observed and compared to models: the mass per unit length of the filaments, $(M/l)_\text{obs}$, and the projected separation between dense cores, $\lambda_\text{prj}$.

First, we obtain the two parameters from observations. We first estimate $(M/l)_\text{obs}$ based on the SMA dust emission. Following \autoref{equ:mass}, the masses of the filaments are calculated. Dust temperatures, under the LTE conditions, are taken to be the mean rotational temperatures inside each filament derived from VLA \amm{} \citep[see \autoref{subsec:cores_temp} and ][]{lu2014}. We do not choose the single-dish \amm{} temperatures \citep[e.g.,][]{sridharan2002}, to avoid confusion due to large velocity gradients inside a single-dish beam. The projected lengths of the filaments are those of the shadowed bars in \autoref{fig:cores}. $(M/l)_\text{obs}$ are then the masses divided by the projected lengths. These results are listed in \autoref{tab:filaments}. However, due to the missing flux issues of interferometers, the masses are deemed to be underestimated. $(M/l)_\text{obs}$ derived in such way should be a lower limit.

Therefore, additionally, we make use of the single-dish bolometer array observations from MAMBO \citep{beuther2002mambo} or SCUBA2 \citep{eden2017} to estimate $(M/l)_\text{obs}$, as shown in \autoref{fig:bolometer}. These data provide sufficiently high angular resolutions (11\arcsec{}--14\arcsec{}) to resolve the filaments, and recover diffuse emission in the filaments missed by the SMA. Dust emission fluxes are measured in each filament above 5$\sigma$ levels. The adopted dust opacity are interpolated from the tabulated values of \citet{ossenkopf1994}, assuming MRN model with thin ice mantles after 10$^5$ years of coagulation at 10$^6$~\cc{}. Then masses of the four filaments are derived using \autoref{equ:mass}, assuming a gas-to-dust mass ratio of 100. These parameters are all listed in \autoref{tab:filaments}. Indeed, the masses derived from single-dish observations are 2 to 3 times larger than those derived from the SMA data. The projected lengths are measured by tracing curvatures of the filaments in the singled-dish continuum images, as marked by the bars in \autoref{fig:bolometer}, starting and ending at the contour of 5$\sigma$ levels. $(M/l)_\text{obs}$ are then the masses of filaments divided by the projected lengths.

The projected separation $\lambda_\text{prj}$ is measured in the SMA continuum maps. Separations between each pair of adjacent dense cores are obtained, and a mean separation for each filament is listed in \autoref{tab:filaments}.

Note that given the projection effect, $(M/l)_\text{obs}$ should be an upper limit while $\lambda_\text{prj}$ should be a lower limit, apart from other uncertainties as discussed in \autoref{subsec:cores_error}.

Then, we derive the critical mass per unit length $(M/l)_\text{crit}$ and fragment separation $\lambda_\text{crit}$ expected from the isothermal cylindric collapse models. Above the critical mass per unit length, the filament will collapse radially and fragmentation along filaments is not expected \citep{inutsuka1992}. Therefore to allow fragmentation along filaments to proceed, the observed critical mass per unit length must not exceed the critical value. We have \citep{inutsuka1992,jackson2010,wang2014}:
\begin{equation}\label{equ:ml}
(M/l)_\text{crit}=2\sigma_v^2/G=465(\frac{\sigma_v}{1~\kms{}})^2~\msol{}\,\text{pc}^{-1},
\end{equation}
and
\begin{equation}\label{equ:lambda}
\begin{split}
\lambda_\text{crit}&=22v(4\pi G\rho_c)^{-0.5}\\
&=0.39(\frac{\sigma_v}{1~\kms{}})(\frac{n_c}{10^6~\cc{}})^{-0.5}~\text{pc}.
\end{split}
\end{equation}

Here $n_c$ is the gas density at the center of the filament. Ideally, $n_c$ is the undisturbed density before gravitational collapse starts, which is inaccessible in these already star-forming clouds. It is inappropriate to use the density between dense cores as an approximation either, as pointed out by \citet{heigl2016}. Therefore, we take a range of densities detected in dense cores, 2$\times$10$^5$--2$\times$10$^6$~\cc{}, to approximate $n_c$. This is larger than the mean density of the filaments, $\sim$5$\times$10$^4$~\cc{}, assuming a cylinder geometry parallel to the plane of the sky and dividing the mass of filaments by their volumes. This uncertainty of $n_c$ only affects the derivation of $\lambda_\text{crit}$, which is actually not as sensitive to the choice of density as linewidth.

In the case of thermal pressure support of filamentary fragmentation, as for the low-mass star formation filaments, the velocity $\sigma_v$ in Equations~\ref{equ:ml} \& \ref{equ:lambda} is substituted by the isothermal sound speed $c_s$, derived from the mean \amm{} rotational temperatures. If additional turbulent support is taken into account, $\sigma_v$ is substituted by the total linewidth $\sigma_\text{in}$ including both thermal and non-thermal components, derived following \autoref{equ:sigma_in}, which is also based on the VLA \amm{} data to resolve out velocity gradients inside filaments. Note that the non-thermal component in $\sigma_\text{in}$ includes not only turbulent pressure, but also possible systematic motions inside the telescope beam, such as infall, which would not support the fragmentation (see discussion in the next subsection). On the other hand, the use of the spatially resolved VLA \amm{} data may filter out large scale turbulence. Linewidths from single dish \amm{} data \citep[e.g.,][]{wienen2012} can be used as an upper limit of turbulent strength which may include significant contribution from unresolved systematic motions.

The critical masses per unit length and fragment separations are listed in \autoref{tab:filaments}. For thermal support, the critical mass per unit length is always an order of magnitude smaller than what is observed. If we assume a high density $n_c$ up to 2$\times$10$^6$~\cc{}, the fragment separations are smaller than observed ones by a factor of 3. However, a low density $n_c$ of 2$\times$10$^5$~\cc{} leads to fragment separations consistent with the observed values.

Taking additional support from turbulent pressure into account, we find that observed masses per unit length are still a factor of 2.5--6 times larger than critical values. Even applying the single dish linewidths from \citet{wienen2012} to represent the turbulent pressure, which is certainly an upper limit, two filaments, I18308 and I18337, are still super-critical with observed masses per unit length 1.3--1.4 times larger than critical values. The fragment separations between the model and the observation are in general consistent assuming $n_c$ of 2$\times$10$^6$~\cc{}. If a low gas density $n_c$ of 2$\times$10$^5$~\cc{} is assumed, then the fragment separations from the model are larger than the observed ones by a factor of $>$3.

Therefore, from the point of the mass per unit length, thermal and turbulent pressure combined is still not sufficient to support the fragmentation of these filaments. This is in contrast to several studies toward early evolutionary phases of filaments in massive infrared dark clouds, which found that turbulence is capable of stabilizing filaments \citep[e.g.,][]{wang2011,wang2014,beuther2015,feng2016c}. These studies either derived lower limits for the masses by using the interferometer data \citep{wang2011,wang2014}, or obtained upper limits for the critical mass per unit length by adopting linewidths from single-dish observations  \citep[usually $\gtrsim$1~\kms{};][]{beuther2015,feng2016c}. In either case, the discrepancy between observed masses per unit length and critical values is underestimated. Our results demonstrate the necessity of using zero-spacing data for the estimate of filament masses and using interferometer spectral lines for resolving out systematic motions in order to reliably trace turbulent linewidth.

The fragment separation in filaments, which is less well constrained observationally due to very uncertain initial densities, could favor turbulent support fragmentation if the initial central density $n_c$ is to the high end of 2$\times$10$^6$~\cc{}, or thermal support fragmentation if $n_c$ is to the low end of 2$\times$10$^5$~\cc{}.

The large masses per unit length observed in these filaments may suggest that they are not in dynamic equilibrium, but are unstable to radial gravitational collapse. In such case, the fragmentation observed in these filaments may take place at an earlier time when they are close to equilibrium, e.g., when the masses per unit length are smaller or turbulence is stronger. Observations of filamentary infrared dark clouds in early evolutionary phases find no stronger, but often weaker turbulence when being resolved to 0.1 pc scales \citep[e.g.,][]{ragan2015,henshaw2016b,sanhueza2017}. It is more likely that these filaments evolve from a radial equilibrium state when their masses per unit length are smaller. Evidence of accretion into filaments from a greater environment that is able to double the filament mass in a few free-fall time scales has been found toward several massive filaments \citep[e.g.,][]{schneider2010,peretto2013}.

Alternatively, if we assume these filaments to be in radial equilibrium, one possibility is that they have large inclination angles to the plane of the sky, in which case the observed masses per unit length are overestimated. An inclination angle of 60\arcdeg{} would decrease the observed values by a factor of 2. Another possibility is additional support from magnetic field \citep{fiege2000,kirk2015}. An axial magnetic field of 1~mG would increase the critical mass per unit length by a factor of 2--3. Magnetic field in these filaments has not been studied, although similar strong axial magnetic field has been found in other massive filamentary clouds \citep[e.g.,][]{pillai2015}. A measurement of magnetic field in these filaments will be essential to address this question.

\begin{figure*}[!t]
\begin{tabular}{p{7cm}p{8cm}}
\raisebox{1.0em}{\includegraphics[width=0.42\textwidth]{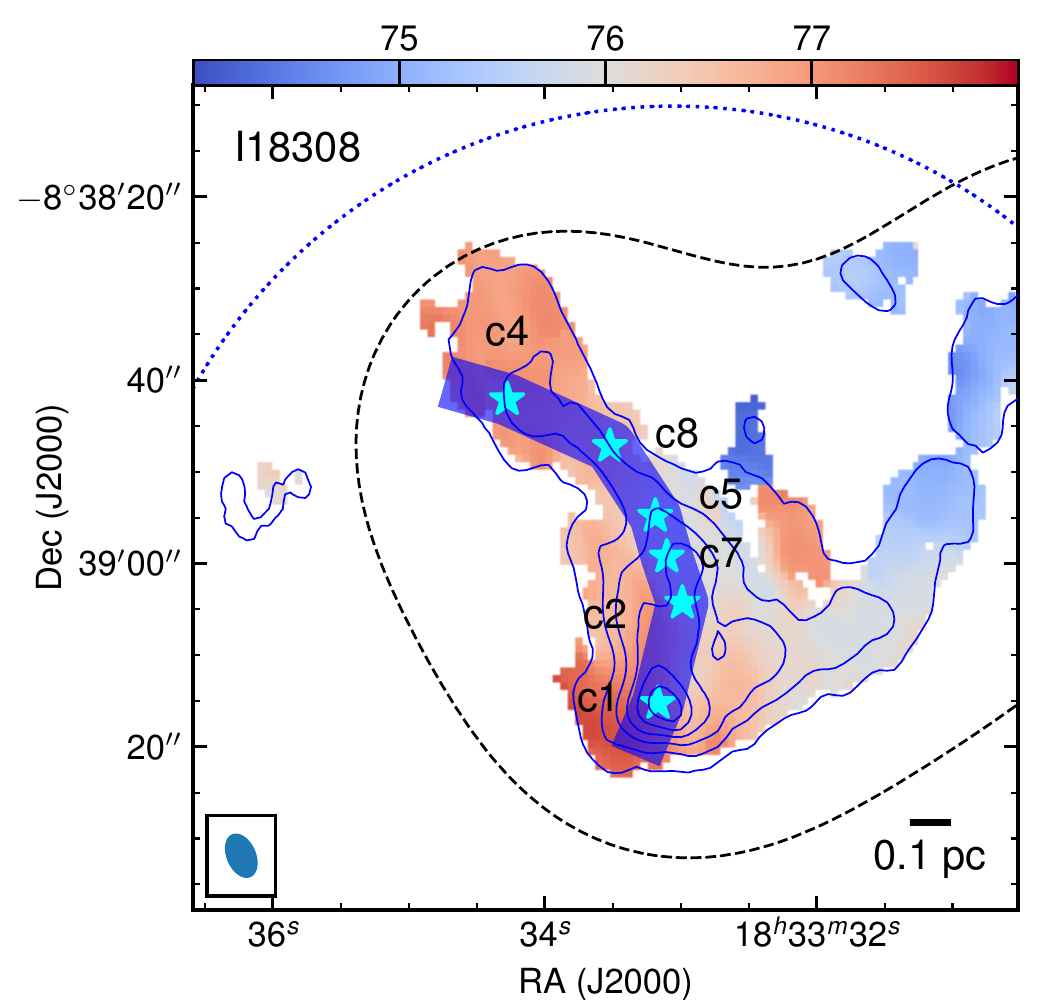}} & \includegraphics[width=0.48\textwidth]{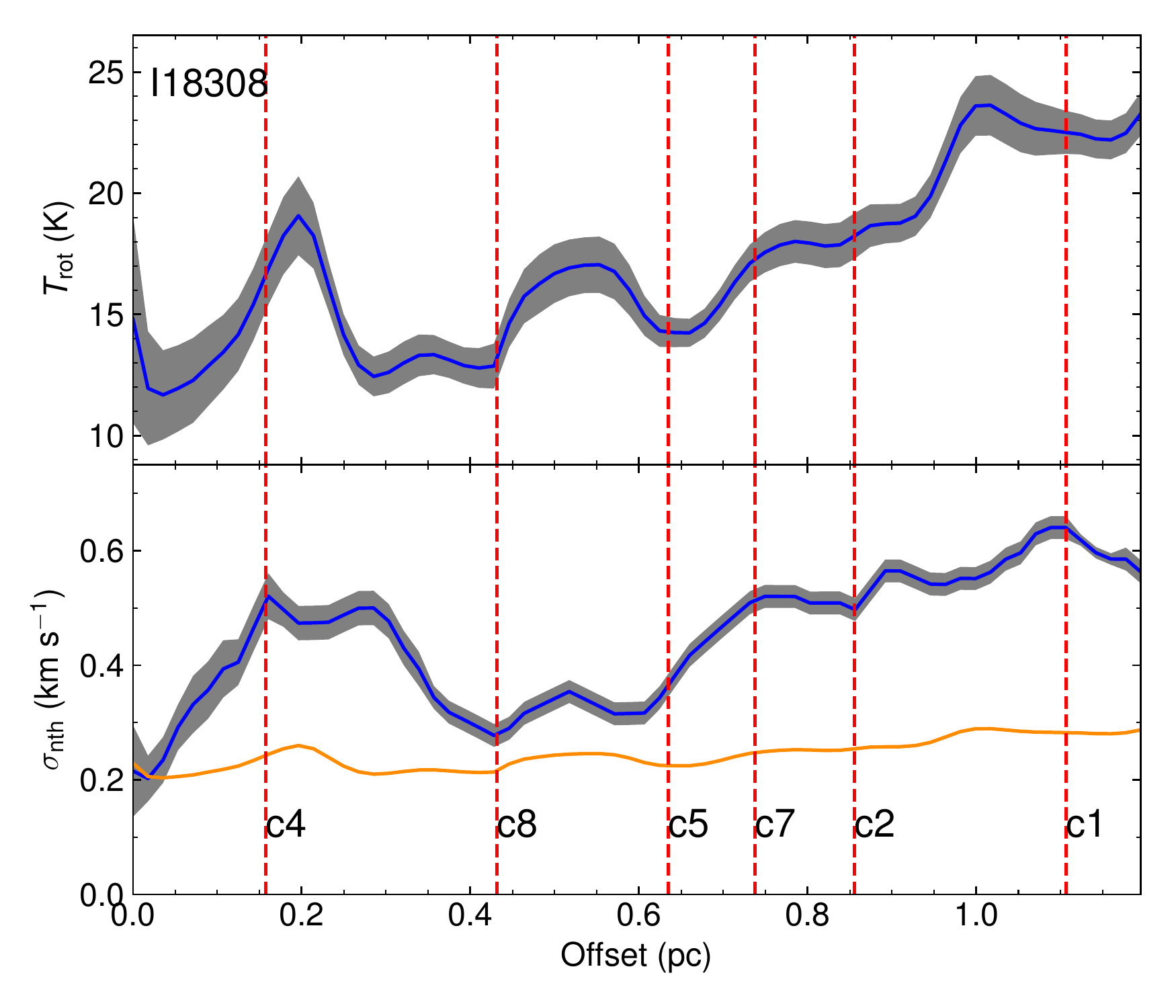} \\ [-1em]
\raisebox{1.0em}{\includegraphics[width=0.42\textwidth]{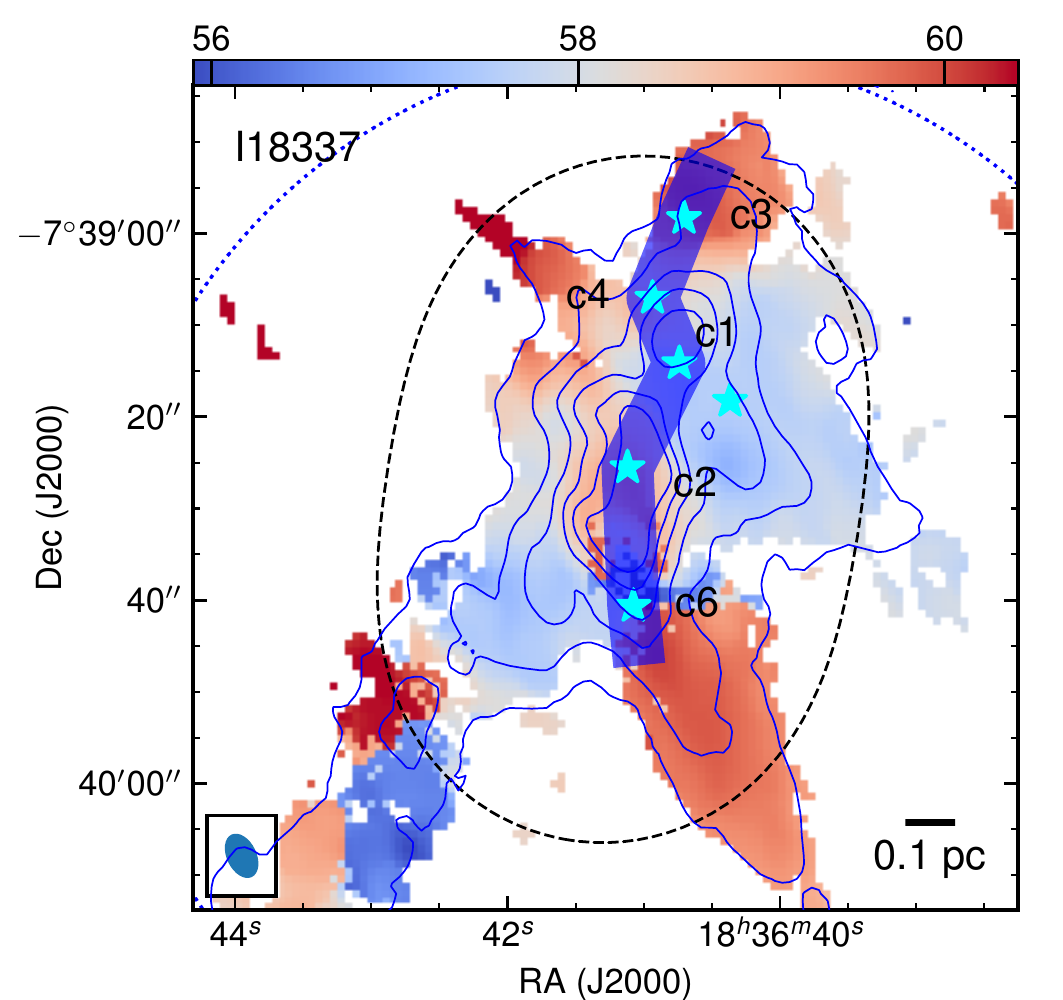}} & \includegraphics[width=0.48\textwidth]{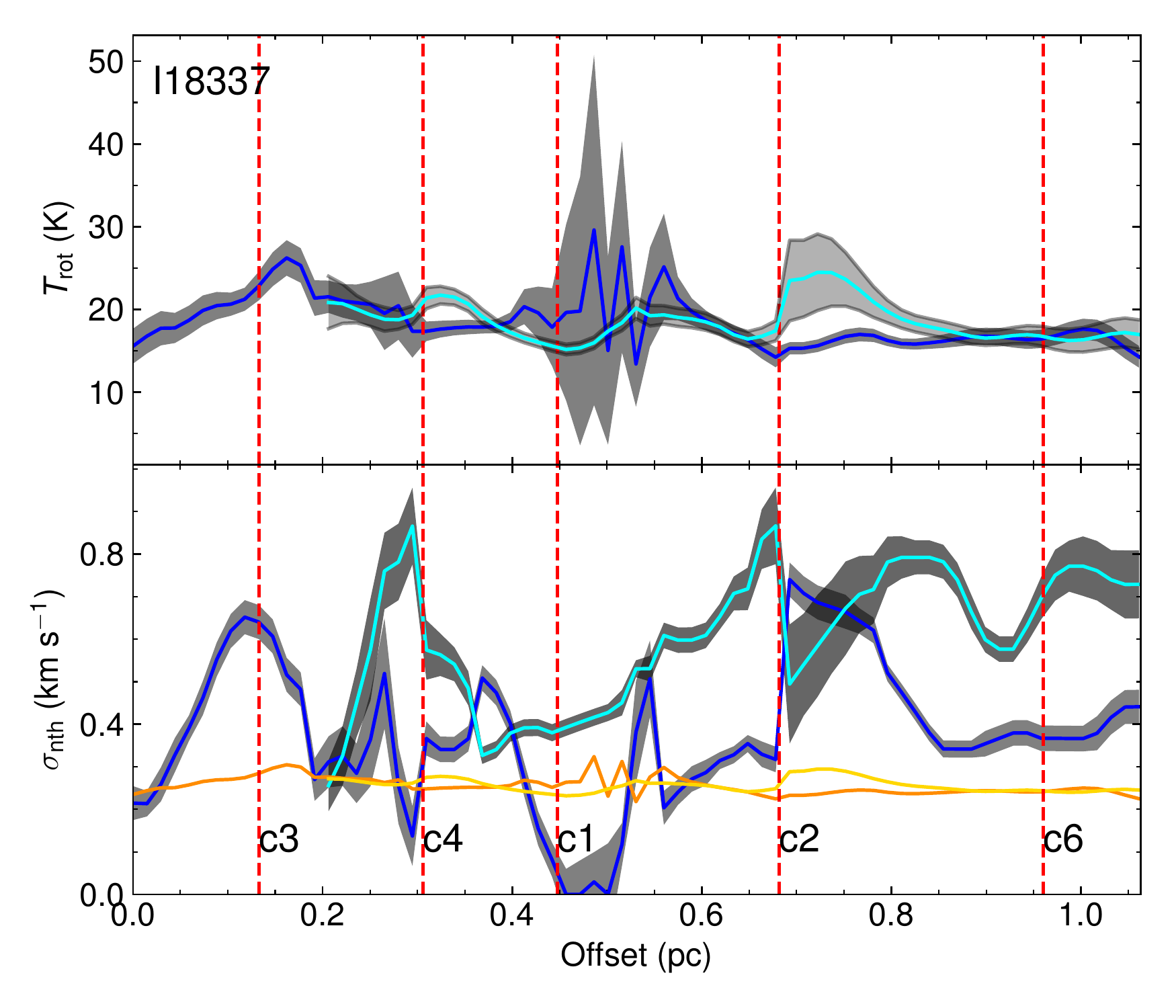} \\ [-1em]
\end{tabular}
\caption{Distributions of gas temperatures and non-thermal linewidths along the filaments. Plots in the left column show the paths from which we extract the \amm{} spectra, marked by blue shadowed bars, with a fixed width of 5\arcsec{}. The \amm{} images are the same as in \autoref{fig:nh3}. Dense cores are marked by cyan stars and labelled. Plots in the right column show gas properties along the paths defined in the left column. For each filament, blue curves in the upper and lower panels show temperatures and non-thermal linewidths along the filament, respectively. Grey shades around the curves mark the 1$\sigma$ errors. For I18337 and I19220 where the second velocity component is seen, the corresponding temperatures and non-thermal linewidths are shown in cyan curves. In the lower panels, the isothermal sound speed is also shown in orange curves, or golden curves for the second velocity component when present. Positions of dense cores along the filaments are marked by vertical dashed lines with core names labelled.}\label{fig:filprop}
\end{figure*}

\addtocounter{figure}{-1}

\begin{figure*}[!t]
\begin{tabular}{p{7cm}p{8cm}}
\raisebox{1.0em}{\includegraphics[width=0.42\textwidth]{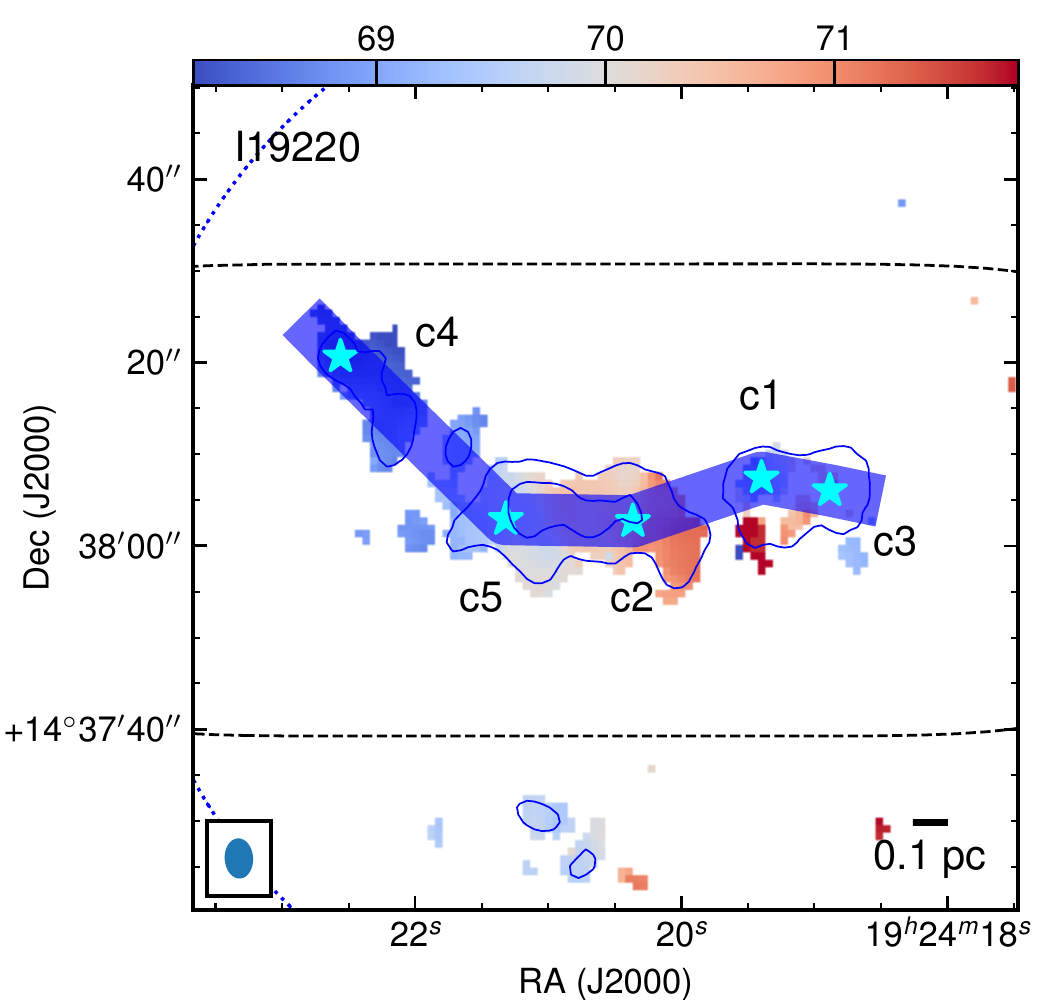}} & \includegraphics[width=0.48\textwidth]{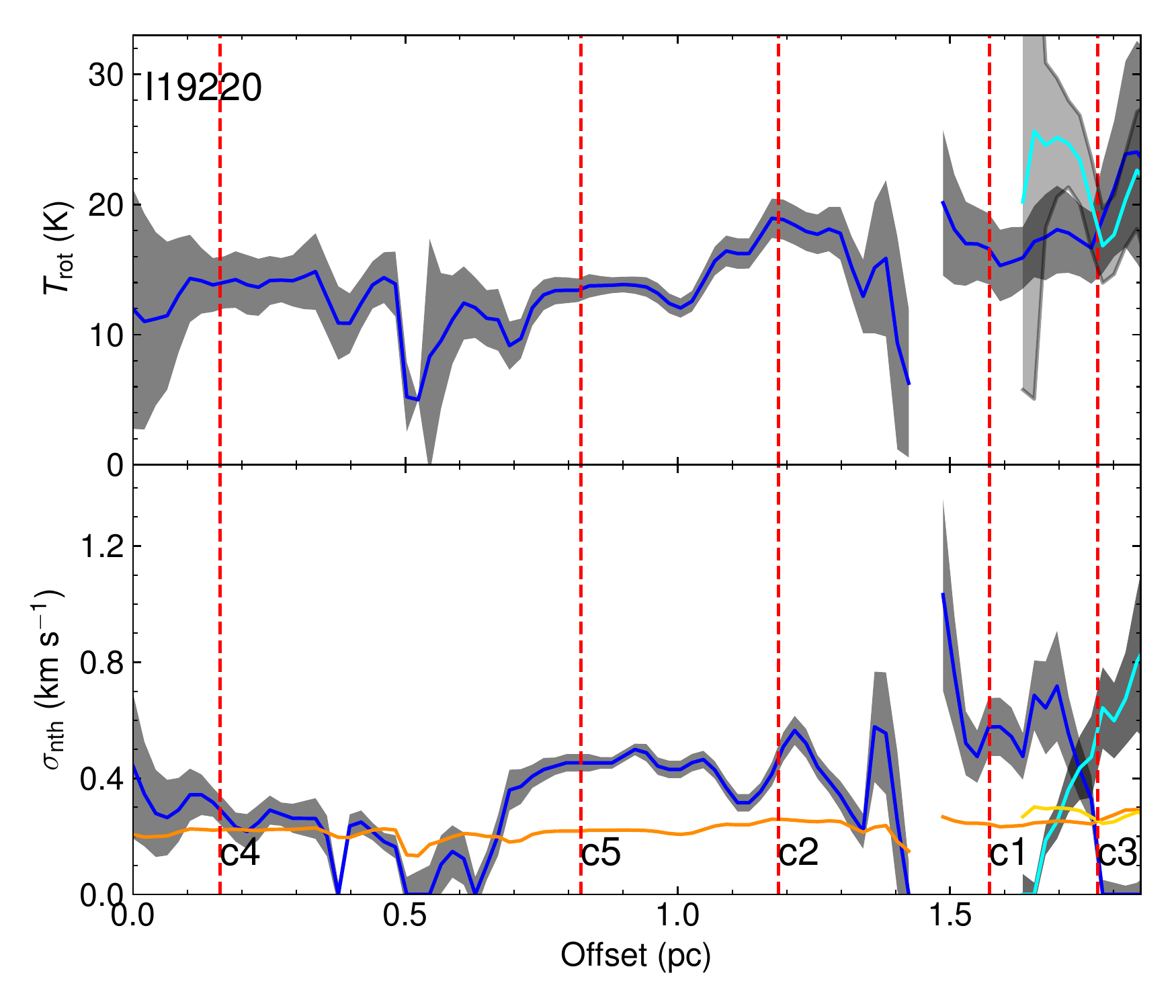} \\ [-1em]
\raisebox{1.0em}{\includegraphics[width=0.42\textwidth]{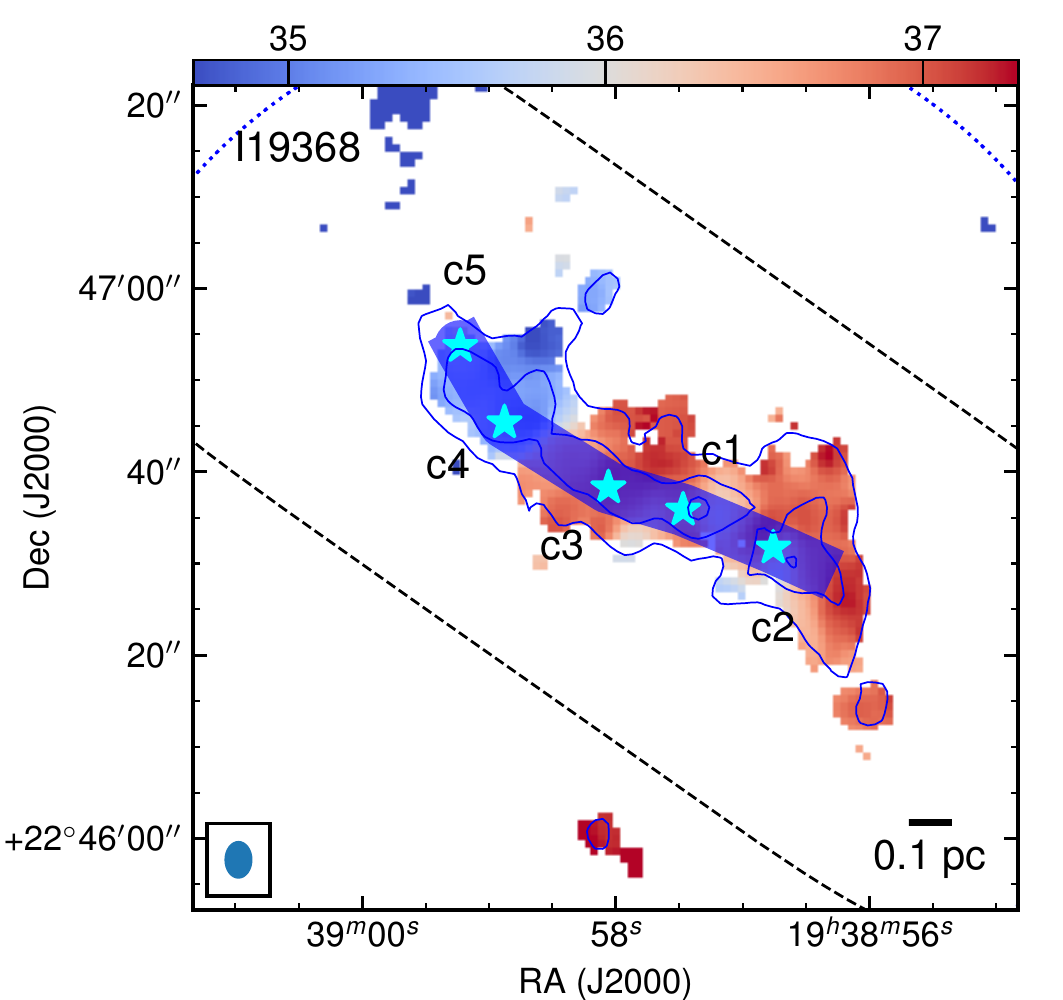}} & \includegraphics[width=0.48\textwidth]{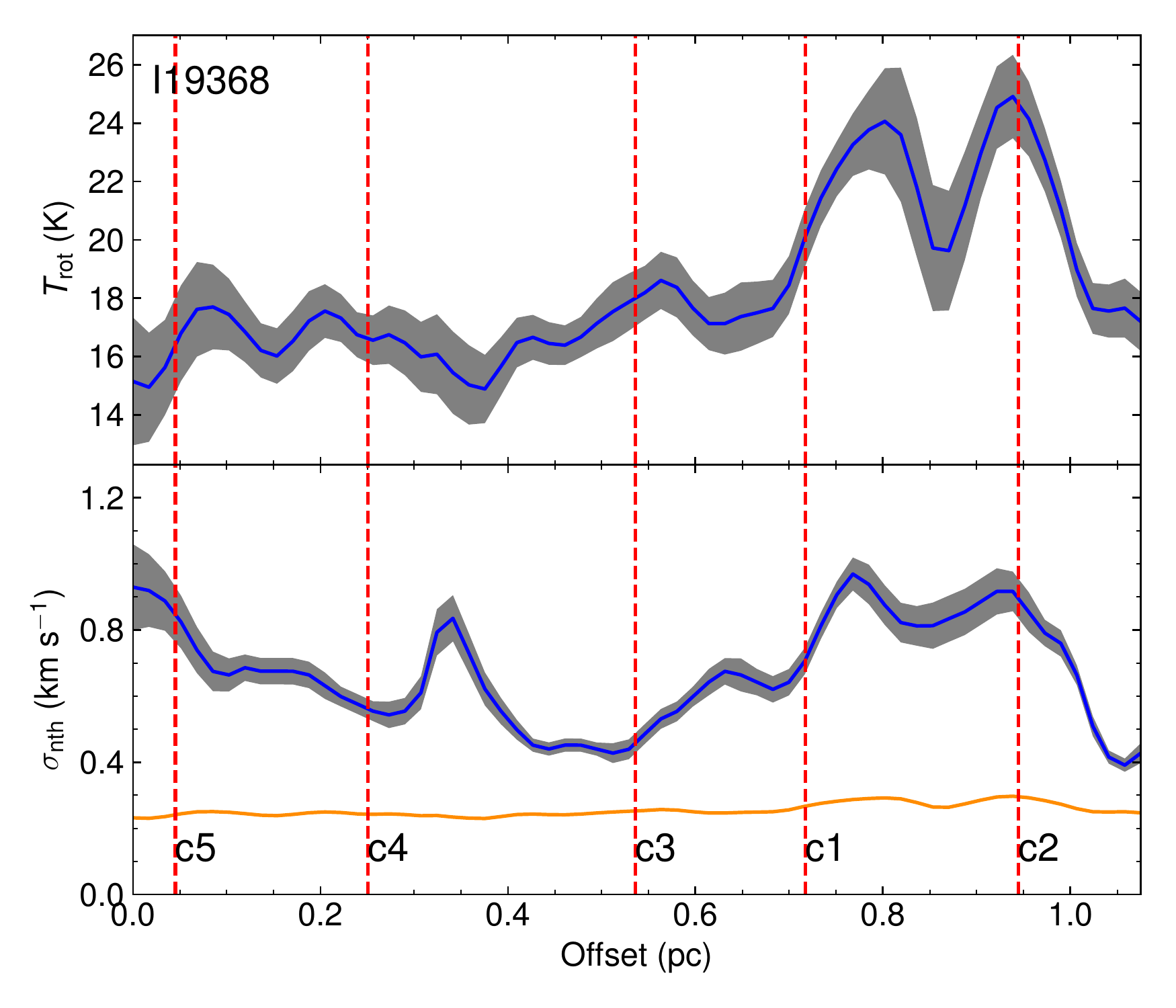} \\ [-1em]
\end{tabular}
\caption{Continued.}
\end{figure*}

\subsection{Variation of Gas Properties along Filaments}\label{subsec:filaments_filprop}
Gas motions in massive filaments have been found to be supersonic, with Mach numbers of $\gtrsim$2 \citep[e.g.,][]{henshaw2014,beuther2015}. Whether this suggests strong turbulent motions in filaments or simply a matter of spatial resolution (e.g., unresolved subsonic sub-filaments) is unclear.
Here, we make use of the VLA \amm{} lines to investigate the distribution of thermal and non-thermal motions along massive filaments.

We focus on the same four filaments as above. To derive gas properties along the filaments, we extract the mean \amm{} (1,~1) and (2,~2) spectra along paths defined by the shadowed bars in \autoref{fig:filprop}, averaged within a width of 5\arcsec{} along the paths. These paths cut through the dense cores embedded in the filaments. The \amm{} spectra are fit with the same routine as in \autoref{subsec:cores_temp}, including those with two velocity components. The results are displayed in \autoref{fig:filprop}.

With gas temperatures of 20--30~K, the isothermal sound speed is typically 0.2--0.3~\kms{}. Among the four clouds, I18308 and I19368 present a single velocity component hence their linewidths are better constrained. Their non-thermal linewidths are typically 0.4--0.8~\kms{}, resulting in Mach numbers of 2--3. I18337 and I19220 contain multiple velocity components. As a result, the fitting of \amm{} spectra are less certain, as indicated by the large errors in \autoref{fig:filprop}. Nevertheless, similarly supersonic motions with large Mach numbers of $\gtrsim$2 are found in these two filaments.

With the angular resolution of 3\arcsec{} ($\sim$0.06~pc at a distance of 4~kpc) of the VLA observations, we do not spatially resolve sub-filaments in the four filaments. The SMA dust and spectral line data do not reveal any sub-filaments either. Although in I18337 and I19220 the \amm{} lines present multiple velocity components that may trace multiple substructures, these velocity components are supersonic. Therefore, we do not find evidence of subsonic sub-filaments with our data and the 4 filaments are indeed supersonic structures.

We then compare the distributions of temperatures and non-thermal linewidths to the star formation activities in the dense cores. Peaks in temperatures along the filaments tend to be offset from the dense cores, which are most evident toward I18308 and I19368 where temperatures are better determined. Most of these dense cores are protostellar, where protostars have formed and started to warm the ambient gas. The temperature distributions in \autoref{fig:filprop} hence suggest that protostellar heating through radiative feedback does not seem to affect gas in the filaments at $>$0.1~pc scales. The increased temperatures in the filaments may instead be related to outflows from the dense cores. For example, the temperature peaks next to I18308-c1 and I18308-c4 may be attributed to heating of strong outflows launched from these two dense cores (\autoref{subsec:cores_sf_outflows}).

The non-thermal linewidths appear to be better correlated with the star-forming dense cores. For example, I18308-c1/c4 and I18337-c3 are associated with strong SiO outflows (\autoref{subsec:cores_sf_outflows}). They are all coincident with large non-thermal linewidths in \autoref{fig:filprop}.
On the other hand, as discussed in \autoref{subsec:cores_sf_evolution}, I18308-c5/c8 and I19220-c4 are prestellar core candidates. Their non-thermal linewidths are smaller with Mach numbers $\sim$1, i.e., they are transonic.

The above results suggest that linewidths in the four filaments are dominated by non-thermal motions (Mach number $\gtrsim$ 2), which seem to be correlated with star formation activities of the dense cores along the filaments. However, whether this non-thermal component is mainly micro-turbulence is unclear. Simulations have suggested that accretion in clouds could also contribute to the non-thermal linewidths \citep{vazquez2009}. Therefore, to account for the correlation between the non-thermal linewidths and star formation in these filaments, two possible interpretations exist depending on which component dominates the non-thermal linewidths, i) turbulence, which suggests that protostellar feedback (outflows) alters the environment and determines the turbulent strength, or ii) accretion, which is the result of stronger accretion around more active star-forming cores. In the second scenario, the smaller non-thermal linewidths of the prestellar core candidates then indicate weaker accretion around them than around the protostellar dense cores.

\begin{figure*}[!t]
\centering
\includegraphics[width=0.306\textwidth]{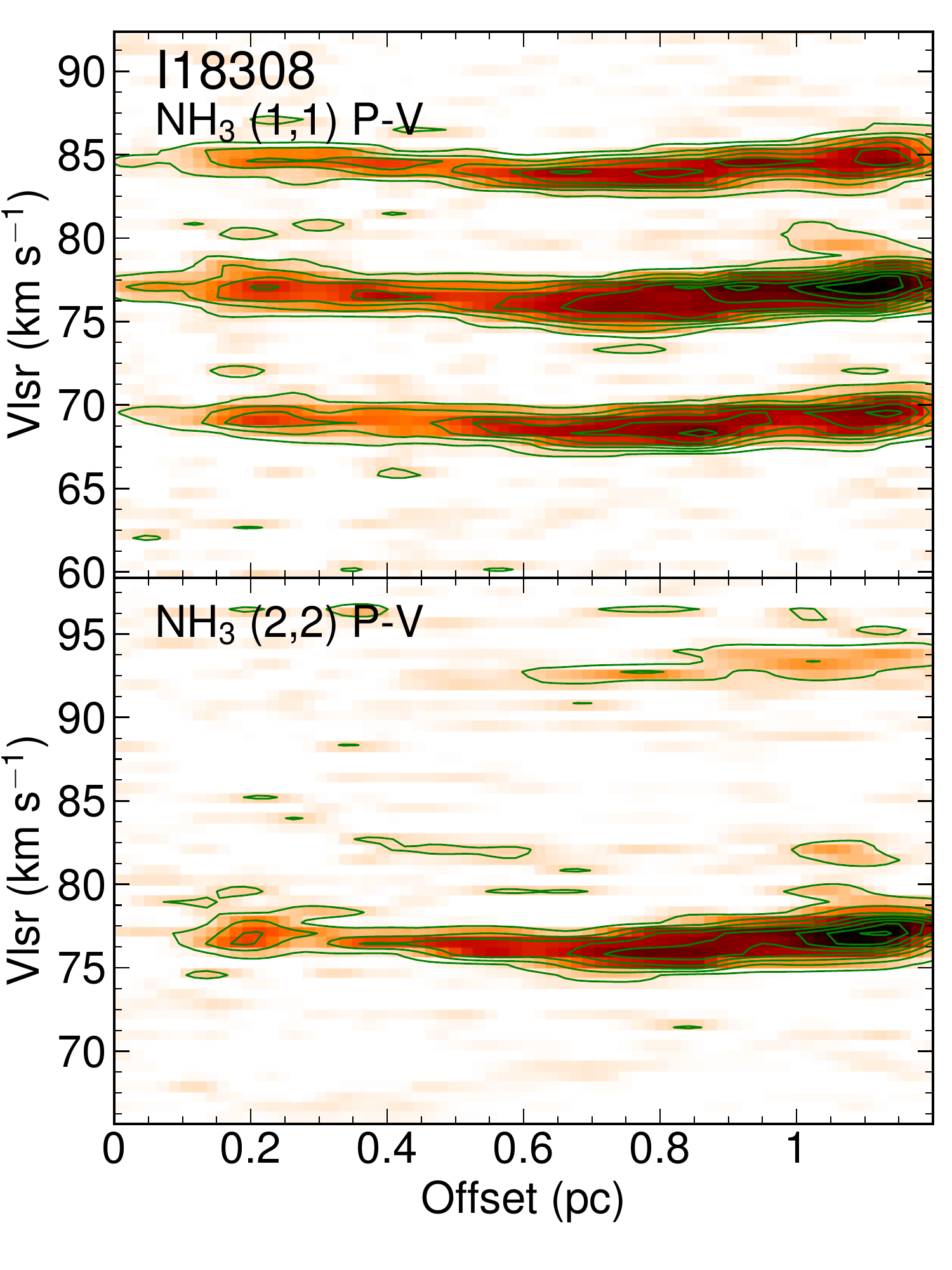} \includegraphics[width=0.48\textwidth]{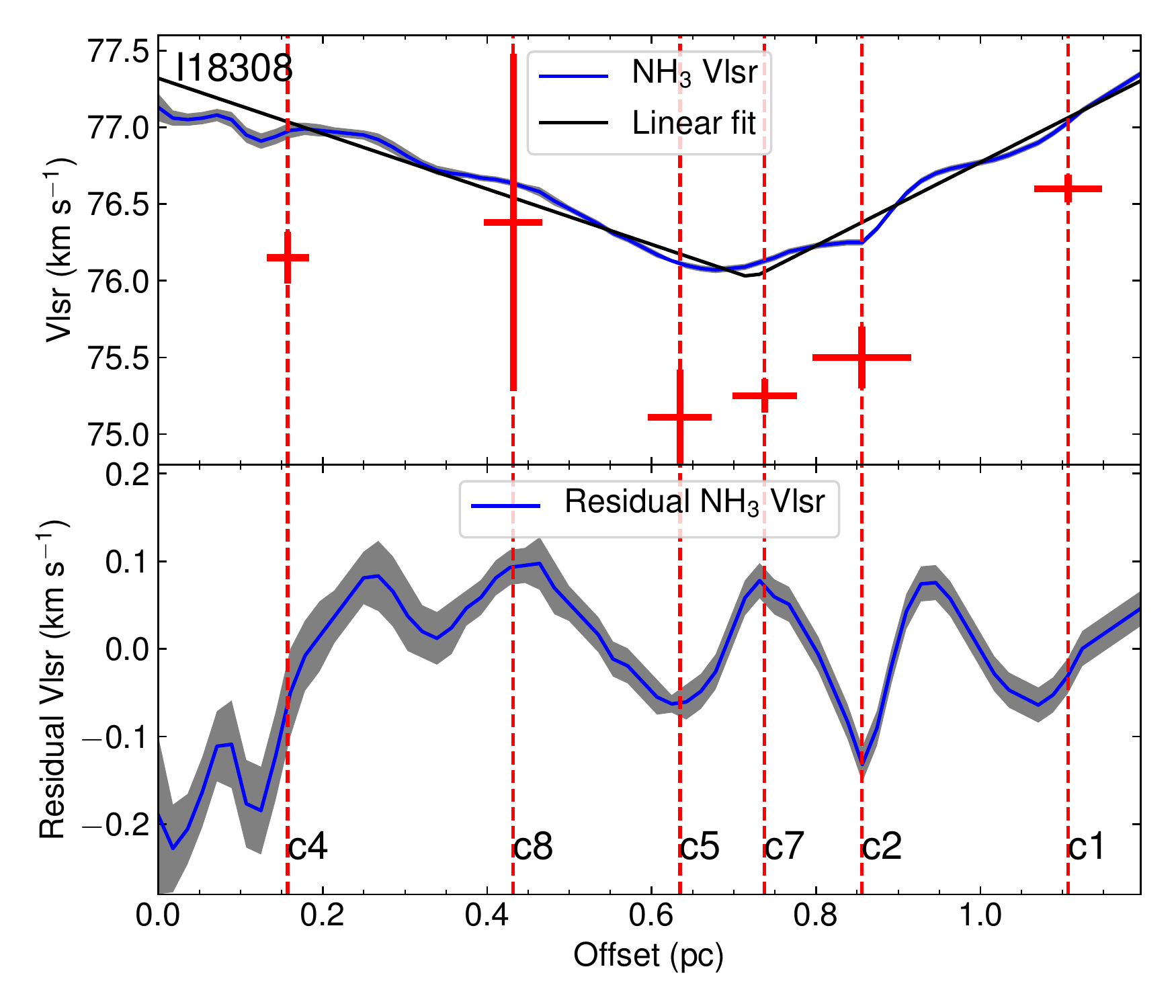} \\
\vspace{-0.8em}
\includegraphics[width=0.306\textwidth]{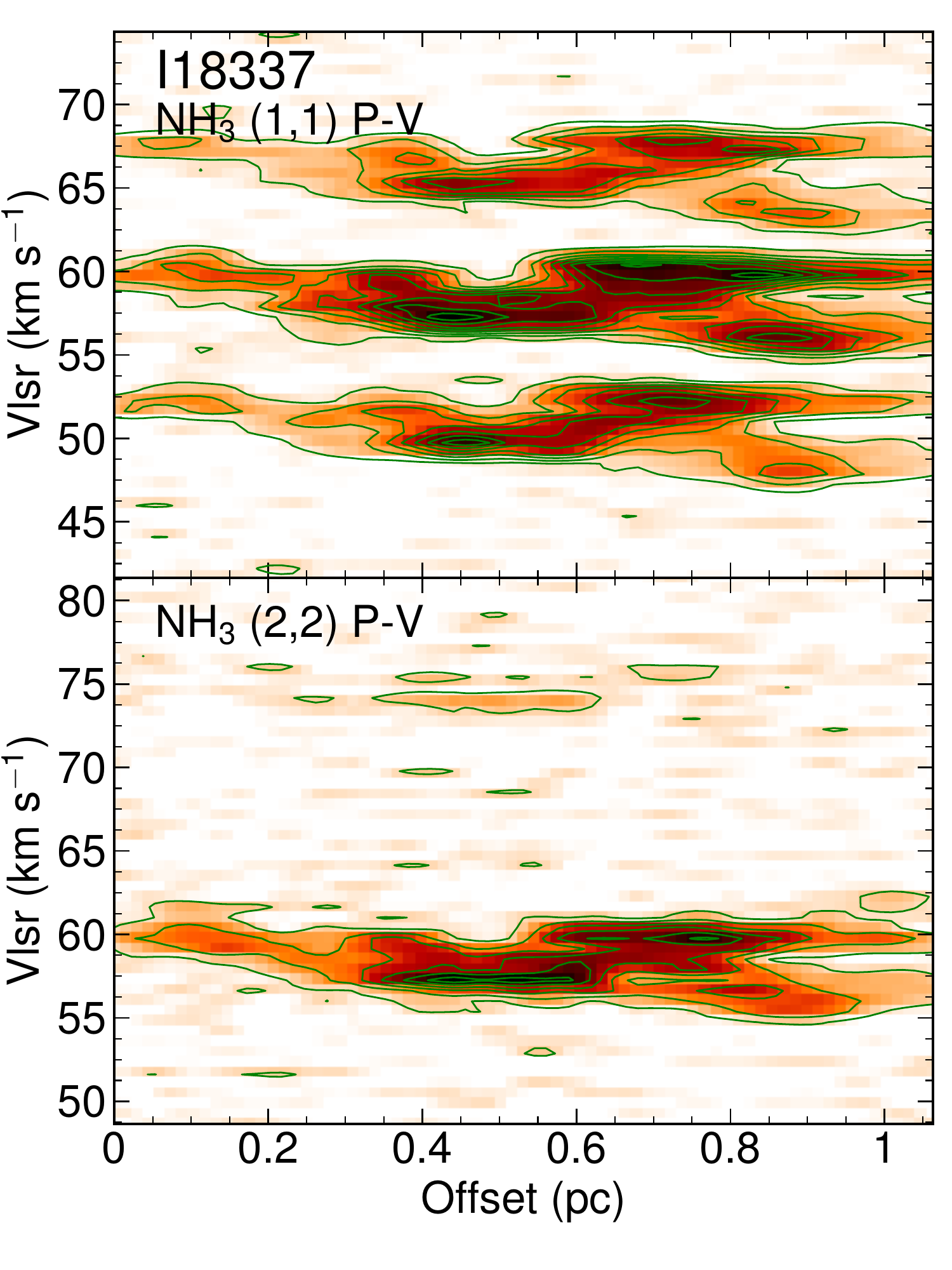} \includegraphics[width=0.48\textwidth]{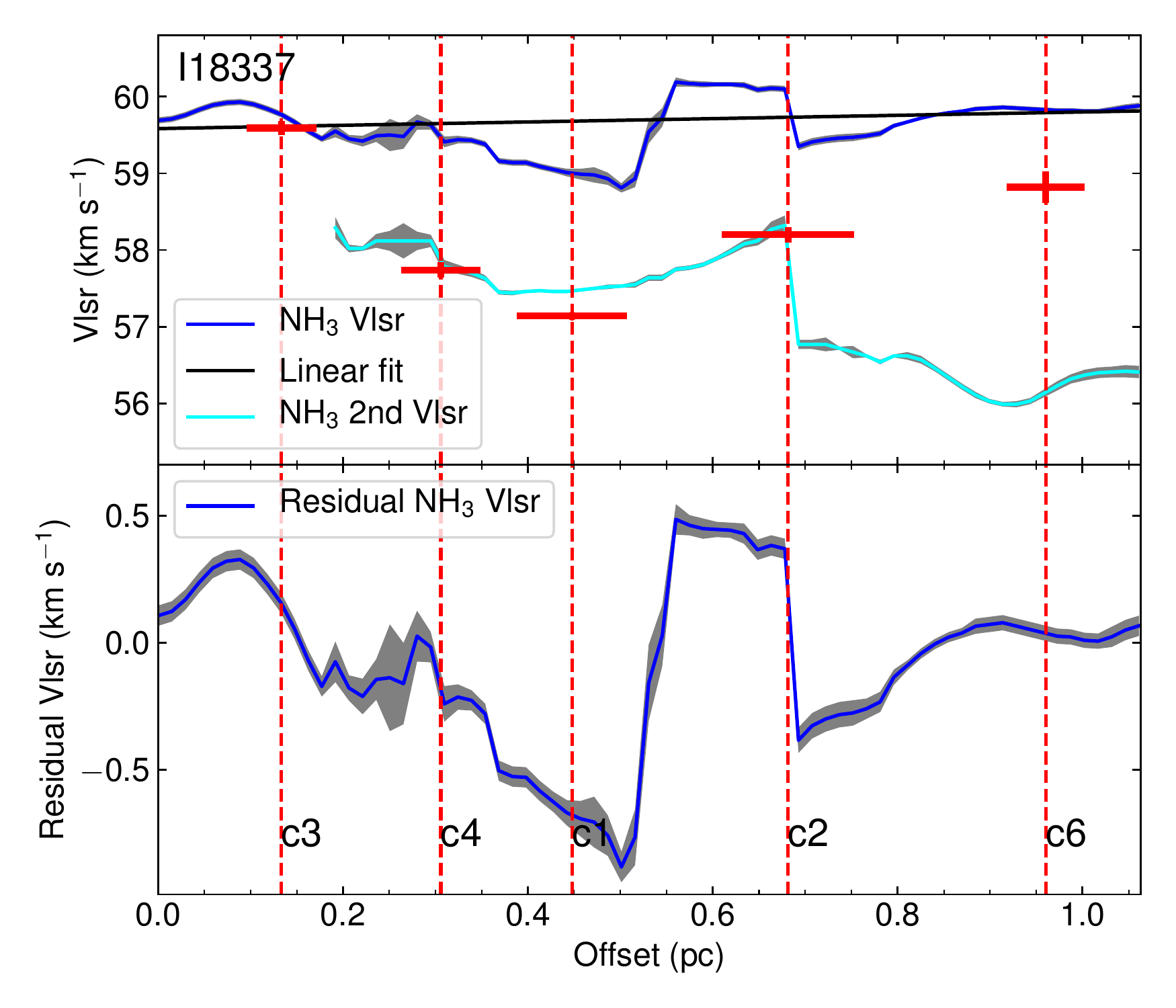} \\
\vspace{-1em}
\caption{Position-velocity diagrams of filaments, extracted along the paths defined in \autoref{fig:filprop}. Plots in the left column show the data, with contours starting from 5~\mjypbm{} in steps of 10~\mjypbm{}. Plots in the right column show the best-fit \vlsr{} using \amm{} (1,~1) and (2,~2) lines: the upper panels show the best-fit \vlsr{} at each pixel and a linear fit to the global velocities, while the lower panels show the residual velocities after subtracting the linear fit. Grey shades around velocities mark the 1$\sigma$ error. For I18337 and I19220 where the second velocity component is seen, the best-fit \vlsr{} of this component is shown in cyan curves in the upper panels. Vertical dashed lines mark positions of dense cores along the paths with core names labelled. Red crosses mark intrinsic radii of the dense cores in the horizontal, and 1$\sigma$ errors in core velocities in the vertical as defined by dense gas tracers including \fmh{}, \methanol{}, or \ceighteeno{}.}
\label{fig:pv}
\end{figure*}

\addtocounter{figure}{-1}

\begin{figure*}[!t]
\centering
\includegraphics[width=0.306\textwidth]{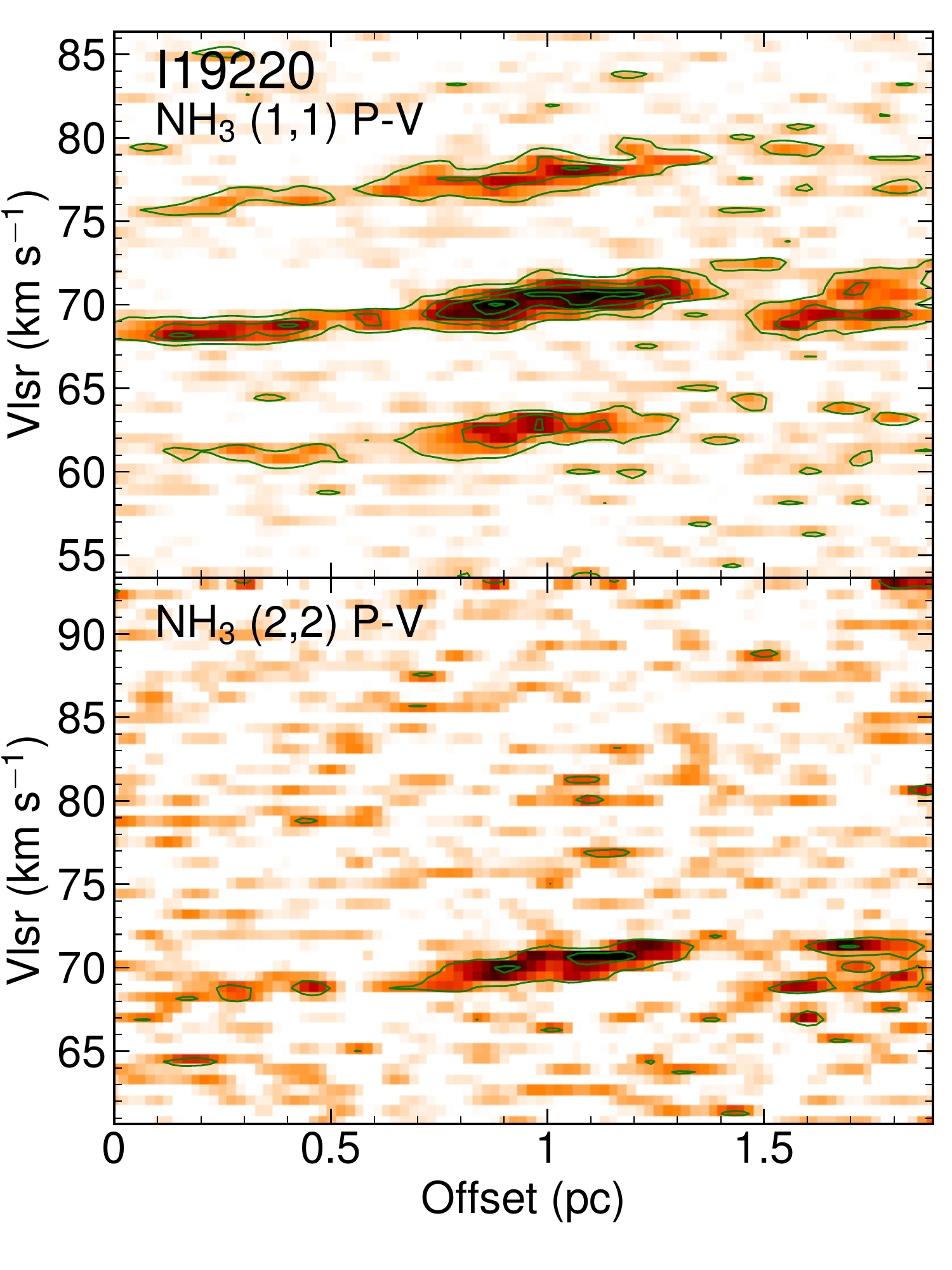} \includegraphics[width=0.48\textwidth]{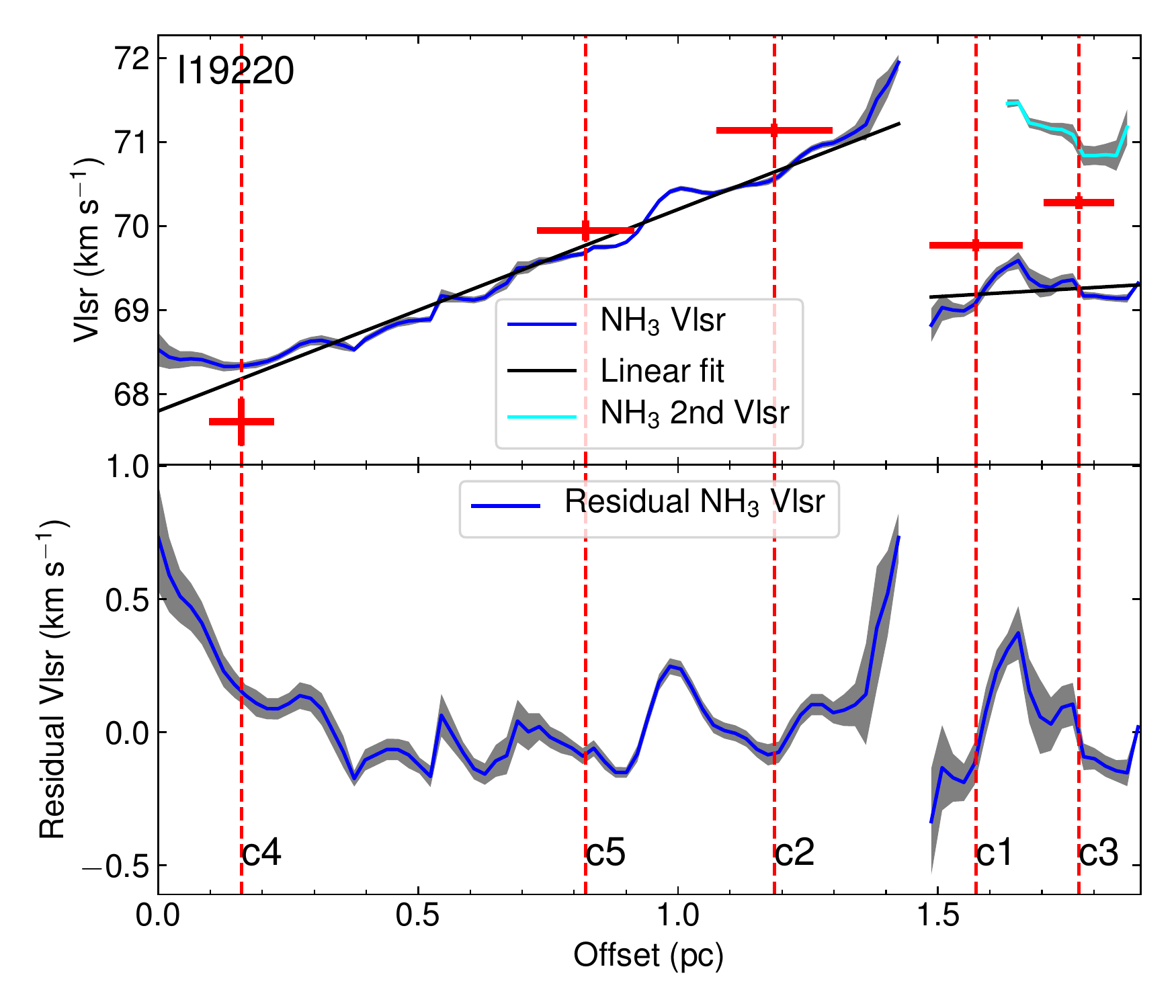} \\
\vspace{-0.8em}
\includegraphics[width=0.306\textwidth]{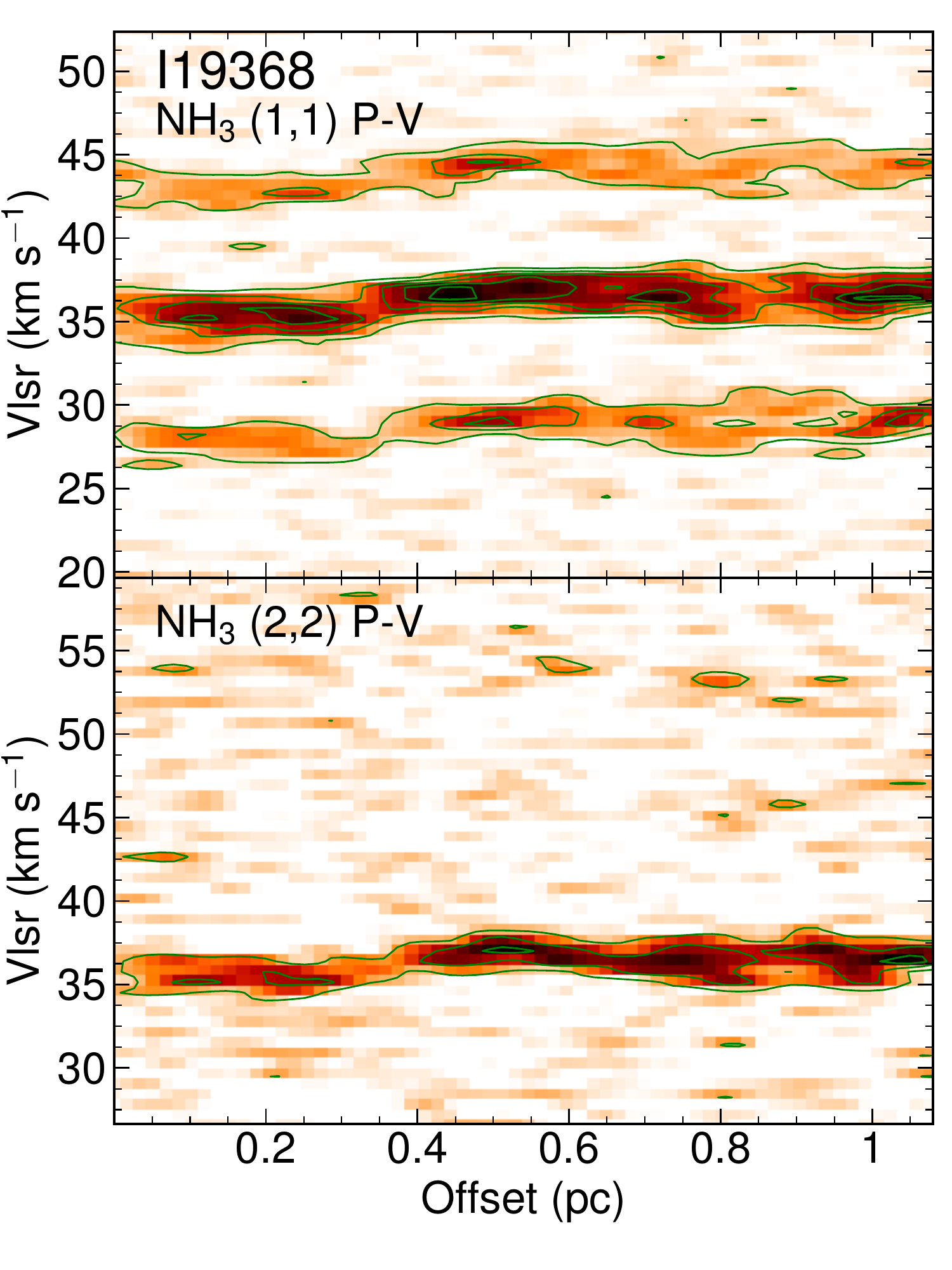} \includegraphics[width=0.48\textwidth]{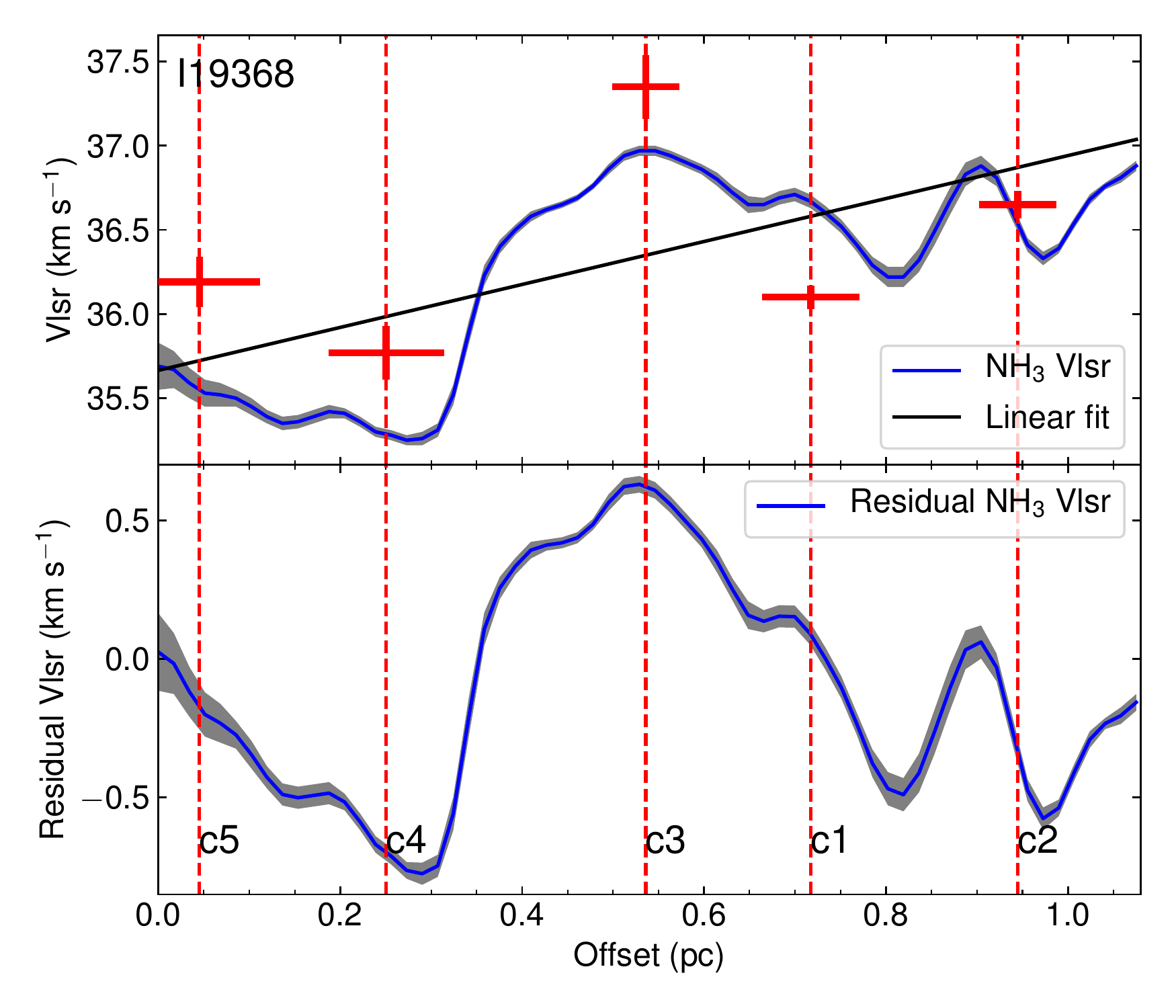} \\
\vspace{-1em}
\caption{Continued.}
\end{figure*}

\subsection{Accretion into Dense Cores along Filaments}\label{subsec:filaments_filaccretion}
Velocity gradients along filaments have been observed in both nearby low-mass star-forming clouds \citep{hacar2011,kirk2013} and massive clouds \citep{schneider2010,peretto2014,tackenberg2014,zhang2015}. They have been interpreted as flows along filaments, feeding gas into dense cores \citep{smith2009}. As shown in \autoref{fig:nh3}, the centroid velocities derived from VLA \amm{} lines in the 8 clouds also reveal velocity gradients along filaments. Typical velocity gradient is 1--2~\kmspc{}. However, whether it is suggestive of accretion along filaments is unclear. The global velocity gradient in the filaments spanning a spatial scale of 1--2~pc may be attributed to the motion of the filaments themselves (e.g., rotation or oscillation along the line of sight) rather than accretion flows.

To investigate the accretion along filaments, we select the same four filaments discussed above. The centroid velocities in \autoref{fig:filprop} are not reliable when multiple velocity components exist, therefore we take a more sophisticated approach: we extract the position-velocity (PV) diagrams along paths defined by the shadowed bars in \autoref{fig:filprop} from the VLA \amm{} (1,~1) and (2,~2) data, using the \textit{pvextractor}\footnote{\href{http://pvextractor.readthedocs.io}{http://pvextractor.readthedocs.io}} program. These paths are the same as in the previous subsection. The best-fit \vlsr{} at each pixel has been derived in the previous subsection simultaneously with gas temperatures and linewidths. Some pixels clearly show multiple velocity components, with which we fit two components and derive two sets of parameters. The resulting \vlsr{} along the paths and corresponding fitting errors are shown in \autoref{fig:pv}. All four filaments present clear global velocity gradients along the paths. 

Then we subtract the global velocity gradients in the filaments, whatever they may be, and use the residual velocities along the filaments that trace local gas motions to reveal gas flows. For I18337 and I19368, we fit the global velocity gradients with a linear function. For I18308 and I19220, however, the global velocity gradients appear to be segmented. The break point in I19220 is naturally chosen to the spatial offset where \amm{} emission is not detected. For I18308, we fit a segmented function containing two linear functions in which the break point is also a free parameter. In all cases, the functions are fit to the data using the least-square method taking errors of \vlsr{} into account. The linear-function-subtracted residual velocities are shown in \autoref{fig:pv}, with positions of dense cores marked by vertical lines.

As mentioned in \autoref{subsec:filaments_filprop}, I18337 and the last 0.3~pc of I19220 present multiple velocity components, hence the fitting is less well constrained. For I18308, the first 1.5~pc of I19220, and I19368, where \amm{} lines can be robustly fit by a single velocity component, the dense cores appear to preferably locate at local maxima or minima of velocities. Meanwhile, several dense cores are also found to be offset from any local velocity maxima or minima (e.g., I18308-c4).

The most intriguing case is found in I18308, where the velocities show a sinuous pattern around dense cores. Velocities around I18308-c7 may be less reliable, because this core is next to the break point of the segmented linear function, which may introduce an artificial change of velocities. Velocities around the dense cores I18308-c1/c2/c5/c8, however, are more robust and the velocity maxima or minima are coincident with dense cores. Similar patterns have been found in low-mass and high-mass star-forming filaments \citep{hacar2011,zhang2015}. \citet{hacar2011} argued that sinuous velocities are resulted from gas accretion around dense cores along the filament, however in their model there is a quarter phase shift between the densities (i.e., dense cores) and velocities. In I18308, such phase shift is not seen. Interestingly, recent simulations by \citet{gritschneder2017} show that an oscillating and fragmenting filament with small inclination angles (30\arcdeg{}) produces such observed velocities. The change of velocities around dense cores is then a result of geometrical bent of the filament along the line of sight \citep[also see discussion in][]{henshaw2014}. Another intriguing feature seen in \autoref{fig:pv} is that \vlsr{} of dense cores in I18308, defined by \methanol{} or \fmh{} lines and marked by red crosses, are always blue-shifted by $\sim$1~\kms{} with respect to those based on \amm{}. \vlsr{} of \methanol{} and \fmh{} agree with each other. Whether this velocity shift is the result of chemical differentiation or a denser gas component moving in a diffuse envelope is unclear. A comparison to detailed radiative transfer models is necessary to interpret the velocities in I18308.

If the local velocity gradients around dense cores in \autoref{fig:pv} are interpreted as infall toward dense cores along filaments, the infall velocities range from $\sim$0 (e.g., I18337-c6) to 2~\kmspc{} (e.g., cores in I18308). The small velocities in I18337-c6 may not necessarily rule out the possibility of infall, but may be due to vanishing velocity gradients when the filament is parallel to the plane of the sky.

The accretion rates along filaments can be estimated assuming a simple cylinder model \citep{kirk2013}:
\begin{equation}\label{equ:accretion}
\dot{M}=\Delta vM/\tan{\theta},
\end{equation}
where $\Delta v$ is the velocity gradient, $M$ is the mass of the filament in consideration, and $\theta$ is the inclination angle of the filament to the plane of the sky (0 when parallel to the plane of the sky).

We assume that accretion along filaments for each dense core only happens in the part of the filament around that core within a length of $\sim$0.2~pc, which is the observed fragment separation. Gas further away in the filament is influenced more by other dense cores. Typical masses are 50--100~\msol{} using the masses per unit lengths listed in \autoref{tab:filaments}.

The inclination angle $\theta$ is difficult to determine observationally. We tentatively constrain $\theta$ by comparing the velocity dispersions along the line of sight and the velocity gradients in the plane of the sky, assuming that velocities are isotropic. A typical velocity dispersion, including both thermal and non-thermal components, is 0.6~\kms{} (\autoref{fig:filprop}). The observed velocity gradients of $\sim$2~\kmspc{} within 0.2~pc lead to a velocity of 0.4~\kms{}. Hence the inclination angle is arcsin(2/3), or 42\arcdeg{}. The observed velocity gradients vary along the filaments, therefore we allow for a larger range of 30\arcdeg{}--60\arcdeg{} for the inclination angle, corresponding to gradients of 1.5--2.5~\kmspc{}.

With the above assumptions, typical accretion rates along these filaments, given a gradient of 2~\kmspc{}, is 1--2$\times$10$^{-4}$~\msol{}\,yr$^{-1}$, or 0.5--3.5$\times$10$^{-4}$~\msol{}\,yr$^{-1}$ after accounting for the variation of inclination angles. With such an accretion rate, typical core masses of 10--50~\msol{} (\autoref{subsec:cores_prop}) can be doubled in several free-fall time scales ($\sim$10$^5$~years, assuming a density of 10$^{5\text{--}6}$~\cc{}).

At last, given that the 3D geometry of filaments is unclear, the same velocity gradients around dense cores may also be interpreted as expansion of gas \citep[see discussion in][]{henshaw2014}. However, the virial parameters of the dense cores are all smaller than 2 (\autoref{subsec:cores_prop}), hence the core are likely gravitationally bound. In addition, these filaments as a whole are supercritical, i.e., they are gravitationally bound or even collapsing, if magnetic field or large inclination angles are not taken into account (\autoref{subsec:filaments_fragmentation}). Therefore the expansion scenario is less likely.

\section{CONCLUSIONS}\label{sec:conclusions}
We use the SMA 1.3~mm continuum and spectral line observations and the VLA \amm{} line and continuum observations to study high-mass star formation in dense cores and fragmentation and gas accretion of filaments in 8 filamentary molecular clouds. The main results are:
\begin{itemize}
\item 50 dense cores are identified in 8 filamentary clouds using the SMA 1.3~mm continuum emission. Their properties, including equivalent radii, gas temperatures, linewidths, masses, \amm{} abundances, and virial parameters are derived.
\item Signatures of high-mass star formation in the dense cores, including \hii{} regions, outflows, hot molecular cores, and infrared emission, are investigated. Evolutionary phases of dense cores are determined based on these signatures: 5 are devoid of any star formation signatures hence are prestellar core candidates, 11 are associated with free-free emission hence are embedded with (UC) \hii{} regions. The remaining 34 dense cores are classified as protostellar. 3 out of the 5 prestellar core candidates may be massive enough to form high-mass stars. All the cores except for 2 (UC) \hii{} regions have virial parameters smaller than 2, while prestellar core candidates show virial parameters of $\lesssim$0.5, suggesting that they are strongly self-gravitating. Linewidths, gas temperatures, \amm{} abundances, and virial parameters show dependency on the evolutionary phases.
\item We focus on 4 morphologically well-defined filamentary clouds with regularly spaced dense cores along major axes to study what mechanisms are responsible for supporting their fragmentation. Assuming an isothermal gas cylinder model, the observed masses per unit length of these filaments are larger than critical values from the model even if both thermal and turbulence pressure support are included, suggesting that these filaments are unstable to radial collapse if not considering large inclination angles of $>$60\arcdeg{} or additional magnetic field support.
\item Gas temperatures and non-thermal linewidths along the 4 filaments are derived. Non-thermal motion in the filaments is in general supersonic, with Mach numbers larger than 2, although around prestellar core candidates it becomes transonic. No evidence of subsonic sub-filaments is found. A likely correlation between the non-thermal linewidths and star formation activities in dense cores along the filaments indicates that feedback or accretion during star formation enhances the non-thermal motion.
\item Possible signatures of gas flows along filaments are found in \amm{} lines toward the 4 filaments, corresponding to accretion rates of order 10$^{-4}$~\msolpyr{}. In one of the filaments, I18308, we observe oscillatory velocities around dense cores, which need to be compared to radiative transfer models to be explained.
\item Based on our analysis, we suggest that fragmentation and accretion in these massive filaments are closely related to high-mass star formation. Dense cores in these massive filaments can be formed through cylindric fragmentation supported by thermal pressure and turbulence in very early evolutionary phases. Then dense cores may remain accreting from the environment through accretion flows along filaments at a high rate. Feedback from star-forming cores may inject energy into filaments and increase the non-thermal linewidths, resulting in supersonic motions in filaments.
\end{itemize}

\acknowledgments
We thank the anonymous referee for helpful comments. XL and QZ acknowledge the support by the National Natural Science Foundation of China (No.\ 11629302). HAS acknowledges partial support from NASA grant NNX12AI55G. QG is supported by the National Key Research and Development Program of China (No.\ 2017YFA 0402703), and by the National Natural Science Foundation of China (No.\ 11733002). The operation staff and postdocs of the SMA are gratefully acknowledged for their help with the SMA observations. This research made use of Astropy, a community-developed core Python package for Astronomy \citep{astropy2013}, APLpy, an open-source plotting package for Python \citep{aplpy2012}, and Mayavi \citep{ramachandran2011}. Data analysis was in part carried out on the open use data analysis computer system at the Astronomy Data Center (ADC) of the National Astronomical Observatory of Japan. Part of this work is based on archival data, software or online services provided by the Space Science Data Center - ASI. This research has made use of the NASA/IPAC Infrared Science Archive, which is operated by the Jet Propulsion Laboratory, California Institute of Technology, under contract with the National Aeronautics and Space Administration. This research has made use of NASA's Astrophysics Data System, and the SIMBAD database operated at CDS, Strasbourg, France.

\facilities{SMA, VLA, IRAM 30~m MAMBO, JCMT SCUBA2, IRSA, Spitzer, Herschel}

\software{MIR, MIRIAD \citep{sault1995}, CASA \citep{mcmullin2007}, APLpy \citep{aplpy2012}, Astropy \citep{astropy2013}, pyAmor, lmfit, Mayavi, TDViz, pvextractor}

\bibliographystyle{aasjournal}


\clearpage

\begin{deluxetable}{cccccc}
\tabletypesize{\scriptsize}
\tablecaption{Sample.\label{tab:sample}}
\tablewidth{0pt}
\tablehead{
\multirow{2}{*}{Target} & R.A. & Decl. & Distance\tablenotemark{a} & $L_\text{bol}$\tablenotemark{b} & $M$\tablenotemark{c} \\ 
 & (J2000) & (J2000) & (kpc) & ($L_{\odot}$) & ($M_{\odot}$)
}
\startdata
IRAS 18308$-$0841 & 18:33:33.00 & $-$08:39:10.0 & 4.6 & 2.0$\times$10$^4$ & 4300 \\ 
IRAS 18310$-$0825 & 18:33:47.50 & $-$08:23:45.0 & 4.8 & 1.6$\times$10$^4$ & 2700 \\
IRAS 18337$-$0743 & 18:36:41.00 & $-$07:39:15.0 & 3.8 & 2.1$\times$10$^4$ & 5100 \\
IRAS 18460$-$0307 & 18:48:39.20 & $-$03:03:55.0 & 4.8 & 1.6$\times$10$^4$ & 800  \\
IRAS 18530$+$0215 & 18:55:33.00 & $+$02:19:04.1 & 4.6 & 7.9$\times$10$^4$ &1740 \\
IRAS 19074$+$0752 & 19:09:54.51 & $+$07:57:22.8 & 3.8 & 1.6$\times$10$^4$ & 200 \\
IRAS 19220$+$1432 & 19:24:21.16 & $+$14:38:02.8 & 5.4 & 4.3$\times$10$^4$ &1640\\
IRAS 19368$+$2239 & 19:38:58.33 & $+$22:46:40.1 & 4.4 & 6.6$\times$10$^3$ & 907 \\
\enddata
\tablenotetext{a}{Kinematic distances derived using the rotation curve model in \citet{reid2009} (see \citealt{lu2014} for detail). The near distance is taken when there is distance ambiguity.}
\tablenotetext{b}{The luminosities are based on \textit{IRAS} fluxes \citep{prusti1992}.}
\tablenotetext{c}{The masses are derived from MAMBO 1.2 mm dust emission \citep{beuther2002mambo}, except for IRAS 19368$+$2239 whose mass is derived from SCUBA2 850~\micron{} dust emission \citep{eden2017} because the MAMBO data are unavailable.}
\end{deluxetable}

\begin{deluxetable}{ccccccccccc}
\tabletypesize{\scriptsize}
\tablecaption{Summary of the SMA observations. \label{tab:smaobs}}
\tablewidth{0pt}
\tablehead{
\multirow{2}{*}{Project ID} & \multirow{2}{*}{Config.} & \multirow{2}{*}{Date} & Number of & \multirow{2}{*}{$\tau_\text{225~GHz}$} & $T_\text{sys}$ & \multirow{2}{*}{Pointings\tablenotemark{a}} & \multirow{2}{*}{Correlator\tablenotemark{b}} & \multicolumn{3}{c}{Calibrators\tablenotemark{c}} \\
\cline{9-11} & & & antennas & & (K) & &  & Bandpass & Flux & Gain
}
\startdata
2012B-S052                          & COM   & 2013 Apr 22  & 7 & 0.06 & 80--160            & F1/F2/F3/F4-P1 & ASIC  & Q1 & Titan & Q4, Q5 \\ \hline
2012B-S100                          & SUB   & 2014 Mar 07 & 7 & 0.26 & 100--250          & F1/F2-P2/P3/P4  & ASIC  & Q1 & Mars & Q4 \\ \hline
\multirow{2}{*}{2013B-S053} & SUB    & 2014 Feb 21 & 5 & 0.12--0.22 & 100--180 & F1/F2-P1, F3/F4-P1/P2 & ASIC & Q1 & Titan, Mars & Q4 \\
                                              & SUB    & 2014 Feb 24 & 6 & 0.03           & 60--90     & F1/F2-P1, F3/F4-P1/P2 & ASIC & Q1 & Titan, Mars & Q4 \\ \hline
2014A-S091                          & SUB    & 2014 Jul 28  & 7 & 0.09   & 80--120           & F7/F8-P1/P2/P3              & ASIC & Q2 & Neptune & Q6, Q7 \\ \hline
\multirow{3}{*}{2014B-S037} & COM   & 2015 Apr 10 & 6 & 0.14    & 200--600        & F7/F8-P1/P2/P3              & ASIC & Q1 & Callisto & Q6 \\
                                              & COM   & 2015 Apr 13 & 6\tablenotemark{d} & 0.13    & 200--700         & F5/F6-P1              & ASIC+2SWARM & Q1 & Titan & Q5, Q8 \\
                                              & SUB    & 2015 Jun 23 & 6 & 0.09--0.15 & 200--390  & F5/F6-P1/P2/P3, F7/F8-P2/P4/P5   & ASIC+2SWARM* & Q1 & Titan & Q8, Q9 \\ \hline
\multirow{3}{*}{2015B-S053} & COM   & 2016 Feb 04 & 6 & 0.03   & 140--260        &  F1/F2-P2/P3/P4, F3/F4-P2 & ASIC+2SWARM* & Q3 & Ganymede & Q4 \\
                                              & COM   & 2016 May 16 & 7 & 0.20--0.30 & 300--1300 & F5/F6-P2/P3, F7/F8-P4/P5       & ASIC\tablenotemark{e} & Q3 & Titan & Q8, Q9 \\
                                              & COM   & 2016 May 19 & 7 & 0.10--0.15 & 200--700 &   F5/F6-P2/P3, F7/F8-P4/P5       & ASIC+4SWARM   & Q3  & Titan & Q8, Q9 \\
\enddata
\tablenotetext{a}{F1--F8 stand for the 8 clouds listed in \autoref{tab:sample}. Coordinates of pointing centers: F1-P1: 18$^\text{h}$33$^\text{m}$33\fs0, $-$8\arcdeg39\arcmin10\farcs{0}; F1-P2: 18$^\text{h}$33$^\text{m}$30\fs0, $-$8\arcdeg38\arcmin36\farcs{5}; F1-P3: 18$^\text{h}$33$^\text{m}$31\fs4, $-$8\arcdeg38\arcmin55\farcs{0}; F1-P4: 18$^\text{h}$33$^\text{m}$33\fs9, $-$8\arcdeg38\arcmin45\farcs{5}; F2-P1: 18$^\text{h}$33$^\text{m}$47\fs5, $-$8\arcdeg23\arcmin45\farcs{0}; F2-P2: 18$^\text{h}$33$^\text{m}$46\fs4, $-$8\arcdeg24\arcmin08\farcs{0}; F2-P3: 18$^\text{h}$33$^\text{m}$47\fs0, $-$8\arcdeg23\arcmin18\farcs{0}; F2-P4: 18$^\text{h}$33$^\text{m}$48\fs8, $-$8\arcdeg24\arcmin08\farcs{0}; F3-P1: 18$^\text{h}$36$^\text{m}$41\fs0, $-$7\arcdeg39\arcmin15\farcs{0}; F3-P2: 18$^\text{h}$36$^\text{m}$41\fs3, $-$7\arcdeg39\arcmin43\farcs{0}; F4-P1: 18$^\text{h}$48$^\text{m}$39\fs2, $-$3\arcdeg3\arcmin55\farcs{0}; F4-P2: 18$^\text{h}$48$^\text{m}$37\fs4, $-$3\arcdeg3\arcmin48\farcs{0}; F5-P1: 18$^\text{h}$55$^\text{m}$33\fs0, $+$2\arcdeg19\arcmin4\farcs{0}; F5-P2: 18$^\text{h}$55$^\text{m}$31\fs5, $+$2\arcdeg18\arcmin51\farcs{0}; F5-P3: 18$^\text{h}$55$^\text{m}$34\fs5, $+$2\arcdeg19\arcmin17\farcs{0}; F6-P1: 19$^\text{h}$9$^\text{m}$54\fs0, $+$7\arcdeg57\arcmin10\farcs{0}; F6-P2: 19$^\text{h}$9$^\text{m}$53\fs5, $+$7\arcdeg57\arcmin35\farcs{0}; F6-P3: 19$^\text{h}$9$^\text{m}$54\fs5, $+$7\arcdeg56\arcmin45\farcs{0}; F7-P1: 19$^\text{h}$24$^\text{m}$17\fs2, $+$14\arcdeg38\arcmin5\farcs{0}; F7-P2: 19$^\text{h}$24$^\text{m}$19\fs0, $+$14\arcdeg38\arcmin5\farcs{0}; F7-P3: 19$^\text{h}$24$^\text{m}$20\fs8, $+$14\arcdeg38\arcmin5\farcs{0}; F7-P4: 19$^\text{h}$24$^\text{m}$22\fs6, $+$14\arcdeg38\arcmin5\farcs{0}; F7-P5: 19$^\text{h}$24$^\text{m}$24\fs4, $+$14\arcdeg38\arcmin5\farcs{0}; F8-P1: 19$^\text{h}$38$^\text{m}$54\fs7, $+$22\arcdeg46\arcmin10\farcs{0}; F8-P2: 19$^\text{h}$38$^\text{m}$56\fs3, $+$22\arcdeg46\arcmin25\farcs{0}; F8-P3: 19$^\text{h}$38$^\text{m}$57\fs8, $+$22\arcdeg46\arcmin40\farcs{0}; F8-P4: 19$^\text{h}$38$^\text{m}$59\fs3, $+$22\arcdeg46\arcmin55\farcs{0}; F8-P5: 19$^\text{h}$39$^\text{m}$0\fs9, $+$22\arcdeg47\arcmin10\farcs{0}.}
\tablenotetext{b}{IFs of the correlator setups: ASIC --- 4.0--8.0~GHz; ASIC+2SWARM --- 4.0--9.0~GHz, 11.0--12.0~GHz; ASIC+2SWARM* --- 4.0--9.5~GHz, 10.5--12.0~GHz; ASIC+4SWARM --- 4.0--12.0~GHz.}
\tablenotetext{c}{Quasar calibrators: Q1: 3C279; Q2: 3C454.3; Q3: 3C273; Q4: 1743$-$038; Q5: 1830$+$063; Q6: 2025$+$337; Q7: 1925$+$211; Q8: 1751$+$096; Q9: 2015$+$371.}
\tablenotetext{d}{The raw data contain 7 antennas. One of them was flagged due to its unusual position in the array that leads to unexpected long baselines.}
\tablenotetext{e}{The raw data contain 4 SWARM chunks, which were all flagged due to erroneous amplitudes.}
\end{deluxetable}

\begin{deluxetable}{cccccccc}[!t]
\tabletypesize{\scriptsize}
\tablecaption{Properties of the images. \label{tab:images}}
\tablewidth{0pt}
\tablehead{
\multirow{3}{*}{Target} & \multicolumn{3}{c}{Continuum} & & \multicolumn{3}{c}{Spectral lines}  \\
\cline{2-4} \cline{6-8} & Bandwidth & Beam & RMS & & Channel width & Beam & RMS \\
& (GHz) & (\arcsec$\times$\arcsec) & (\mjypbm{}) & & (\kms{}) & (\arcsec$\times$\arcsec) & (\mjypbm{})
}
\startdata
I18308, SMA 1.3~mm & 8--14 & 3.6$\times$3.1 & 1.0 & & 1.1 & 3.9$\times$3.6 & 40 \\
I18308, VLA K band & 0.0017 & 5.1$\times$3.2 & 0.3 & & 0.62 & 5.1$\times$3.2 & 2.5 \\
I18310, SMA 1.3~mm & 8--14 & 3.7$\times$3.4 & 1.0 & & 1.1 & 3.9$\times$3.8 & 40 \\
I18310, VLA K band & 0.0016 & 5.1$\times$3.2 & 0.4 & & 0.62 & 5.1$\times$3.2 & 2.5 \\
I18337, SMA 1.3~mm & 8--14 & 3.7$\times$3.4 & 1.0 & & 1.1 & 4.2$\times$3.8 & 40 \\
I18337, VLA K band & 0.0018 & 5.1$\times$3.2 & 0.4 & & 0.62 & 5.1$\times$3.2 & 2.6 \\
I18460, SMA 1.3~mm & 8--14 & 3.7$\times$3.5 & 1.0 & & 1.1 & 3.9$\times$3.9 & 40 \\
I18460, VLA K band & 0.0023 & 5.1$\times$3.2 & 0.3 & & 0.62 & 5.1$\times$3.2 & 2.0 \\
I18530, SMA 1.3~mm & 12--16 & 3.4$\times$2.6 & 1.2 & & 1.1 & 3.5$\times$2.7 & 80 \\
I18530, VLA K band & 0.0015 & 4.0$\times$3.0 & 0.9 & & 0.62 & 3.9$\times$2.7 & 1.6 \\
I19074, SMA 1.3~mm & 12--16 & 3.3$\times$2.6 & 1.1 & & 1.1 & 3.4$\times$2.6 & 80 \\
I19074, VLA K band & 0.0020 & 4.1$\times$3.1 & 0.5 & & 0.62 & 4.2$\times$3.0 & 1.6 \\
I19220, SMA 1.3~mm & 8--16 & 4.4$\times$2.8 & 1.0 & & 1.1 & 4.4$\times$3.0 & 80 \\
I19220, VLA K band & 0.0024 & 4.5$\times$3.1 & 0.5 & & 0.62 & 4.4$\times$3.1 & 2.7 \\
I19368, SMA 1.3~mm & 8--16 & 4.4$\times$2.8 & 1.0 & & 1.1 & 4.4$\times$3.0 & 80 \\
I19368, VLA K band & 0.0023 & 4.2$\times$3.1 & 0.4 & & 0.62 & 4.1$\times$3.1 & 2.6 \\
\enddata
\tablecomments{All images are CLEANed with Biggs weighting (robustness = 0.5). Listed beams and RMS of the SMA 1.3~mm spectral line images are measured for line-free channels of  \ceighteeno{} 2--1 images, but they slightly vary between different lines. Beams and RMS of the VLA \amm{} images are measured for line-free channels of  \amm{} (1,1) images.}
\end{deluxetable}

\clearpage

\startlongtable
\begin{deluxetable}{rccccccccccc}
\tabletypesize{\scriptsize}
\tablecaption{Properties of dense cores. \label{tab:cores}}
\tablewidth{0pt}
\tablehead{
\multirow{2}{*}{Core} & R.A.~\& Decl. & Deconvl.~size \& PA & $R$ & Flux & $T_\text{rot}$\tablenotemark{a} & $M_\text{core}$ & $n$(H$_2$) & $N$(H$_2$) & $X$(\amm{}) & $\sigma_\text{in}$ & $\alpha_\text{vir}$ \\
 & (J2000) & (\arcsec$\times$\arcsec, \arcdeg) & (10$^3$~AU) &  (mJy) & (K) & (\msol{}) & (10$^6$~\cc) & (10$^{23}$~\sqc) & (10$^{-8}$) & (\kms) &
}
\startdata
I18308-c1 & 18:33:33.17, $-$08:39:15.23 & 4.3$\times$3.2, 96   & 8.5   & 337(335) & 22.7 & 137  & 6.7 & 8.9 & 1.9 & 0.70 & 0.17 \\
            c2 & 18:33:32.99, $-$08:39:04.30 & 6.5$\times$4.5, 4     & 12.4 & 119 & 18.6 & 63    & 1.0 & 1.9 & 8.8 & 0.59 & 0.39 \\
            c3 & 18:33:32.70, $-$08:39:14.50 & 8.7$\times$3.2, 148 & 12.1 & 102 & 19.6 & 50    & 0.9 & 1.6 & 4.2 & 0.63 & 0.54\\
            c4 & 18:33:34.28, $-$08:38:42.10 & 2.9$\times$1.8, 36   & 5.3   & 65 & 16.6   & 40    & 8.3 & 6.9 & 1.2 & 0.52 & 0.20 \\
            c5 & 18:33:33.19, $-$08:38:54.82 & 4.6$\times$2.6, 45   & 8.0   & 47 & 14.7   & 34    & 2.1 & 2.6 & 6.7 & 0.43 & 0.24 \\
            c6 & 18:33:32.40, $-$08:39:08.66 & 7.1$\times$1.5, 89   & 7.5   & 39 & 20.9   & 18    & 1.3 & 1.5 & 6.7 & 0.52 & 0.66 \\
            c7 & 18:33:33.10, $-$08:38:59.27 & $<$4.3$\times$2.9, 124 & $<$8.1 & 18 & 16.5 & 11 & $>$0.6 & $>$0.8 & $<$13.7 & 0.56 & $<$1.27 \\
            c8 & 18:33:33.52, $-$08:38:47.20 & 4.4$\times$2.3, 31   & 7.3   & 26 & 12.7   & 23    & 1.8 & 2.1 & 7.4 & 0.33 & 0.19 \\
            c9 & 18:33:31.76, $-$08:39:04.94 & 7.9$\times$2.0, 135 & 9.1   & 30 & 15.0   & 21    & 0.8 & 1.2 & 5.7 & 0.55 & 0.74 \\
          c10 & 18:33:30.05, $-$08:38:36.17 & $<$4.6$\times$3.4, 8     & $<$9.1 & 15 & 15.3 & 10 & $>$0.4 & $>$0.6 & $<$2.2 & 0.63 & $<$1.96 \\
I18310-c1 & 18:33:47.88, $-$08:23:54.46 & 8.5$\times$4.9, 96   & 15.5 & 194& 16.1 & 136 & 1.1 & 2.7 & 4.2 & 0.60 & 0.23 \\
            c2 & 18:33:48.46, $-$08:23:52.57 & 5.8$\times$3.1, 143 & 10.2 & 79  & 19.2 & 44   & 1.3 & 2.0 & 3.4 & 0.60 & 0.47 \\
            c3 & 18:33:47.79, $-$08:23:35.82 & 8.5$\times$4.1, 156 & 14.2 & 72(69)  & 33.2 & 19  & 0.2 & 0.5 & 1.1 & 0.75 & 2.32 \\
            c4 & 18:33:47.68, $-$08:23:43.38 & 4.6$\times$3.7, 175 & 9.9   & 33  & 20.2 & 17   & 0.5 & 0.8 & 3.7 & 0.53 & 0.92 \\
            c5 & 18:33:48.14, $-$08:23:41.98 & 8.3$\times$2.2, 31   & 10.3 & 42  & 18.8 & 24   & 0.7 & 1.1 & 3.6 & 0.55 & 0.72 \\
            c6 & 18:33:47.46, $-$08:23:47.69 &$<$6.7$\times$3.6, 10&$<$11.8&23&22.9& 10  &$>$0.2&$>$0.4& $<$5.8 & 0.54 & $<$1.91 \\
            c7 & 18:33:47.23, $-$08:23:42.73 & 4.9$\times$2.5, 53   & 8.4  & 23  & 23.5 & 10   & 0.5 & 0.7 & 5.2 & 0.43 & 0.90 \\
            c8 & 18:33:47.45, $-$08:23:28.38 & 5.7$\times$3.6, 162 &10.9 & 26  & 17.5 & 16   & 0.4 & 0.7 & 3.2 & 0.31 & 0.35 \\
I18337-c1 & 18:36:40.74, $-$07:39:14.17 & 8.5$\times$4.9, 165 & 12.3 & 314     & 16.0 & 139 & 2.3 & 4.4 & 2.2 & 0.44 & 0.10 \\
            c2 & 18:36:41.12, $-$07:39:25.53 & 12.2$\times$5.0, 173&14.8 & 326     & 17.6 & 127 & 1.2 & 2.7 & 4.3 & 1.09 & 0.78 \\
            c3 & 18:36:40.71, $-$07:38:58.33 & 5.6$\times$3.0, 157 & 7.8   & 99       & 23.1 &   27 & 1.7 & 2.1 & 1.5 & 0.68 & 0.76 \\
            c4 & 18:36:40.94, $-$07:39:07.04 & 7.6$\times$2.9, 145 & 8.9   & 58       & 21.0 &   18 & 0.8 & 1.1 & 2.0 & 0.61 & 1.06 \\
            c5 & 18:36:40.37, $-$07:39:18.27 &$<$7.9$\times$3.2, 176&$<$9.6& 30 & 22.0 &   9   & $>$0.3 & $>$0.5 & $<$10.8 & 0.49 & $<$1.45 \\
            c6 & 18:36:41.08, $-$07:39:40.65 &$<$5.8$\times$3.6, 30 & $<$8.7& 20 & 16.5 &   9   & $>$0.4 & $>$0.5 & $<$3.4 & 0.42 & $<$1.00 \\
I18460-c1 & 18:48:40.16, $-$03:03:53.97 & 5.0$\times$2.3, 96  & 8.1 & 79 & 27.0   & 28 & 1.6 & 2.0 & 0.3 & 0.81 & 1.07 \\
            c2 & 18:48:39.27, $-$03:03:55.21 & 1.6$\times$1.0, 35  & 3.0 & 31 & [30]    & 10 &11.0& 5.1 & \nodata & (0.89) & 1.39 \\
            c3 & 18:48:39.72, $-$03:04:04.57 & 5.0$\times$3.2, 82  & 9.6 & 48 & 18.4   & 28 & 1.0 & 1.4 & 1.7 & 0.33 & 0.21 \\
            c4 & 18:48:39.37, $-$03:04:07.68 & 3.9$\times$2.4, 170& 7.3 & 19 & 17.6   & 12 & 0.9 & 1.0 & 2.8 & 0.36 & 0.45 \\
            c5 & 18:48:37.59, $-$03:03:46.98 & 6.6$\times$6.0, 99  &15.1& 27 & 14.6   & 22 & 0.2 & 0.5 & 5.2 & 0.43 & 0.71 \\
I18530-c1 & 18:55:33.69, $+$02:19:06.74 & 4.3$\times$2.9, 84  & 8.1   & 273(271)   & 25.7 & 95 & 5.4 & 6.8 & 0.7 & 0.74 & 0.26 \\
            c2 & 18:55:33.99, $+$02:19:10.56 & 5.0$\times$4.0, 49  & 10.3 & 307(210)   & 42.7 & 41 & 1.1 & 1.8 & 0.3 & 0.73 & 0.77 \\
            c3 & 18:55:34.33, $+$02:19:12.67 & 4.7$\times$1.5, 130& 6.1   & 62(26)       &[30.0]& 8 & 1.0 & 1.0 & \nodata & 0.87 & 3.48 \\
            c4 & 18:55:34.13, $+$02:19:06.12 & $<$4.0$\times$2.9, 69&$<$7.8&32(19)  & 27.5 & 6 & $>$0.4 & $>$0.5 & $<$1.4 & 0.48 & $<$1.64 \\
            c5 & 18:55:32.99, $+$02:19:02.69 & 6.1$\times$2.5, 63  & 9.0   & 61             & 15.2 & 42 & 1.8 & 2.5 & 7.2 & 0.44 & 0.23 \\
            c6 & 18:55:34.46, $+$02:19:07.33 & 4.6$\times$3.5, 60  & 9.2   & 39(28)       & 27.9 & 9 & 0.3 & 0.5 & 1.5 & 0.34 & 0.69 \\
            c7 & 18:55:31.31, $+$02:18:49.49 & 6.8$\times$1.6, 76  & 7.6   & 36             & 10.8 & 42 & 2.9 & 3.4 & 1.0 & 0.37 & 0.14 \\
I19074-c1 & 19:09:53.55, $+$07:57:14.87 & 4.3$\times$3.4, 63  & 7.3   & 114(107)  & 20.6 & 34 & 2.7 & 3.0 & 0.4 & 0.75 & 0.69 \\
            c2 & 19:09:53.68, $+$07:56:53.96 & 5.5$\times$3.1, 145 & 7.8  & 49            & 14.5 & 25 & 1.6  & 1.9 & 1.3 & 0.33 & 0.19 \\
            c3 & 19:09:53.33, $+$07:57:10.47 & 4.4$\times$1.6, 146 & 5.0  & 24            & 19.6 & 8   & 1.9  & 1.5 & 1.8 & 0.50 & 0.89 \\
            c4 & 19:09:54.48, $+$07:57:39.24 & 3.1$\times$0.3, 165 & 1.8  & 16            & 20.2 & 5 & 26.0& 7.3 & 0.3 & 0.30 & 0.17 \\
            c5 & 19:09:54.43, $+$07:57:23.58 & 4.5$\times$3.3, 58   & 7.3  & 23(22)     & 16.5 & 9 & 0.7 & 0.8 & 1.8 & 0.43 & 0.83 \\
I19220-c1 & 19:24:19.40, $+$14:38:07.44 & 8.9$\times$5.3, 115  & 18.5 & 134(129)  & 16.9 & 107 & 0.5 & 1.5 & 0.5 & 0.88 & 0.76 \\
            c2 & 19:24:20.36, $+$14:38:02.68 & 11.8$\times$6.2, 86  & 23.1 & 176(172)  & 17.1 & 140 & 0.3 & 1.2 & 1.2 & 0.54 & 0.27 \\
            c3 & 19:24:18.88, $+$14:38:06.05 & 7.8$\times$3.5, 148  & 14.1 & 57    & 20.2 & 37   & 0.4 & 0.9 & 0.5 & 0.70 & 1.03 \\
            c4 & 19:24:22.56, $+$14:38:20.65 & 6.9$\times$3.3, 62    & 12.9 & 32    & 14.0 & 34   & 0.5 & 1.0 & 0.9 & 0.36 & 0.27 \\
            c5 & 19:24:21.32, $+$14:38:02.92 & 10.8$\times$4.8, 112&19.4& 43      & 13.3 & 50   & 0.2 & 0.6 & 4.6 & 0.55 & 0.66 \\
I19368-c1 & 19:38:57.47, $+$22:46:35.85 & 7.6$\times$3.3, 69  & 11.0 & 213    & 20.6 & 90 & 2.0 & 3.5 & 1.1 & 0.82 & 0.46 \\
            c2 & 19:38:56.76, $+$22:46:31.72 & 4.7$\times$3.3, 62  & 8.7   & 130    & 23.1 & 48 & 2.2 & 3.0 & 1.2 & 0.93 & 0.88 \\
            c3 & 19:38:58.06, $+$22:46:38.29 & 7.5$\times$1.6, 40  & 7.6   & 81      & 17.4 & 43 & 2.9 & 3.5 & 2.7 & 0.54 & 0.29 \\
            c4 & 19:38:58.88, $+$22:46:45.38 & 9.8$\times$3.6, 64  & 13.1 & 68      & 16.4 & 39 & 0.5 & 1.1 & 2.4 & 0.70 & 0.92 \\
            c5 & 19:38:59.23, $+$22:46:53.71 & 7.6$\times$5.1, 45  & 13.7 & 62      & 16.8 & 34 & 0.4 & 0.9 & 2.3 & 0.86 & 1.66 \\
\enddata
\tablenotetext{a}{Temperatures in square brackets were adopted rather than calculated because \amm{} lines are not detected.}
\tablecomments{Fluxes in parentheses are free-free emission subtracted, based on which corresponding dense cores masses and gas densities are derived. Upper limits are given for deconvolved sizes and equivalent radii of unresolved cores, hence corresponding densities are lower limits and \amm{} abundances and virial parameters are upper limits.}
\end{deluxetable}

\begin{deluxetable*}{lccccccccccccccc}
\tabletypesize{\scriptsize}
\tablecaption{Properties of SiO outflows.\label{tab:outflow}}
\tablewidth{0pt}
\tablehead{
\multirow{2}{*}{Parameters} & \multicolumn{2}{c}{I18308-c1} & \multicolumn{2}{c}{I18308-c4} & \multicolumn{2}{c}{I18310-c1} & \multicolumn{2}{c}{I18337-c2} & \multicolumn{2}{c}{I18337-c3} & I18337-c6 & \multicolumn{2}{c}{I18530-c1} & \multicolumn{2}{c}{I19368-c1}\\
 & Blue & Red & Blue & Red & Blue & Red & Blue & Red & Red & Blue & Red & Blue & Red & Blue & Red
 }
\startdata
Velocity range (\kms{})         & [45,75] & [77,107] & [45,75] & [77,107] & [63,83] & [85,105] & [31,57] & [59,85] & [31,59] & [60,85] & [59,85] & [55,76] & [77,97] & [25,35] & [37,48]\\
Terminal velocity (\kms{})     & 27 & 30 & 32 & 16 & 16 & 5 & 28 &18 & 29 & 25 & 25 & 9 & 3 & 7 & 11 \\
Excitation temperature (K)   & 22.7 & 22.7 & 16.6 & 16.6 & 16.1 & 16.1 & 17.6 & 17.6 & 23.1 & 23.1 & 16.5 & 25.7 & 25.7 & 20.6 & 20.6 \\
Mass (\msol{})                      & 3.5 & 7.4 & 0.6 & 0.7 & 0.7 & 0.1 & 1.0 & 4.9 & 4.9 & 10.4 & 2.1 & 1.0 & 0.8 & 1.1 & 1.2 \\
Momentum (\msol{}\,\kms{}) & 32 & 78 & 6 & 2 & 8 & 0.2 & 9 & 37 & 56 & 66 & 13 & 5 & 1 & 2 & 3 \\
Kinetic energy (\msol\,km$^2$\,s$^{-2}$) & 224 & 612 & 58 & 7 & 50 & 0.4 & 67 & 181 & 427 & 305 & 63 & 16 & 2 & 2 & 9 \\
Projected lobe length (pc)    & 0.16 & 0.11 & 0.09 & 0.16 & 0.16 & 0.09 & 0.20 & 0.18 & 0.18 & 0.26 & 0.29 & 0.11 & 0.09 & 0.16 & 0.17 \\
Dynamic age (10$^3$ yr)     & 5.8 & 3.7 & 2.8 & 9.8 & 9.9 & 19.0 & 7.2 & 9.3 & 6.3 & 10.0 & 11.7 & 12.5 & 25.7 & 21.1 & 14.8 \\
Outflow rate (10$^{-4}$ \msol{}\,yr$^{-1}$) & 6.0 & 20.1 & 2.0 & 0.7 & 0.7 & 0.1 & 1.3 & 5.2 & 5.5 & 7.3 & 1.4 & 0.8 & 0.3 & 0.5 & 0.8 \\
\enddata
\end{deluxetable*}

\clearpage

\startlongtable
\begin{deluxetable*}{rccccccc}
\tabletypesize{\scriptsize}
\tablecaption{Star formation indicators in dense cores. \label{tab:sf}}
\tablewidth{0pt}
\tablehead{
Core & $f$-$f$ emission & Outflows & Hot core lines & 4.5~\micron{} & 24~\micron{} & 70~\micron{} & Evolutionary phase
}
\startdata
I18308-c1 & Yes & SiO, \water{} maser & Yes & bright & bright & bright & (UC) \hii{} \\
            c2 & \nodata  & \nodata & \nodata & bright & bright & bright & Protostellar \\
            c3 & \nodata  & \nodata & \nodata & dark & bright & bright & Protostellar \\
            c4 & \nodata  & SiO & \nodata & dark & bright & bright & Protostellar \\
            c5 & \nodata  & \nodata & \nodata & dark & dark & dark & Prestellar \\
            c6 & \nodata  & \nodata & \nodata & dark & bright & bright & Protostellar \\
            c7 & \nodata  & \nodata & \nodata & dark & bright & bright & Protostellar \\
            c8 & \nodata  & \nodata & \nodata & dark & dark & dark & Prestellar \\
            c9 & \nodata  & \nodata & \nodata & dark & dark & bright & Protostellar \\
I18310-c1 & \nodata  & SiO & \nodata & dark & dark & bright & Protostellar \\
            c2 & \nodata  & \nodata & \nodata & bright & bright & bright & Protostellar \\
            c3 & Yes & \nodata & \nodata & bright & bright & bright & (UC) \hii{} \\
            c4 & \nodata  & \nodata & \nodata & dark & bright & bright & Protostellar \\
            c5 & \nodata  & \nodata & \nodata & dark & bright & bright & Protostellar \\
            c6 & \nodata  & \nodata & \nodata & bright & bright & bright & Protostellar \\
            c7 & \nodata  & \nodata & \nodata & dark  & bright & bright & Protostellar \\
            c8 & \nodata  & \nodata & \nodata & dark  & bright & bright & Protostellar \\
I18337-c1 & \nodata  & \nodata & Yes & bright & bright & bright & Protostellar \\
            c2 &  \nodata  & SiO &      Yes & dark   & bright & bright & Protostellar \\
            c3 &  \nodata  & SiO &      Yes & dark   & bright & bright & Protostellar \\
            c4 &  \nodata  & \nodata & \nodata & bright & bright & bright & Protostellar \\
            c5 & \nodata  & \nodata & \nodata &  bright & bright & bright & Protostellar \\
            c6 & \nodata  & SiO & \nodata       &  dark   & dark & dark & Protostellar \\
I18460-c1 & \nodata  & \nodata & \nodata & bright & bright & bright & Protostellar \\
            c2 & \nodata  & \nodata & Yes       & bright & bright & bright & Protostellar \\
            c3 & \nodata  & \nodata & \nodata & dark & bright & bright & Protostellar \\
            c4 & \nodata  & \nodata & \nodata & dark & bright & bright & Protostellar \\
            c5 & \nodata  & \nodata & \nodata & dark & dark &  dark & Prestellar \\
I18530-c1 & Yes  & SiO        &      Yes & dark &  bright &  bright & (UC) \hii{} \\
            c2 & Yes  & \nodata       & Yes & bright &  bright & bright & (UC) \hii{} \\
            c3 & Yes  & \nodata & \nodata & bright &  bright &  bright & (UC) \hii{} \\
            c4 & Yes  & \nodata & \nodata & bright &  bright &  bright & (UC) \hii{} \\
            c5 & \nodata  & \nodata & \nodata & dark &  bright & bright & Protostellar \\
            c6 & Yes  & \nodata & \nodata & bright &  bright &  bright & (UC) \hii{} \\
            c7 &  \nodata  & \nodata & \nodata &dark & dark &  dark & Prestellar \\
I19074-c1 & Yes  & \nodata & Yes             & bright & bright &  bright & (UC) \hii{} \\
            c2 &  \nodata  & \twelveco{}? & \nodata & dark & dark & dark & Protostellar \\
            c3 & \nodata  & \nodata & \nodata & dark   & bright & bright & Protostellar \\
            c4 & \nodata  & \nodata & \nodata & bright & bright & bright & Protostellar \\
            c5 & Yes  & \nodata & \nodata       & dark   & bright & bright & (UC) \hii{} \\
I19220-c1 & Yes  & \nodata & \nodata       & bright & bright & bright & (UC) \hii{} \\
            c2 & Yes  & \nodata & \nodata       & bright & bright & bright & (UC) \hii{} \\
            c3 & \nodata  & \nodata & \nodata & dark &   bright & bright & Protostellar \\
            c4 & \nodata  & \nodata & \nodata & dark &   dark & dark & Prestellar \\
            c5 & \nodata  & \nodata & \nodata & bright & bright & bright & Protostellar \\
I19368-c1 & \nodata  & SiO &             Yes & bright & bright &  bright & Protostellar \\
            c2 & \nodata  & \nodata &       Yes & bright & bright &  bright & Protostellar \\
            c3 & \nodata  & \nodata & \nodata & bright & bright &  bright & Protostellar \\
            c4 & \nodata  & \nodata & \nodata & bright & bright &  bright & Protostellar \\
            c5 & \nodata  & \nodata & \nodata & bright & bright &  bright & Protostellar \\
\enddata
\end{deluxetable*}

\begin{deluxetable*}{cccccc}[!h]
\tabletypesize{\scriptsize}
\tablecaption{Statistical properties of dense cores.\label{tab:cores_stats}}
\tablewidth{0pt}
\tablehead{
 & $\sigma_\text{in}$ & $T_\text{rot}$ & $M_\text{core}$ & $X$(\amm{}) & $\alpha$ \\
 & (\kms{}) & (K) & (\msol) & (10$^{-8}$) & 
}
\startdata
Prestellar mean       & 0.38 & 13.4 & 31.0 & 4.2 & 0.31 \\
Prestellar median    & 0.37 & 14.0 & 34.2 & 5.2 & 0.24 \\
Protostellar mean    & 0.58 & 19.2 & 37.8 & 3.6 & 0.73 \\
Protostellar median & 0.55 & 18.7 & 27.5 & 3.0 & 0.73 \\
(UC) \hii{} mean        & 0.66 & 25.5 & 55.0 & 1.0 & 1.08 \\
(UC) \hii{} median     & 0.73 & 25.7 & 33.8 & 1.1 & 0.76 \\
Overall mean           & 0.58 & 20.0 & 40.9 & 3.1 & 0.78 \\
Overall median        & 0.55 & 18.7 & 28.2 & 2.1 & 0.70 \\
\enddata
\end{deluxetable*}

\begin{deluxetable*}{ccccccccccccccc}[!h]
\tabletypesize{\scriptsize}
\tablecaption{Properties of the 4 well-defined filaments.\label{tab:filaments}}
\tablewidth{0pt}
\tablehead{
Filament & Device/Wavelength & Flux\tablenotemark{a} & $\kappa_\nu$ & $T_\text{rot}$ & $c_s$ & $\sigma_\text{in}$ & Mass & Length & ($M/l$)$_\text{obs}$ & $\lambda_\text{prj}$ & ($M/l$)$_\text{crit,th}$ & $\lambda_\text{cl,th}$ & ($M/l$)$_\text{crit,turb}$ & $\lambda_\text{cl,turb}$ \\
& & (Jy) & (\sqc{}\,g$^{-1}$) & (K) & (\kms{}) & (\kms{}) & (\msol) & (pc) & (\msol{}\,pc$^{-1}$) & (pc) & (\msol{}\,pc$^{-1}$) & (pc) & (\msol{}\,pc$^{-1}$) & (pc)
}
\startdata
I18308 & SMA/1.3~mm      & 0.63 & 0.899 & 16.1 & 0.24 & 0.48 & 404   & 1.2 & 342  & 0.19 & 26 & 0.07 & 109 & 0.13 \\
 & MAMBO/1.2~mm           & 1.99 & 1.06 & \textquotedbl  & \textquotedbl & \textquotedbl  & 917 & 1.5 & 605 & \textquotedbl & \textquotedbl & \textquotedbl & \textquotedbl & \textquotedbl \\
I18337 & SMA/1.3~mm      & 0.79 & 0.899 & 17.3 & 0.25 & 0.75 & 314 & 1.0 & 315  & 0.21 & 28 & 0.07 & 263 & 0.21 \\
 & MAMBO/1.2~mm           & 4.35 & 1.06 & \textquotedbl & \textquotedbl & \textquotedbl & 1237 & 1.4 & 908  & \textquotedbl & \textquotedbl & \textquotedbl & \textquotedbl & \textquotedbl \\
I19220 & SMA/1.3~mm       & 0.39 & 0.899 & 14.7 & 0.23 & 0.44 & 391 & 2.0 & 199  & 0.40 & 24 & 0.06 & 91 & 0.12 \\
 & MAMBO/1.2~mm           & 1.82 & 1.06 & \textquotedbl & \textquotedbl & \textquotedbl & 1318 & 2.0 & 654 & \textquotedbl & \textquotedbl & \textquotedbl & \textquotedbl & \textquotedbl \\
I19368 & SMA/1.3~mm      & 0.49 & 0.899 & 17.0  & 0.24 & 0.60 & 267 & 1.2 & 228 & 0.23 & 28 & 0.07 & 169 & 0.17 \\
 & SCUBA2/850~\micron{} & 6.81 & 1.97 & \textquotedbl & \textquotedbl & \textquotedbl & 845  & 2.0 & 417  & \textquotedbl &\textquotedbl & \textquotedbl & \textquotedbl & \textquotedbl \\
\enddata
\tablenotetext{a}{References for bolometer array data: MAMBO/1.2~mm -- \citet{beuther2002mambo}; SCUBA2/850~\micron{} -- \citet{eden2017}.}
\end{deluxetable*}

\end{CJK}
\end{document}